\documentstyle[12pt,amssymbols]{report}
\def\ie{\mbox{{\it i.e.} }}

\def\eg{\mbox{{\it e.g.} }}
\def\cf{\mbox{{\it cf.} }}

\def\tr{\mbox{tr}}
\def\id{\mathop{\rm id}}
\def\real{{\Bbb R}}
\def\integer{{\Bbb Z}}
\def\R{{\cal R}}
\def\Rhat{\hat{R}}
\def\matrix#1#2#3{#1 ^{#2}{}_{#3}}
\def\lie{\mbox{\boldmath $\cal L$}}
\def\partder#1#2{{\partial #1\over\partial #2}}
\def\bra#1{\left\langle #1\right|}
\def\ket#1{\left| #1\right\rangle}
\def\bracket#1#2{\left\langle #1 | #2 \right\rangle}
\def\half{{1\over 2}}
\def\inv#1{{1\over #1}}
\def\comm#1#2{\left[#1, #2\right]}
\def\acomm#1#2{\left\{#1, #2\right\} }
\def\gcomm#1#2{\left[#1, #2\right]_{\pm}}
\def\inprod#1#2{\left\langle #1, #2\right\rangle}
\def\complex{{\Bbb C}}
\def\trg{{\triangleright}}
\def\tlg{{\triangleleft}}
\def\A{{\cal A}}
\def\B{{\cal B}}
\def\C{{\cal C}}
\def\D#1{{{\cal D}}^#1}
\def\F{{\cal F}}
\def\V{{\cal V}}
\def\G{{\Gamma (\A)}}
\def\UDE{{\Omega (\A)}}
\def\slR{\Re}
\def\M{{\cal M}}
\def\N{{\cal N}_{\M}}
\def\U{{\cal U}}
\def\T{{\cal T}_{\M}}
\def\K{{\cal K}}
\def\KM{\K / \M}
\def\GN{\Gamma_{\M}}
\def\GDE{\Omega_{\M}}
\def\IA{{1_{\A}}}
\def\IU{{1_{\U}}}
\def\Ik{{1_k}}
\def\IS{{1_{\smash}}}
\def\DA{{\Delta_{\A}}}
\def\AD{{{}_{\A}\Delta}}
\def\I{{\mbox{\boldmath $i$}}}
\def\dinprod#1#2{\left\langle \inprod{#1}{#2} \right\rangle}
\def\dual#1#2{\left( #1,#2\right)}
\def\bid{\mbox{\bf id}}
\def\dg{{\mbox{\boldmath $\delta$}}}
\def\ad{{\stackrel{\mbox{\scriptsize ad}}{\triangleright}}}
\def\tad{{\stackrel{\mbox{\tiny ad}}{\triangleright}}}
\def\Ad{{\Delta^{\mbox{\scriptsize Ad}}}}
\def\killing#1#2#3{\eta^{(#1)}( #2 \otimes #3)}
\def\metric#1#2{\eta^{(#1)}_{#2}}
\def\invmet#1#2{\eta^{(#1)\,{#2}}}
\def\trace#1#2{\tr_{#1}\left( u #2 \right)}
\def\g{{\mbox{$\bf g$}}}
\def\O#1#2{{{\cal O}}_{#1}{}^{#2}}
\def\bigR#1#2{\matrix{\hat{\real}}{#1}{#2}}
\def\bigA#1#2{\matrix{{\Bbb A}}{#1}{#2}}
\def\struc#1#2{f_{#1}{}^{#2}}
\def\gen#1{\chi_{#1}}
\def\det#1{\mbox{det}_q #1}
\def\Det#1{\mbox{Det} #1}
\def\smash{\A\rtimes\U}
\def\J{{\cal J}}
\def\Irep#1#2{{{\cal I}}^{({#1})}_{#2}}
\def\H{\mbox{\boldmath $H_0$}}
\def\Y{\mbox{\boldmath ${\cal Y}$}}
\def\Z{\mbox{\boldmath ${\cal Z}$}}
\def\quint#1#2{\left[ #1 \right] _{#2}}
\def\vaca{\Omega_{\A}}
\def\vacu{\Omega_{\U}}
\def\germ{\stackrel{S}{\cong}}
\def\Fun{\mbox{Fun}}
\def\Chi{{\cal X}}
\def\cas#1{Q_2^{( #1 )}}

\begin{document}
\begin{titlepage}
\begin{center}
December 15, 1994\hfill LBL-36537\\
\hfill UCB-PTH-94/35\\
\vskip .2in
{\large \bf Differential Geometry on Hopf Algebras and Quantum
Groups}\\
(Ph.D.\ Thesis)\footnote{This work was supported in part by the
Director, Office of Energy Research, Office of High Energy and Nuclear
Physics, Division of High Energy Physics of the U.S.  Department of
Energy under Contract DE-AC03-76SF00098 and in part by the National
Science Foundation under grant PHY-90-21139.}
\vskip .2in
Paul Watts\footnote{e-mail: watts@cpt.univ-mrs.fr}
\vskip .2in
{\it Department of Physics\\
University of California\\
and\\
Theoretical Physics Group\\
Lawrence Berkeley Laboratory\\
University of California\\
Berkeley, CA 94720}
\vskip .5in
{\bf Abstract}
\end{center}

\noindent The differential geometry on a Hopf algebra is constructed,
by using the basic axioms of Hopf algebras and noncommutative
differential geometry.  The space of generalized derivations on a Hopf
algebra of functions is presented via the smash product, and used to
define and discuss quantum Lie algebras and their properties.  The
Cartan calculus of the exterior derivative, Lie derivative, and inner
derivation is found for both the universal and general differential
calculi of an arbitrary Hopf algebra, and, by restricting to the
quasitriangular case and using the numerical R-matrix formalism, the
aforementioned structures for quantum groups are determined.

\end{titlepage}
\pagenumbering{roman}
\setcounter{page}{1}
\tableofcontents

\chapter*{Acknowledgements}

First and foremost, I wish to thank my research advisor Bruno Zumino,
whose guidance and encouragement has been invaluable to me; it is
literally impossible to thank him enough for all his help, and this
work could not exist without his many contributions.  I must also
recognize my fellow partners-in-crime Peter Schupp and Chryss
Chryssomalakos, without whom I would be considerably poorer in both
friendship and knowledge.

There are naturally many others who I must thank; among my fellow
physics grad students here at Berkeley, I have benefitted immeasurably
from conversations with Wati Taylor and Scott Hotes.  Rich Lebed
certainly deserves special mention for being able to share an office
with my whining and pedantry for several years, and Steve Johnson and
Peter Grudberg should get medals for being able to live with me for
what to them probably seems like an eternity.  Among the postdocs,
Markus Luty, Michael Schlieker and Joanne Cohn, as well as Markus
Pflaum in the Math Department, all deserve my gratitude for their
friendship and encouragement.

I have gained quite a lot from my interactions with many of the
faculty here at Cal, but I would particularly like to thank Orlando
Alvarez for numerous fascinating discussions, and to Korkut
Bardak\c{c}i and Nicolai Reshetikhin for agreeing to read and approve
this work.

There is no doubt that I have survived grad school largely due to the
diligence and assistance of the LBL Theory Group staff, Betty Moura
and Luanne Neumann, as well as their counterparts on campus, Anne
Takizawa and Donna Sakima.  They have been infinitely helpful in their
efforts to make the LBL and UCB bureaucracies (almost) tolerable.

Last, but by no stretch of the imagination least, I must thank my
parents Jean and Bob Watts, for being very atypical military parents;
their willingness to let me follow my own muse and their support of
the path I have chosen has given me the utmost respect for them.  They
did a damn fine job with me and my siblings.

This work was supported in part by the Director, Office of Energy
Research, Office of High Energy and Nuclear Physics, Division of High
Energy Physics of the U.S. Department of Energy under Contract
DE-AC03-76SF00098 and in part by the National Science Foundation under
grant PHY-90-21139.

\newpage
\pagenumbering{arabic}
\setcounter{page}{1}
\chapter{Introduction}

\section{A Brief Outline}

In the present chapter, we hope to present to the reader the
motivation behind this work, and what we hope to accomplish herein.
The next two chapters, ``Hopf Algebras'' and ``Actions, the Smash
Product, and Coactions'', are largely introductory, and serve to
establish the language, notations, and methods which we use
throughout; once we have these foundations, we are able to build upon
them in the subsequent chapters to obtain the more advanced and
original results which form the core of this work.  Chapter 4,
``Quantum Lie Algebras'', deals with a particular class of Hopf
algebras and examines their structures, with an emphasis finding the
deformed analogues of classical concepts (\eg the Killing metric).
The following chapter, ``Cartan Calculus on Hopf Algebras and Quantum
Lie Algebras'' examines in depth the differential geometry of the
titular objects, by introducing an algebra of generalized derivations,
and using many of the concepts of noncommutative geometry to examine
the structure of this algebra.  In Chapter 6, ``The Linear Lie Groups
$GL_q(N)$ and $SL_q(N)$'', we apply the results of the preceding
chapters to these two specific cases.  Finally, in ``Conclusions'', we
mention what other problems might be constructively pursued using the
ideas here presented, and what difficulties might arise in so doing.

There are also three appendices, ``Numerical R-Matrix Relations'',
``Classical Differential Geometry'', and ``Differential Calculus on
Hopf Algebras''; these cover topics which the reader may already be
familiar with, but which serve as introductions to the relevant
material if he/she is not.

\section{Why Quantum Groups?}

The reader may wonder why any of the material contained herein would
be of any interest or use to a physicist.  In the following
subsections, we hope to give some possible instances where the results
of this work might prove to be useful.

\subsection{Mathematical and Physical Motivations}

Perhaps most importantly, Hopf algebras (and quantum groups, which are
specific types of Hopf algebras) provide us with a generalization of
many of the algebras which are common in physics.  (For instance, as
we describe in one of the examples in Chapter \ref{chap-Hopf}, any Lie
algebra is actually a Hopf algebra, albeit with a somewhat trivial
structure in many ways.)  Furthermore, they also allow us to
generalize many ``classical'' physical ideas to ``deformed'' versions
in a completely self-consistent manner.  The new versions are most
often specified by one or more parameters, and the classical case is
recovered by setting these parameters to some fixed values.  In this
sense, the situation is much like quantum mechanics as a ``deformed''
version of classical mechanics described by Planck's constant $h$,
with the latter being recovered in the $h\rightarrow 0$ limit (in
fact, this was the motivation for the term ``quantum group'').  The
reader will encounter many of these generalizations throughout this
work, such as the quantum group $SL_q(N)$, which is a ``deformed''
version of the ``classical'' $SL(N)$, which is recovered in the limit
$q\rightarrow 1$.

The language of Hopf algebras also gives us a way of ``rephrasing''
many of the ideas and concepts which are used extensively in physics
in more mathematical terms.  In many cases, we feel that this not only
eases computations, but also provides some insight into a more general
structure of the problem in question.  For instance, in Chapter
\ref{chap-coactions}, we describe how the common physical concepts of
finite and infinitesimal transformations may be recast in terms of
``coactions'' and ``actions'' of a Hopf algebra on a vector space.  We
also show how the commutation relations between differential operators
and functions, \ie how the former ``act on'' and ``move through'' the
latter, also have a straightforward mathematical interpretation in
terms of the so-called ``smash product''.

\subsection{Quantum Group Symmetry in Field Theory}

In physics, we often encounter systems which have certain global
symmetries, and when we formulate a way of describing these systems,
our formulation must respect these symmetries.  However, we might
consider a theory in which one of the symmetry groups is not a {\em
classical} group, but rather a {\em quantum} group.  Such a theory
will contain the classical theory as a specific case, of course, but
will have more degrees of freedom to play with, namely the
parameter(s) of deformation which characterize the quantum group.

The first thing we could do with a theory like this is to try to
measure these extra parameters experimentally.  Suppose we have a
lagrangian which is a scalar under transformations in some quantum
group, \eg $SU_q(2)$.  We could use standard techniques to calculate
various physical processes, such as scattering amplitudes or decay
rates; these would then have some dependence on the parameter $q$, and
by actually running experiments which would test the predictions of
this theory, its value could be determined.  In this sense, it would
be just another constant in the theory.

However, consider the case where we believe the lagrangean to be
invariant under the action of the {\em classical} group being
considered (\eg $SU(2)$).  Then the interpretation of the quantum
symmetry would be as a method of explicitly breaking the symmetry of
the lagrangean, with the degree of breaking parametrized by $q-1$.
This quantity would then serve as a measure of how much the actual
symmetry deviates from the expected symmetry (in this sense, it is
much like the parameter $\epsilon$ which characterizes CP-violation).

\subsection{Lattices and Regularization}

Consider a space whose points are described by some set of
coordinates, and a set of coordinate transformations described by a
group of matrices; we normally assume that these coordinates commute
with each other, and therefore, if the transformations are to respect
this commutativity, the entries of the transformation matrices will
commute as well.  However, if we now want to generalize to the case
that the matrices may be representations of a quantum group, their
entries will in general no longer be commutative, and therefore the
coordinates on the space will not remain so either.  It is no surprise
that the derivatives with respect to these coordinates also no longer
commute.

However, what may be surprising is that when one of these differential
operators acts on a function (\ie a sum of ordered monomials), the
result is not a normal derivative, but rather a {\em finite}
difference between the function evaluated at two discrete points.
(See Chapter \ref{chap-coactions-euclidean}, as well as \cite{WZ}, for
an explicit example of this.)  Therefore, requiring that the
transformations on the coordinates be given by a quantum group rather
than a classical one also implies that the space itself can be thought
of as a lattice of discrete points, with the spacing depending on the
deformation parameter(s) characterizing the quantum group.

One of the uses of putting any field theory on a lattice is that the
lattice spacing $a$ becomes a parameter which incorporates the
small-scale behavior of the theory.  Since this behavior is often
responsible for divergences which arise, $a$ is often used as a
regularization parameter: explicit $a$-dependent counterterms are put
into the theory so as to cancel any infinities in the $a\rightarrow 0$
limit.  We see that we have an exact analogue of this; a lattice which
in a certain limit ($a\rightarrow 0$ versus quantum group
$\rightarrow$ classical group) becomes ``real space'' (\ie a
continuous space for lattices, commuting coordinates for quantum
groups).  Therefore, deforming a classical coordinate transformation
group provides a natural lattice with which to work, and therefore
perhaps a way of handling small-scale behavior of a theory.

\subsection{Quantum Gravity}

An intriguing possibility is that noncommutive algebras like the ones
we consider in this work may hold the key to dealing with the age-old
(well, decades-old) problem of how to incorporate gravity into a
quantum field theory.  There are two main reasons for thinking this to
be the case:

First, an obstacle in quantizing gravity has always been that all our
previous field theories place space-time on a special footing.
Generally, when we consider a field theory, we introduce an algebra of
objects with given commutation relations, and a Hilbert space of
states on which they have an action.  The operators themselves are
taken to live in various representation of some particluar set of
symmetry groups, one of which is the group of diffeomorphisms on the
underlying manifold which describes the space-time.  Therefore, we
talk about scalar particles, vector particles, {\it etc.}, depending
on how they transform under a given coordinate transformation.
However, this formulation gives the space-time of the theory the mere
status of an index space labelling the operators.

This is a big problem if we now want to consider the geometry of
space-time as itself described by an operator in the algebra, because
we can no longer use the points of the manifold to label the fields.
Therefore, it would be nice to eliminate {\em any} reference to the
underlying space-time, and describe all fields as algebraic objects.
There does in fact seem to be a way to do this; a theorem of Gel'fand
states (roughly) that an associative algebra with unit is isomorphic
to the algebra of functions over some topological space.  This is
precisely what we want, since it allows us to switch topological
considerations like general coordinate invariance into purely
algebraic language; we merely specify all the commutation relations of
our field operators and how they act on the Hilbert space, and
determine what space-times have this as a function algebra.  Since the
operators will in general not commute, this obviously allows us to use
the ideas and techniques of noncommutative geometry.

The second reason that the noncommutative structures we consider may
be relevant to quantizing gravity stems from the discussion on
lattices in the previous subsection, namely, allowing for noncommuting
coordinates on the space in question in effect discretizes it.  There
have always been problems with the nonrenormalizability of gravity due
to Planck-scale effects; as soon as one starts to consider distances
less than this, all hell breaks loose, and the theory becomes
divergent.  One solution to this would be to propose that space-time
itself is discretized, with the distance between points being around
the Planck length.  We would want some sort of mechanism which would
explain such a structure rather than taking it as an {\it ad hoc}
assumption, and the noncommutativity of coordinates does this very
naturally.

\section{Our Approach}

The main emphasis of this work will be on developing a constructive
method for introducing differential geometric structures on Hopf
algebras using, at first, {\em only} the basic axioms of Hopf
algebras.  This implies that we do not {\it \`a priori} assume any
particular multiplicative structure on the Hopf algebras in question,
and therefore use the techniques developed by Connes \cite{Connes} for
constructing the universal differential calculus of a unital
associative algebra.  However, since we consider Hopf algebras, we
have additional structure, and may use the results of Woronowicz
\cite{Woronowicz1} in dealing with the differential geometry.
However, when we consider the physically interesting case where the
Hopf algebra is actually a quantum group with a given numerical
R-matrix, there are, in fact, given commutation relations, and the
work of Reshetikhin, Takhtadzhyan, and Faddeev \cite{RTF} becomes
extremely useful.

The reason for taking this approach lies in the eventual goal, namely,
the formulation of a quantum field theory with a quantum group as a
gauge symmetry.  Since gauge theories are equivalent to looking at a
fiber bundle whose connection is the gauge field and where matter
fields are merely sections, finding a way of deforming said bundles
seems like the most promising way to specify a deformed gauge theory.
Since a connection is a 1-form over a bundle, and actions are by
definition intergals of forms over the base space of the bundle, it
becomes paramount to analyze the differential geometry before
proceeding.  In particular, by using sheaf theoretic techniques in
which the topological properties of the manifolds involved become
algebraic properties of their function spaces (\cf Appendix
\ref{chap-classical} and references therein), we find that the
structure group is described by a quantum group, so determining the
corresponding Lie algebra and Cartan calculus is necessary for any
discussion involving connections.  We hope that the results we obtain
will further us toward this goal.

\chapter{Hopf Algebras}\label{chap-Hopf}

In this chapter, we look at many of the basic properties of Hopf
algebras and quantum groups.  This is not only to introduce the
concepts needed in the manipulations of these mathematical objects,
but to also establish much of the notation which will appear
throughout this work.  (For the interested reader, much more
information about Hopf algebras and their properties is readily
available in \cite{Sweedler,Abe,Majid1}.)

\section{Basic Definitions}\label{chap-Hopf-defs}

An {\em algebra} is a vector space $\A$ over a field $k$ such that the
algebra multiplication $m:\A\otimes \A \rightarrow \A$ is a bilinear
map satisfying
\begin{eqnarray}
m (a \otimes (b+c))&=&m (a \otimes b)+m (a \otimes c),\nonumber\\
m ((a+b) \otimes c)&=&m (a \otimes c)+m (b \otimes c),
\end{eqnarray}
for all $a,b,c \in \A$.  (In general, we will suppress $m$ for
purposes of brevity, writing $ab$ instead of $m(a \otimes b)$.)

A {\em unital algebra} is an algebra which contains an element
$\IA$ having the properties
\begin{equation}
\IA a=a\IA=a.
\end{equation}

An {\em associative algebra} is an algebra in which $m$
satisfies the further condition
\begin{equation}
(ab)c=a(bc).
\end{equation}

A {\em coalgebra} is a vector space $\A$ over a field $k$, together
with linear maps $\Delta:\A \rightarrow \A\otimes \A$ and $\epsilon
:\A\rightarrow k$ (the coproduct and counit, respectively) which
satisfy
\begin{eqnarray}
(\Delta \otimes \id)\Delta (a)&=&(\id \otimes \Delta)\Delta
(a),\nonumber \\
(\epsilon \otimes \id)\Delta (a)&=&(\id \otimes\epsilon)\Delta (a)=a.
\end{eqnarray}
In analogy with the associative algebra case, the first of these is
often referred to as {\em coassociativity}.

A {\em bialgebra} is both a unital associative algebra and a
coalgebra, with the compatibility conditions that $\Delta$ and
$\epsilon$ are both algebra maps with $\Delta (\IA )=\IA \otimes
\IA$ and $\epsilon (\IA) =\Ik$.

A {\em Hopf algebra} is a bialgebra together with a linear map $S:
\A\rightarrow\A$, the antipode, which satisfies
\begin{equation}
m((S \otimes \id)\Delta (a))=m((\id \otimes S)\Delta
(a))=\epsilon(a)\IA.
\end{equation}
It follows that the antipode is an antialgebra map, \ie $S(ab)=S(b)
S(a)$\footnote{We always make the further assumption that $S$ is
bijective, so that the inverse map $S^{-1}$ exists.  It too is an
antialgebra map.}.

A {\em *-Hopf algebra} is a Hopf algebra with involution $\theta : \A
\rightarrow\A$ which satisfies
\begin{eqnarray}
\theta(\alpha a)&=&\alpha^* \theta(a),\nonumber\\
\theta^2 (a)&=&a,\nonumber\\
\theta (ab)&=&\theta(b)\theta(a),\nonumber \\
\Delta(\theta(a)) &=& (\theta \otimes \theta)\Delta(a),\nonumber \\
\epsilon(\theta(a))&=& \epsilon(a)^* ,\nonumber\\
\theta(S(\theta(a)))&=& S^{-1}(a),
\end{eqnarray}
$\alpha\in k$.  Here ${}^*$ is the involution on $k$, \eg complex
conjugation when $k=\complex$.

A {\em quasitriangular Hopf algebra} \cite{Drinfeld} is a Hopf algebra
together with an invertible element $\R=r_{\alpha}\otimes r^{\alpha}$
(summation implied) in $\A\otimes\A$ which must satisfy the relations
\begin{eqnarray}
(\Delta \otimes \id)(\R)&=&\R_{13}\R_{23},\nonumber\\
(\id \otimes \Delta)(\R)&=&\R_{12}\R_{23},\nonumber\\
(\tau \circ \Delta )(a) &=&\R \Delta (a) \R^{-1},
\end{eqnarray}
where $\tau :\A\otimes\A\rightarrow\A\otimes\A$ is the permutation map
$a\otimes b \mapsto b\otimes a$,
\begin{eqnarray}
\R_{12}&=&r_{\alpha}\otimes r^{\alpha}\otimes\IA,\nonumber \\
\R_{13}& =&r_{\alpha}\otimes\IA\otimes r^{\alpha},\nonumber\\
\R_{23}&=&\IA\otimes r_{\alpha}\otimes r^{\alpha},
\end{eqnarray}
and the multiplication map $m$ has been suppressed on the right-hand
side.  $\R$ is called the {\em universal R-matrix of} $\A$, and, as a
consequence of these relations, satisfies the {\em quantum Yang-Baxter
equation (QYBE)}
\begin{equation}
\R_{12}\R_{13}\R_{23}=\R_{23}\R_{13}\R_{12}.\label{QYBE}
\end{equation}

\section{Dually Paired Hopf Algebras}\label{chap-Hopf-dual}

Two *-Hopf algebras $\U$ and $\A$ are said to be {\em dually paired}
if there exists a nondegenerate inner product $\inprod{\,}{\,}:\U
\otimes \A\rightarrow k$ such that
\begin{eqnarray}
\inprod{xy}{a}&=&\inprod{x \otimes y}{\Delta (a)}, \nonumber\\
\inprod{\IU}{a}&=&\epsilon (a),\nonumber \\
\inprod{\Delta (x)}{a \otimes b}&=&\inprod{x}{ab},\nonumber \\
\epsilon (x) &=& \inprod{x}{\IA},\nonumber\\
\inprod{S(x)}{a}&=&\inprod{x}{S(a)},\nonumber\\
\inprod{\theta (x)}{a}&=&\inprod{x}{\theta (S(a))}^{*},
\end{eqnarray}
where $x,y\in\U$ and $a,b\in\A$.  It is easily shown that all the
relevant consistency relations between the various operations are
satisfied.

Note that the relations above may be used constructively, \ie given
one *-Hopf algebra, one can construct a dually paired *-Hopf algebra;
this is the method usually employed when the Drinfel'd double $D(\A)$
of a Hopf algebra $\A$ is found \cite{Drinfeld}.

\section{Representations of Hopf Algebras and Quantum Groups}
\label{chap-Hopf-reps}

Let $\U$ be a Hopf algebra, and suppose $\rho :\U\rightarrow M_N (k)$
is a $N\times N$ faithful matrix representation, with entries in $k$,
of $\U$.  This representation can be used to define another Hopf
algebra dually paired with $\U$; we take this new Hopf algebra $\A$ to
be that which is generated by the $N^2$ elements $\matrix{A}{i}{j}$
defined by \cite{Woronowicz3}
\begin{equation}
\matrix{\rho}{i}{j}(x)\equiv\inprod{x}{\matrix{A}{i}{j}}
\end{equation}
for $x\in\U$.  The faithfulness of the representation ensures that
this inner product is nondegenerate, and thus the elements of the
matrix $A$ are uniquely determined; furthermore, the fact that $\rho$
is an algebra map immediately gives
\begin{equation}
\Delta(\matrix{A}{i}{j})=\matrix{A}{i}{k}\otimes\matrix{A}{k}{j},\,
\epsilon(\matrix{A}{i}{j})=\delta^i _j ,\, S(\matrix{A}{i}{j})=
\matrix{(A^{-1})}{i}{j}.\label{QG-Hopf}
\end{equation}
The multiplication on $\A$ will of course depend upon the form of the
coproduct in $\U$, respectively.  However, in the case where $\U$ is
quasitriangular with universal R-matrix $\R$, a rather famous result
follows; let $x\in\U$, and $\Delta' = \tau\circ\Delta$.  Using the
last of the properties of the universal R-matrix from above, we see
\begin{eqnarray}
0&\equiv&\inprod{\R\Delta(x)-\Delta '(x)\R}{\matrix{A}{i}{k}\otimes
\matrix{A}{j}{\ell}}\nonumber\\
&=&\inprod{\R}{\matrix{A}{i}{m}\otimes\matrix{A}{j}{n}}
\inprod{\Delta(x)}{\matrix{A}{m}{k}\otimes\matrix{A}{n}{\ell}}\nonumber\\
&&-\inprod{\Delta(x)}{\matrix{A}{j}{n}\otimes\matrix{A}{i}{m}}
\inprod{\R}{\matrix{A}{m}{k}\otimes\matrix{A}{n}{\ell}}\nonumber\\
&=&\inprod{x}{\matrix{R}{ij}{mn}\matrix{A}{m}{k}\matrix{A}{n}{\ell}-
\matrix{A}{j}{n}\matrix{A}{i}{m}\matrix{R}{mn}{k\ell}}\label{RAA-comp}
\end{eqnarray}
where
\begin{equation}
\matrix{R}{ij}{k\ell}:=\inprod{\R}{\matrix{A}{i}{k}\otimes
\matrix{A}{j}{\ell}}
\end{equation}
is the $N^2 \times N^2$ dimensional {\em numerical R-matrix of} $\A$.
Since $x$ was arbitrary, the vanishing of (\ref{RAA-comp}) implies
that
\begin{equation}
RA_1 A_2 =A_2 A_1 R\label{RAA}
\end{equation}
where the indices have been suppressed, and the subscripts refer to
the indices in an obvious way.  This is the noted ``RAA equation''
\cite{RTF}, and gives explicitly the commutation relations between
elements of $\A$.  It is immediate that the QYBE has the numerical
counterpart, simply referred to as the Yang-Baxter equation (YBE):
\begin{equation}
R_{12}R_{13}R_{23}=R_{23}R_{13}R_{12}.\label{YBE}
\end{equation}
This leads to the following important definition: a Hopf algebra $\A$
which is dually paired with a quasitriangular Hopf algebra $\U$ is a
{\em quantum group} \cite{Drinfeld}.  However, we often take the
opposite view, saying that a quantum group is a Hopf algebra where the
$N^2$ generators $\matrix{A}{i}{j}$ satisfy (\ref{QG-Hopf}) and
(\ref{RAA}), and $R$ satisfies the YBE.

(\ref{YBE}) was obtained from (\ref{QYBE}) by taking the
representation in all three spaces of $\U\otimes\U\otimes\U$, \eg
\begin{equation}
(\rho\otimes\rho\otimes\rho )\R_{12}\R_{13}\R_{23}=R_{12}R_{13}
R_{23}.
\end{equation}
It is also useful to consider the case where we take the
representation in only one or two of the tensor product spaces.  To
see this, we define the $N\times N$ matrices $L^{\pm}$ with entries in
$\U$ by
\begin{eqnarray}
L^+ &:=& (\id\otimes\rho)\R\equiv r_{\alpha}\rho (r^{\alpha}),
\nonumber\\
L^- &:=& (\rho\otimes\id )\R^{-1}\equiv\rho (S(r_{\alpha}))r^{\alpha}.
\end{eqnarray}
{}From the properties of $\R$, we then find that
\begin{eqnarray}
\Delta(L^{\pm})=L^{\pm}\dot{\otimes}L^{\pm},&\epsilon
(L^{\pm})=I,&\nonumber\\
S(L^+ )=(L^+ )^{-1}=(\id\otimes\rho)\R^{-1},& S(L^- )=(L^- )^{-1}=
(\rho\otimes\id)\R &
\end{eqnarray}
(where we use the notation $\matrix{(M\dot{\otimes}N)}{i}{j}:=
\matrix{M}{i}{k}\otimes\matrix{N}{k}{j}$).  Now, suppose we apply
$\id\otimes\rho\otimes\rho$ to (\ref{QYBE}); the left-hand side is
\begin{eqnarray}
\lefteqn{(\id\otimes\matrix{\rho}{i}{k}\otimes\matrix{\rho}{j}{\ell})
\R_{12}\R_{13}\R_{23}=}\nonumber\\
& & (\id\otimes\matrix{\rho}{i}{m})(\R)(\id\otimes\matrix{\rho}{j}{n})
(\R)(\matrix{\rho}{m}{k}\otimes\matrix{\rho}{n}{\ell})(\R)
=\matrix{(L^+_1 L^+_2 R)}{ij}{k\ell}.
\end{eqnarray}
The right-hand side is computed similarly, and the resulting identity
is
\begin{equation}
L^+_1 L^+_2 R=RL^+_2 L^+_1 .\label{LL1}
\end{equation}
By writing (\ref{QYBE}) in various ways using $\R^{-1}$, we find two
more independent equations:
\begin{eqnarray}
L^-_1 L^-_2 R=RL^-_2 L^-_1 ,&L^-_2 L^+_1 R=RL^+_2 L^-_1 .& \label{LL2}
\end{eqnarray}
The matrix representations for $L^{\pm}$ are easily found:
\begin{eqnarray}
\matrix{\rho}{i}{j}(\matrix{(L^+ )}{k}{\ell})&=&\matrix{R}{ik}{j\ell},
\nonumber\\
\matrix{\rho}{i}{j}(\matrix{(L^- )}{k}{\ell})&=&
\matrix{(R_{21}^{-1})}{ik}{j\ell}.
\end{eqnarray}
As we shall see, these matrices will figure very prominently in the
construction of quantum Lie algebras.

It often becomes convenient to use the permutation matrix
$\matrix{P}{ij}{k\ell}\equiv\delta^i_{\ell}\delta^j _k$ to define the
matrix $\Rhat$:
\begin{equation}
\matrix{\Rhat}{ij}{k\ell}:=\matrix{(PR)}{ij}{k\ell}\equiv
\matrix{R}{ji}{k\ell}.
\end{equation}
$\Rhat$, not $R$, is the matrix which appears in knot theory; we will
not rely upon this interpretation of $\Rhat$ in this work, although
the fact that it satisfies a characteristic (``skein'') equation will
be used extensively.  (See Appendix \ref{knot} for more details.)

\section{Examples}\label{chap-Hopf-ex}

\subsection{Classical Lie Algebras}\label{chap-Hopf-ex-class}

There is a very straightforward way to turn a classical
finite-di\-men\-sional Lie algebra $\g$ into a quasitriangular Hopf
algebra; let $\{ T_a |a=1,\ldots,N \}$ be a basis for $\g$, and
$\struc{ab}{c}$ the structure constants in this basis.  Let $\A$ be
the universal enveloping algebra of $\g$ modulo the commutation
relations $\comm{T_a }{T_b }=T_a T_b -T_b T_a =\struc{ab}{c}T_c$,
denoted $U(\g)$.  We can then give $\A$ a Hopf algebra structure by
defining $\Delta$ and $\epsilon$ to be linear algebra maps and $S$ to
be a linear antialgebra map whose actions on the basis elements of
$\A$ are given by
\begin{eqnarray}
\Delta (T_a )=T_a \otimes\IA +\IA\otimes T_a,&\epsilon(T_a )=0,&S(T_a
)=-T_a .
\end{eqnarray}
Furthermore, $\A$ is quite obviously quasitriangular, since $\R=
\IA\otimes\IA$ satisfies all the appropriate relations trivially.

\subsection{The Hopf Algebra $U_q (su(2))$}

A nontrivial example of a quasitriangular *-Hopf algebra can be
obtained from $su(2)$; let $\A$ be the universal enveloping algebra of
the three generators $H$, $X_+$, and $X_-$ modulo the Jimbo-Drinfel'd
commutation relations \cite{Jimbo,Drin}
\begin{eqnarray}
\comm{H}{X_{\pm}}&=&\pm 2X_{\pm},\nonumber\\
\comm{X_+ }{X_- }&=&\frac{q^H - q^{-H}}{q-q^{-1}},
\end{eqnarray}
where $q\in\real$.  This unital associative algebra is usually denoted
by $U_q (su(2))$, the ``deformed'' universal enveloping algebra of
$su(2)$.  The coproducts, counits, antipodes and involutions are given
by
\begin{eqnarray}
\Delta(H)=H\otimes\IA+\IA\otimes H,&\Delta(X_{\pm})=X_{\pm}\otimes
q^{\half H}+ q^{-\half H}\otimes X_{\pm},&\nonumber\\
\epsilon(H)=\epsilon(X_{\pm})=0,&&\nonumber\\
S(H)=-H,&S(X_{\pm})=-q^{\pm 1}X_{\pm},&\nonumber\\
\theta (H)=H,&\theta (X_{\pm})=X_{\mp}.&
\end{eqnarray}
Notice that in the limit $q\rightarrow 1$, we recover the familiar
classical $su(2)$ Hopf algebra described in the previous subsection.
The universal R-matrix for $\A$ is given in terms of the above
generators, and has the form
\begin{equation}
\R=\sum_{n=0}^{\infty}\frac{(1-q^{-2})^n}{\quint{n}{q}!}q^{\half(H\otimes
H+nH\otimes\IA -n\IA\otimes H)}X_+^n \otimes X_-^n ,
\end{equation}
where we use the standard notations for the ``quantum number''
\begin{equation}
\quint{x}{q}:=\frac{q^{2x}-1}{q^2 -1}
\end{equation}
and the ``quantum factorial''
\begin{equation}
\quint{n}{q}!:=\left\{ \begin{array}{ll}
1&n=0,\\
\prod_{m=1}^n \quint{m}{q}&n=1,2,\ldots
\end{array}\right.
\end{equation}
The fundamental repesentations for both the deformed and undeformed
cases coincide, \ie the matrices
\begin{eqnarray}
H=\left( \begin{array}{cc}
-1&0\\0&1\end{array}\right) ,&
X_+ =\left( \begin{array}{cc}
0&0\\-1&0\end{array}\right) ,&
X_- =\left( \begin{array}{cc}
0&-1\\0&0\end{array}\right) ,
\end{eqnarray}
satisfy the Jimbo-Drinfel'd commutation relations for any value of
$q$.  When we express the universal R-matrix in this representation,
we obtain
\begin{equation}
R=q^{-\half}\left( \begin{array}{cccc}
q&0&0&0\\0&1&0&0\\0&\lambda &1&0\\0&0&0&q
\end{array}\right) ,\label{R-sl2}
\end{equation}
where $\lambda\equiv q-q^{-1}$.  We can also use this representation
to find the $2\times 2$ matrices $L^{\pm}$, defined in
(\ref{chap-Hopf-reps}):
\begin{eqnarray}
L^+ =\left( \begin{array}{cc}
q^{-\half H}&-q^{-\half}\lambda X_+ \\0&q^{\half H}
\end{array}\right) ,&
L^- =\left( \begin{array}{cc}
q^{\half H}&0\\q^{\half}\lambda X_- &q^{-\half H}
\end{array}\right) .&
\end{eqnarray}

\subsection{The Quantum Group $GL_q (2)$}

The canonical example of a quantum group is the deformed version of
$GL(2)$, denoted $GL_q (2)$.  This is the Hopf algebra generated by
the four elements $\{ a,b,c,d\}$ satisfying
\begin{eqnarray}
ab=qba,&ac=qca,&ad-da=\lambda bc,\nonumber\\
bc=cb,&bd=qdb,&cd=qdc.
\end{eqnarray}
We can express these commutation relations in the form (\ref{RAA}) by
defining
\begin{equation}
A=\left( \begin{array}{cc}
a&b\\c&d
\end{array} \right),\,
R=\left( \begin{array}{cccc}
q&0&0&0\\0&1&0&0\\0&\lambda &1&0\\0&0&0&q
\end{array} \right).
\end{equation}
(Note that this differs from the $U_q (su(2))$ R-matrix only by an
overall factor of $q^{-\half}$; we will explain the reason for this in
a later section.)  For consistency with (\ref{QG-Hopf}), we require
\begin{eqnarray}
\Delta(a)=a\otimes a+b\otimes c,&\Delta(b)=a\otimes b+b\otimes
d,&\nonumber\\
\Delta(c)=c\otimes a+d\otimes c,&\Delta(d)=c\otimes b+d\otimes
d,&\nonumber\\
\epsilon(a)=\epsilon(d)=1,&\epsilon(b)=\epsilon(c)=0,&\nonumber\\
S(a)=(\det{A})^{-1}d,&S(b)=-q^{-1}(\det{A})^{-1}b,&\nonumber\\
S(c)=-q(\det{A})^{-1}c,&S(d)=(\det{A})^{-1}a,&
\end{eqnarray}
where $\det{A}:=ad-qbc$ is the ``quantum determinant'' of $A$, and is
central within the algebra.

\section{Sweedler's Notation}\label{chap-Hopf-Sweedler}

We end this chapter with a discussion of an extremely useful notation
which we will use for the remainder of this work.  It is referred to
as ``Sweedler's notation'' after the man who first introduced it in
\cite{Sweedler}, and is a way of easing the computations involved in
dealing with Hopf algebras.

If $\A$ is a Hopf algebra, then the coproduct $\Delta(a)$ of an
element $a\in\A$ will in general consist of a sum of elements in
$\A\otimes\A$; the examples presented in the previous section show
this fact explicitly.  Thus, we could in theory write
\begin{equation}
\Delta(a)=\sum_i a^i_{(1)}\otimes a_{(2)i},
\end{equation}
where $a^i_{(1)}$ and $a_{(2)i}$ both live in $\A$.  We could easily
adopt the standard Einstein summation convention and drop the
summation sign, realizing that any pair of identical indices, one up
and one down, are to be summed over.  However, Sweedler went further
than that; he also dropped the indices themselves, preferring to write
\begin{equation}
\Delta(a)=a_{(1)}\otimes a_{(2)}.
\end{equation}
Therefore, anytime an algebra element is subscripted with a number in
parentheses, it is understood to be obtained from a coproduct, with
the appropriate summation implied.  But there's more: using this
convention, the coassociativity condition looks like
\begin{equation}
(a_{(1)})_{(1)}\otimes (a_{(1)})_{(2)}\otimes a_{(2)}=a_{(1)}\otimes
(a_{(2)})_{(1)}\otimes (a_{(2)})_{(2)}.\label{swee-coass}
\end{equation}
Compare this to the case of an associative algebra: the analogous
identity to (\ref{swee-coass}) is $(ab)c=a(bc)$, and due to this,
there is no ambiguity in writing $abc$.  In the coassociative case, we
can therefore adopt the unambiguous convention
\begin{equation}
(\Delta\otimes\id)\Delta(a)=(\id\otimes\Delta)\Delta(a)=a_{(1)}\otimes
a_{(2)}\otimes a_{(3)},
\end{equation}
again, with the implied sum.

To further illustrate the use of Sweedler's notation, the identity
$(\epsilon\otimes\id)\Delta (a)=a$ takes the form
\begin{equation}
\epsilon(a_{(1)})a_{(2)}=a.
\end{equation}
Note that the Hopf algebra axioms imply the identity
\begin{equation}
((m(\id\otimes S)\Delta)\otimes\id)\Delta (a) =a;
\end{equation}
this can be written as
\begin{equation}
a_{(1)}S(a_{(2)})a_{(3)}=a.
\end{equation}

For the remainder of this thesis, we will make extensive use of this
notation, and the reader is encouraged to familiarize him/herself with
its use.

\chapter{Actions, the Smash Product, and Coactions}
\label{chap-coactions}

The importance of the three topics in the title above to physics
cannot be overstressed, so it is worthwhile to set aside an entire
chapter to a discussion of them.  We simply rewrite many familiar
concepts in the language of Hopf algebras, thus providing a method of
generalizing the classical case.

\section{Actions and Generalized Derivations}
\label{chap-coactions-actions}

Suppose we have a unital associative algebra $\B$ and a vector space
$\V$; a {\em left action} of $\B$ on $\V$ is a bilinear map $\trg
:\B\otimes\V\rightarrow\V$ satisfying
\begin{eqnarray}
(xy)\trg v&=&x\trg (y\trg v),\nonumber\\
1_{\B}\trg v&=&v,
\end{eqnarray}
for all $x,y\in\B$ and $v\in\V$.  (Note that this is merely another
way of saying that we have a linear representation of $\B$ with right
module $\V$.)  A right action $\tlg$ of $\B$ on $\V$ can be defined
similarly.  In the case where $\B$ is a Hopf algebra and $\V$ is a
unital algebra, we further require that for $x\in\B$ and $a,b\in\V$,
\begin{eqnarray}
x\trg (ab) &=& (x_{(1)}\trg a)(x_{(2)}\trg b),\nonumber\\
x\trg 1_{\V} &=& 1_{\V}\epsilon (x).
\end{eqnarray}
In this case, $\trg$ is called a {\em generalized (left) derivation},
and we can interpret $\B$ as an algebra of differential operators
which act on functions (\ie elements of $\V$), and, as such, may be
thought of as providing a means for generalizing infinitesimal
transformations.  (We will see in a little while that there is a way
of generalizing finite transformations as well.)

There are two extremely important examples of such generalized
derivations, both of which will be relevant for this work:
\begin{itemize}
\item The {\em adjoint action} of a Hopf algebra $\U$ on itself is
defined as the bilinear map $\ad :\U\otimes\U\rightarrow\U$ given by
$x\otimes y\mapsto x\ad y:=x_{(1)}yS(x_{(2)})$\footnote{Note that if
$\U$ is the classical Hopf algebra discussed in Section
\ref{chap-Hopf-ex-class}, the right adjoint action is just the
classical commutator: $T_a \ad T_b = \comm{T_a}{T_b}$.}; it is a left
action as defined above.  Similarly, $y\stackrel{\mbox{\scriptsize
ad}}{\tlg}x:=S(x_{(1)})y x_{(2)}$ defines a perfectly good right
action.
\item If $\U$ and $\A$ are two dually paired Hopf algebras, we can
define the left and right actions of $\U$ on $\A$ respectively as
\begin{eqnarray}
x\trg a:=a_{(1)}\inprod{x}{a_{(2)}},&&a\tlg x:=\inprod{x}{a_{(1)}}
a_{(2)}.\label{diffact}
\end{eqnarray}
As stated above, this allows the interpretation of $\U$ as an algebra
of differential operators which act on elements (``functions'') of
$\A$.  (An explicit example of this interpretation is the familiar
left action of a quantum mechanical Hamiltonian $H$ on some
Schr\"odinger state $\psi$, namely, $H\trg\psi (t)=i\hbar\partder{\psi
(t)}{t}$.)

\end{itemize}

\section{The Smash Product}\label{chap-coactions-smash}

Let $\A$ and $\U$ be two dually paired Hopf algebras.  We can
introduce a unital associative algebra which is denoted $\smash$, the
``smash product'' of $\A$ and $\U$.  (This object is also called the
``cross product'' \cite{Majid1}, and is a Hopf algebra generalization
of the Heisenberg double and the Weyl semidirect product.)  $\smash$
is constructed to be isomorphic to $\A\otimes\U$ as a vector space;
this may be seen explicitly through the definition of the
multiplication on $\smash$:
\begin{eqnarray}
ab&\simeq&ab\otimes\IU ,\nonumber\\
xy&\simeq&\IA\otimes xy,\nonumber\\
ax&\simeq&a\otimes x,\nonumber\\
xa&\simeq&a_{(1)}\otimes x_{(2)}\inprod{x_{(1)}}{a_{(2)}},
\label{mult}
\end{eqnarray}
where $a,b\in\A$, $x,y\in\U$, and the $\simeq$ denotes equivalence
under the aforementioned isomorphism.  Note that this multiplication
is associative, and also that $\smash$ contains subalgebras isomorphic
to both $\IA\otimes\U$ and $\A\otimes\IU$.  However, throughout the
rest of this work we will be glib and refer to these subalgebras of
$\smash$ as $\U$ and $\A$ respectively\footnote{Notice that although
$\U$ and $\A$ are both Hopf algebras, $\smash$ is not, \ie $\smash$ is
an algebra that does not admit a Hopf algebra structure (coproduct,
counit, antipode) even though the subalgebras $\U$ and $\A$ do.}.
With this convention, and the form of the multiplication (\ref{mult}),
we see that $\smash$ is spanned by elements of the form $ax$ with
$a\in\A$, $x\in\U$, and we can specify all linear maps on $\smash$ by
considering their values on such elements.

The physical meaning of the smash product becomes clear when we
realize that the multiplication in the smash product $\smash$ may be
written as
\begin{equation}
xa=(x_{(1)}\trg a)x_{(2)},\label{U-A-comm}
\end{equation}
where the left action is, as it will be for the remainder of this
work, the one given in (\ref{diffact}).  Thus, the multiplication
relations in $\smash$ may be interpreted as the commutation relations
between the differential operators in $\U$ and the elements of $\A$,
namely, how to take a differential operator and ``move it through'' a
function.  This is a very natural interpretation in physics, and is
the one we will adopt.

As an explicit (and important) example of how the smash product works,
consider the case where $\U$ is a quasitriangular Hopf algebra, and
$\rho$ is a representation of $\U$ which defines the dually paired
quantum group $\A$, in the manner of Section \ref{chap-Hopf-reps}.  If
$x\in\U$, then the commutation relation between $x$ and a basis
element $\matrix{A}{i}{j}$ of $\A$ in $\smash$ is
\begin{eqnarray}
x\matrix{A}{i}{j}&=&(\matrix{A}{i}{j})_{(1)}
\inprod{x_{(1)}}{(\matrix{A}{i}{j})_{(2)}}{x_{(2)}}\nonumber\\
&=&\matrix{A}{i}{k}\inprod{x_{(1)}}{\matrix{A}{k}{j}}x_{(2)}
\nonumber\\
&=&\matrix{A}{i}{k}\matrix{\rho}{k}{j}(x_{(1)})x_{(2)}.
\end{eqnarray}
For the case where $x$ is an entry in $L^{\pm}$, we find
\begin{eqnarray}
L^+_1 A_2 =A_2 R_{21}L^+_1 ,&L^-_1 A_2 =A_2 R^{-1}L^-_1 .&
\label{LA}
\end{eqnarray}
These relations will come in handy when we discuss quantum Lie
algebras.

\subsection{Example:  The Haar Measure and the Smash Product}
\label{chap-coactions-Haar}

At this point, it may be instructive to take a slight detour in order
to illustrate how the smash product may be used in computations.  The
example we choose involves the introduction of {\em right-invariant
Haar measure} on $\A$.  This is a linear map $\int :\A\rightarrow k$
which satisfies the two properties
\begin{eqnarray}
\left( \int a_{(1)}\right) a_{(2)}=\left( \int a\right) \IA
,&&\int\IA=1_k .\label{haar}
\end{eqnarray}
(It is readily shown that these conditions uniquely determine $\int$,
and that such a measure is left-invariant as well \cite{Chryss}.)  We
will use the smash product machinery developed above to construct such
a measure in the case where $\A$ is finite-dimensional.  We introduce
the element $E\in\smash$ given by $E:=S^{-1}(f^i )e_i$.  For $a\in\A$,
\begin{eqnarray}
Ea&=&S^{-1}(f^i )e_i a\nonumber\\
&=&S^{-1}(f^i )a_{(1)}\inprod{(e_i )_{(1)}}{a_{(2)}}(e_i )_{(2)}
\nonumber\\
&=&S^{-1}(f^i f^j )a_{(1)}\inprod{e_i }{a_{(2)}}e_j \nonumber\\
&=&S^{-1}(f^j )S^{-1}(a_{(2)})a_{(1)}e_j \nonumber\\
&=&E\epsilon (a).
\end{eqnarray}
and similarly, for $x\in\U$,
\begin{equation}
xE=\epsilon (x)E.
\end{equation}
It is easily shown using these properties that $E^2 =E$, and therefore
\begin{eqnarray}
ExaE=\inprod{x}{a}E,&&EaxE=\epsilon(x)\epsilon(a)E.
\end{eqnarray}
We now assume that there exists a Hilbert bimodule of $\smash$
containing the two vacua $\vacu$ and $\vaca$ which satisfy
\begin{eqnarray}
x\vacu =\vacu x &=&\epsilon(x)\vacu,\nonumber\\
\vaca a=a\vaca &=&\vaca\epsilon(a),\nonumber\\
\bracket{\vaca}{\vacu}&=&1,
\end{eqnarray}
for $x\in\U$, $a\in\A$.  (These should recall the definitions of left
and right vacua introduced in \cite{Leipzig}, denoted by $\langle$ and
$\rangle$ respectively, which satisfy
\begin{eqnarray}
L^+ \rangle =L^- \rangle =I\rangle ,&\langle A=\langle I,&
\bracket{\,}{\,}=1.)
\end{eqnarray}
One consequence of these definitions is that $\inprod{x}{a}\equiv
\bracket{\vaca}{xa\vacu}$.  We may therefore conclude that $E$ may be
represented by $\ket{\vacu}\bra{\vaca}$.  There also exists an object
$\tilde{E}\in\smash$, given by $\tilde{E} =S^2 (e_i)f^i$, which has
properties similar to that of $E$, \eg $\tilde{E}^2 =\tilde{E}$,
$\tilde{E}x=\tilde{E}\epsilon(x)$ and $a\tilde{E}=\epsilon(a)
\tilde{E}$ for $x\in\U$, $a\in\A$; thus, we represent $\tilde{E}$ by
$\ket{\vaca}\bra{\vacu}$.

An equivalent way of formulating the second relation of (\ref{haar})
is by utilizing the left action (\ref{diffact}) of $\U$ on $\A$:
\begin{eqnarray}
\int x\trg a&=&\int a_{(1)}\inprod{x}{a_{(2)}}\nonumber\\
&=&\inprod{x}{\left( \int a_{(1)}\right) a_{(2)}}\nonumber\\
&=&\inprod{x}{\left( \int a\right) \IA}\nonumber\\
&=&\epsilon(x)\int a.\label{haar-adj}
\end{eqnarray}
Since within $\smash$, $x\trg a\equiv x_{(1)}aS(x_{(2)})$, we see that
(\ref{haar-adj}), together with the uniqueness of $\int$, implies
\begin{equation}
\int a\equiv\frac{\bracket{\vacu}{a\vacu}}{\bracket{\vacu}{\vacu}}.
\end{equation}
Note, however, that the Hilbert space representations of $E$ and
$\tilde{E}$ give
\begin{eqnarray}
\tilde{E}aE&=&\ket{\vaca}\bracket{\vacu}{a\vacu}\bra{\vaca}
\nonumber\\
&=&\ket{\vaca}\left( \bracket{\vacu}{\vacu}\int a\right)
\bra{\vaca}\nonumber\\
&=&\tilde{E}E\int a.\label{integral}
\end{eqnarray}
To isolate $\int a$ from this, first we push all the $\A$-elements to
the left using (\ref{mult}); the result for the left-hand side is
\begin{eqnarray}
\tilde{E}aE&=&S^2 (e_i )f^i a S^{-1}(f^j )e_j\nonumber\\
&=&(f^i a S^{-1}(f^j ))_{(1)}\inprod{S^2 ((e_i)_{(1)})}{(f^i a S^{-1}
(f^j ))_{(2)}}S^2 ((e_i)_{(2)})e_j .
\end{eqnarray}
$\tilde{E}E$ is obtained by setting $a=\IA$.  We may then sandwich
this between $\vacu$ on the left and $\vaca$ on the left to obtain
\begin{equation}
\bracket{\vacu}{\tilde{E}aE\vaca}=\inprod{S^2 (e_i )}{f^i a}.
\end{equation}
Comparing this to (\ref{integral}), we see that this is equal to
$\bracket{\vaca}{\vaca}\bracket{\vacu}{\vacu}\int a$.  Therefore, we
find an explicit form for the Haar measure on $\A$:
\begin{equation}
\int a\equiv\frac{\inprod{S^2 (e_i )}{f^i a}}{\inprod{S^2 (e_i
)}{f^i}}.
\end{equation}
The finiteness of $\A$ insures that this expression exists.  For the
case where $\A$ is not finite, the situation is more problematic; the
above argument may not hold, because some of the quantities involved,
\eg $\inprod{S^2 (e_i )}{f^i}$, may not exist.  Furthermore, it is
possible in some cases that we cannot define $\int$ consistently on
the entirety of $\A$, and in particular $\int\IA$ may not exist (this
latter case may be a statement of the ``noncompactness'' of $\A$); for
both of these possibilities, the computation above may run into
problems.  In any case, this should serve as an illustration of how
the smash product may be used to obtain useful results.

\section{Coactions}\label{chap-coactions-coacts}

Suppose we have a coalgebra $\C$ and a vector space $\V$; a {\em right
coaction} of $\C$ on $\V$ is a linear map $\Delta_{\C} :\V
\rightarrow\V\otimes\C$ satisfying
\begin{eqnarray}
(\Delta_{\C}\otimes \id)\Delta_{\C}(v)&=&(\id\otimes\Delta)
\Delta_{\C}(v),\nonumber \\
(\id\otimes\epsilon)\Delta_{\C}(v)&=&v,
\end{eqnarray}
for all $v\in\V$, where $\Delta$ and $\epsilon$ are the coproduct and
counit on $\C$, respectively.  We will often use the Sweedleresque
notation $\Delta_{\C}(v)=v^{(1)}\otimes v^{(2)'}$, where the unprimed
elements live in $\V$, the primed elements in $\C$.  The left coaction
$_{\C}\Delta(v)=v^{(1)'}\otimes v^{(2)}$ is defined similarly.  If
$\C$ and $\C'$ are two coalgebras which coact on $\V$ from the left
and from the right respectively, we will generally require that they
commute, \ie
\begin{equation}
(_{\C}\Delta\otimes\id)\Delta_{\C'}(v)=(\id\otimes\Delta_{\C'})_{\C}
\Delta (v)
\end{equation}
for $v\in\V$.  If $\C$ is a Hopf algebra and $\V$ is a unital algebra,
we impose the further conditions that
\begin{eqnarray}
\Delta_{\C}(ab)&=&\Delta_{\C}(a)\Delta_{\C}(b),\nonumber \\
\Delta_{\C}(1_{\V})&=&1_{\V}\otimes 1_{\C},
\end{eqnarray}
for $a,b\in\V$, \ie $\Delta_{\C}$ must be an algebra homomorphism.  If
$a\in\V$ satisfies $\Delta_{\C}(a)=a\otimes 1_{\C}$, we say that $a$
is {\em right-invariant} (and similarly for left-invariance).

One coaction which will figure prominently in this work is the {\em
adjoint (right) coaction} of a Hopf algebra $\A$ on itself.  This
action $\Ad :\A\rightarrow\A\otimes\A$ is a right coaction in the
first sense above, namely it acts on $\A$ as if it were only a vector
space, {\em not} a unital algebra.  Therefore, it is not a
homomorphism, as is easily seen by its definition:
\begin{equation}
\Ad (a):=a_{(2)}\otimes S(a_{(1)})a_{(3)}.\label{adj-coact}
\end{equation}

A comment on terminology: as the reader may have guessed, the reason
for the term ``coaction'' is because of duality.  If $\B$ is a unital
associative algebra which is dual to a coalgebra $\C$ in the obvious
way, and $\trg$ is a left action of $\B$ on some vector space $\V$,
then there is a natural way to pair it with a right coaction of $\C$
on $\V$ via
\begin{equation}
v^{(1)}\inprod{x}{v^{(2)'}}=x\trg v,
\end{equation}
for $x\in\B$, $v\in\V$.  Similarly, a right action will induce a left
coaction.

The interpretation of the coaction is straightforward: to illustrate
this, let $\C$ be a coalgebra with elements $\matrix{g}{i}{j}$ which
satisfy $\Delta (g)=g\dot{\otimes}g$ and $\epsilon (g)=I$.  Define the
right coaction of $\C$ on a basis element $e_i \in\V$ via
\begin{equation}
\Delta_{\C}(e_i ):=e_j \otimes\matrix{g}{j}{i}.
\end{equation}
This looks a lot like a simple transformation of the basis elements,
which is how we interpret it.  If we coact on the first space once
more we obtain
\begin{equation}
(\Delta_{\C}\otimes\id )\Delta_{\C}(e_i )=e_k \otimes\matrix{g}{k}{j}
\otimes\matrix{g}{j}{i},
\end{equation}
which is simply two successive ``rotations'' of the basis.  The tensor
product between the two indicates that the two transformations are
independent of each other.  This illustrates the fact that the
coaction is the generalization of a {\em finite} transformation of an
element of $\V$, as opposed to the {\em infinitesimal} transformation
provided by the action.

\section{Actions and Coactions on the Smash Product}

We have already noted that we can interpret the smash product as the
algebra of differential operators and the functions which they act on,
with the multiplication within this algebra being interpreted as the
commutation relations between the two types of elements, \ie how the
differential operators first act on, then are moved through, the
functions (\ref{U-A-comm}).  In this section, we discuss how to define
actions and coactions on this algebra consistent with this
interpretation.

\subsection{Bicovariance of the Smash Product}

We now introduce specific actions and coactions in the case where we
have the two dually paired Hopf algebras $\U$ and $\A$ and the
associated smash product $\smash$.  The left and right actions of $\U$
on $\smash$ are defined to be
\begin{eqnarray}
x\trg\sigma &\equiv & x_{(1)}\sigma S(x_{(2)}), \nonumber \\
\sigma\tlg x &\equiv & S(x_{(1)})\sigma x_{(2)},
\end{eqnarray}
for $x\in\U$, $\sigma\in\smash$.  Note that for the case where
$\sigma\in\U$, these are the left and right adjoint actions, and when
$\sigma\in\A$, we reobtain the usual right action of a differential
operator $x$ on a function $\sigma$ given by (\ref{diffact}).  Since
all elements of $\smash$ have the form $ax$, $a\in\A$ and $x\in\U$,
this gives $\trg$ and $\tlg$ on all $\smash$.

Keeping in mind that the coaction should describe the transformation
properties of the elements of $\smash$, we make the following choices:
$\A$ left coacts on $\smash$ so as to leave $\U$ invariant, \ie
\begin{equation}
\AD (x)\equiv\IA\otimes x,
\end{equation}
$x\in\U$.  Furthermore, $\A$ left and right coacts on $\A$ via the
coproduct:
\begin{equation}
\AD (a)=\DA(a)=\Delta (a),\label{coprod}
\end{equation}
for $a\in\A$, so on a element $ax\in\smash$,
\begin{equation}
\AD (ax)=a_{(1)}\otimes a_{(2)}x.
\end{equation}
The right coaction of $\A$ on $\U$ is taken to be the natural one
induced by the left adjoint action, namely $\DA(x)=x^{(1)}
\otimes x^{(2)'}$ with
\begin{equation}
y\ad x\equiv x^{(1)}\inprod{y}{x^{(2')}}\label{adjoint-coact}
\end{equation}
for $y\in\U$.  We can find a more explicit form of $\DA(x)$ by
introducing $\{ e_i |i\in\J\}$ as a basis for $\U$ ($\J$ is the
appropriate index set, assumed to be countable), and $\{ f^i|i\in\J\}$
as the basis for $\A$ such that $\inprod{e_i}{f^j} =\delta^i _j$.  We
now write $\DA(x)$ as
\begin{equation}
\DA(x)\equiv x_i \otimes f^i ,
\end{equation}
where $x_i \in\U$.  Therefore,
\begin{equation}
e_j \ad x = x_i \inprod{e_j }{f^i } = x_j ,
\end{equation}
giving
\begin{equation}
\DA(x)=(e_i \ad x)\otimes f^i .\label{coact}
\end{equation}
All of the above definitions are consistent with the conditions
necessary for $\DA$ to be a right coaction on $\smash$:
\begin{equation}
\DA (ax)=a_{(1)}(e_i \ad x)\otimes a_{(2)}f^i .
\end{equation}
As required, the left and right coactions of $\A$ on $\smash$ commute.

Since $\smash$ is an algebra on which $\A$ left and right coacts such
that the commutation relations (\ref{mult}) transform into themselves,
we will often say that $\smash$ is {\em bicovariant}, or, more
specifically, left-invariant and right-covariant \cite{Woronowicz1}.

Going back to the case where $\U$ and $\A$ are a quasitriangular Hopf
algebra and its associated quantum group respectively, we see that
\begin{equation}
\AD (A)= \DA (A)=A\stackrel{\cdot}{\otimes}A.
\end{equation}
The requirement that the coactions respect the commutation relations
(\ref{LA}) requires that $L^{\pm}$ be left-invariant.  Unfortunately,
without further information about $\U$, the right coactions cannot be
given more explicitly than through (\ref{coact}).  However, as we will
see in Chapter \ref{chap-QLA}, this will not be a major problem.

In a similar fashion to $\DA$, we can define a left coaction of $\U$
on $\smash$, $_{\U}\Delta : \smash\rightarrow\U\otimes\smash$, as
$\sigma\mapsto {}_{\U}\Delta (\sigma ):=\sigma^{(\bar{1})}\otimes
\sigma^{(2)}$.  On $\U$, $_{\U}\Delta$ is the coproduct:
\begin{equation}
_{\U}\Delta (x)\equiv x^{(\bar{1} )}\otimes x^{(2)}=x_{(1)}\otimes x_{(2)}
\equiv\Delta (x).
\end{equation}
On $\A$, $_{\U}\Delta$ is defined again implicitly via
\begin{equation}
ab=b_{(1)} \inprod{a^{(\bar{1} )}}{b_{(2)}}a^{(2)}.
\end{equation}
Using the right action of a function $b$ on another function $a$ given by
\begin{equation}
a \triangleleft b \equiv S(b_{(1)}) a b_{(2)},
\end{equation}
one can easily show that
\begin{equation}
_{\U}\Delta (a) = e_i \otimes (a \triangleleft f^i ),
\end{equation}
so for $ax\in\smash$,
\begin{equation}
_{\U}\Delta (ax)=e_i x_{(1)}\otimes (a\triangleleft f^i )x_{(2)}.
\end{equation}

\subsection{The Canonical Element of
$\smash$}\label{chap-coactions-canonical}

We are now in a position to introduce the canonical element $C$, which
lives in $\U\otimes\A$:
\begin{equation}
C:= e_i \otimes f^i.
\end{equation}
$C$ satisfies several relations; for instance, note that
\begin{eqnarray}
(\Delta\otimes\id)(C)&=&\Delta(e_i )\otimes f^i \nonumber \\
&=&(e_i )_{(1)}\otimes (e_i )_{(2)}\otimes f^i \nonumber \\
&=&e_i \otimes e_j \otimes f^i f^j \nonumber\\
&=&(e_i \otimes\IU\otimes f^i )(\IU\otimes e_j \otimes f^j )
\nonumber\\
&=&C_{13}C_{23}
\end{eqnarray}
(where in going from the second to the third line we have used the
duality between $\U$-comultiplication and $\A$-multiplication).
Similar calculations also give $(\id\otimes\Delta)(C)=C_{12}C_{13}$,
as well as the following:
\begin{eqnarray}
(S\otimes\id)(C)=(\id\otimes S)(C)&=&C^{-1},\nonumber \\
(\epsilon\otimes\id)(C)=(\id\otimes\epsilon)(C)&=&\IU\otimes\IA.
\end{eqnarray}
So far, $C$ does does not seem to be very interesting; however, to see
that it is indeed a useful quantity, we now compute the right coaction
of $\A$ on a basis vector in $\U$: using (\ref{coact}),
\begin{eqnarray}
\DA (e_i )&=&(e_j \ad e_i )\otimes f^j \nonumber\\
&=&(e_j )_{(1)}e_i S((e_j )_{(2)})\otimes f^j \nonumber \\
&=&e_m e_i S(e_n )\otimes f^m f^n \nonumber \\
&=&(e_m \otimes f^m )(e_i \otimes \IA)(S(e_n )\otimes f^n )\nonumber \\
&=&C(e_i \otimes\IA)(S\otimes\id)(C),
\end{eqnarray}
so for any $x\in\U$,
\begin{equation}
\DA (x)=C(x\otimes\IA)C^{-1}.
\end{equation}
A similar calculation shows that for $a\in\A$,
\begin{equation}
_{\U}\Delta (a)=C^{-1}(\IU\otimes a)C.
\end{equation}

So far in this section, we have not made any reference to the smash
product; however, when we think of $C$ as living in $(\smash)\otimes
(\smash)$, with $e_i$ and $f^i$ as the bases for the subalgebras $\U$
and $\A$ of $\smash$ respectively, $C$ takes on a much expanded role.
The first thing we notice is that for $a\in\A$,
\begin{eqnarray}
C(a\otimes\IS)C^{-1}&=&e_i aS(e_j )\otimes f^i f^j \nonumber\\
&=&\left( a_{(1)}(e_i )_{(2)}\inprod{(e_i )_{(1)}}{a_{(2)}}\right)
S(e_j )\otimes f^i f^j \nonumber\\
&=&a_{(1)}\inprod{(e_k )_{(1)}}{a_{(2)}}(e_k )_{(2)}S((e_k )_{(3)})
\otimes f^k \nonumber\\
&=&a_{(1)}\otimes\inprod{e_k }{a_{(2)}}f^k \nonumber\\
&=&a_{(1)}\otimes a_{(2)},
\end{eqnarray}
(where $\IS\simeq\IA\otimes\IU$ is the unit in $\smash$) so that
\begin{equation}
C(a\otimes\IS)C^{-1}=\Delta (a).
\end{equation}
Since this is just the right coaction of $\A$ on itself, we can
therefore write $\DA$ on all of $\smash$ as
\begin{equation}
\DA (\sigma )=C(\sigma\otimes\IS)C^{-1} \label{gencoact}
\end{equation}
for any $\sigma\in\smash$.  (This expression shows explicitly that
$\DA$ is an algebra homomorphism.)  We can continue doing calculations
along these lines, and we find that for $x\in\U$, $\Delta (x)=C^{-1}
(\IS\otimes x)C$, so that the left coaction of $\U$ on $\smash$ is
\begin{equation}
_{\U}\Delta (\sigma)=C^{-1}(\IS\otimes\sigma )C
\end{equation}
for $\sigma\in\smash$.  Using these results, together with the
coproduct relations for $C$, we obtain the equation
\begin{equation}
C_{23}C_{12}=C_{12}C_{13}C_{23}.\label{pentagon}
\end{equation}
Alternatively, this equation can be viewed as giving the
multiplication on $\smash$ as defined in (\ref{mult}).

In the case where $\U$ is a quasitriangular Hopf algebra with universal
R-matrix $\R$, the coproduct relations involving $C$ imply the following
consistency conditions:
\begin{eqnarray}
\R_{12}C_{13}C_{23}&=&C_{23}C_{13}\R_{12},\nonumber\\
\R_{23}C_{12}&=&C_{12}\R_{13}\R_{23}, \nonumber\\
\R_{13}C_{23}&=&C_{23}\R_{13}\R_{12}.\label{quasi}
\end{eqnarray}
To see the added significance of these equations, note that
\begin{equation}
\inprod{C}{a\otimes\id}=a,
\end{equation}
where $a\in\A$\footnote{We use the convention that the inner product
of any object with the identity map returns that object, \eg
$\inprod{x}{\id}=x.$}.  Now, let $\rho :\U\rightarrow M_N (k)$ be an
$N\times N$ matrix representation of $\U$, and $A$ the matrix of basis
elements of $\A$, as in Section \ref{chap-Hopf-reps}.  We see
immediately that $(\rho\otimes\id)(C)=A$.  Now let us apply
$\matrix{\rho}{i}{k}\otimes\matrix{\rho}{j}{\ell}\otimes\id$ to the
first of equations (\ref{quasi}); the left-hand side gives
\begin{eqnarray}
(\matrix{\rho}{i}{k}\otimes\matrix{\rho}{j}{\ell}\otimes\id)
\R_{12}C_{13}C_{23}&=&
(\matrix{\rho}{i}{m}\otimes\matrix{\rho}{j}{n})(\R)
(\matrix{\rho}{m}{k}\otimes\id)(C)(\matrix{\rho}{n}{\ell}\otimes\id)
(C)\nonumber\\
&=&\matrix{(RA_1 A_2)}{ij}{k\ell}.
\end{eqnarray}
The right-hand side gives $\matrix{(A_2 A_1 R)}{ij}{k\ell}$, so we
obtain (\ref{RAA})!  Doing similar gymnastics with the other two
equations in (\ref{quasi}), equations (\ref{LA}) can be obtained.
Thus, we recover all the commutation relations between the elements of
$A$ and between $L^{\pm}$ and $A$.

The physical content of the canonical element formulation presented
here has been discussed in \cite{Majid2}: the fact that $C$ generates
co\-ac\-tions on $\smash$, \ie trans\-for\-ma\-tions of operators and
functions, suggests a possible interpretation of $C$ as a
time-evolution operator for certain Hamiltonian systems which may be
formulated in a Hopf algebraic manner.

\section{Example:  The 2-Dimensional Quantum Euclidean Group}
\label{chap-coactions-euclidean}

In this section, we present an example which will serve to illustrate
many of the concepts we discussed in this chapter.  We begin by
presenting a review of Woronowicz's deformation of the 2-dimensional
Euclidean group $E(2)$ \cite{Woronowicz2}: he introduces the *-Hopf
algebra $\A$ generated by elements $n$, $v$, and $\bar{n}$ which
satisfy
\begin{eqnarray}
vn=qnv,&v\bar{n}=q\bar{n}v,&n\bar{n}=\bar{n}n,\nonumber\\
\Delta (n)=n\otimes v^{-1}+v\otimes n,&\Delta (\bar{n})=\bar{n}
\otimes v+v^{-1}\otimes\bar{n},&\Delta (v)=v\otimes v,\nonumber\\
\epsilon (n)=\epsilon (\bar{n})=0,&\epsilon (v)=1,&\nonumber\\
S(n)=-q^{-1}n,&S(\bar{n})=-q\bar{n},&S(v)=v^{-1},\nonumber\\
\theta (n)=\bar{n},&\theta (v)=v^{-1},&
\end{eqnarray}
with $q\in\real$.  For the calculations which follow, it is convenient
to introduce the elements $\gamma$, $m$ and $\bar{m}$, defined by
\begin{eqnarray}
\gamma:=-2i\ln v,&m:=nv,&\bar{m}:=v^{-1}\bar{n},
\end{eqnarray}
All of these new elements have vanishing counit, and have commutation
relations, coproducts, antipodes, and conjugates given by
\begin{eqnarray}
\comm{\gamma}{m}=-2i\ln q\, m,&\comm{\gamma}{\bar{m}}=-2i\ln q\, \bar{m},
&m\bar{m}=q^2 \bar{m}m,\nonumber\\
\Delta (m) =m\otimes\IA +e^{i\gamma}\otimes m,&\Delta (\bar{m})=\bar{m}
\otimes\IA  +e^{-i\gamma}\otimes\bar{m},&\nonumber\\
\Delta (\gamma)=\gamma\otimes\IA+\IA\otimes\gamma,& &\nonumber\\
S(m)=-e^{-i\gamma}m,&S(\bar{m})=-e^{i\gamma}\bar{m},&S(\gamma)=-\gamma
,\nonumber\\
\theta (m)=\bar{m},&\theta (\gamma)=\gamma .&
\end{eqnarray}

Note that the $2\times 2$ matrices $E$ and $\bar{E}=\theta (E)$ given
by
\begin{eqnarray}
E=\left( \begin{array}{cc}
e^{i\gamma}&m\\0&\IA \end{array} \right) ,&\bar{E}=\left(
\begin{array}{cc} e^{-i\gamma}&\bar{m}\\0&\IA \end{array} \right)
\end{eqnarray}
satisfy the relations (\ref{QG-Hopf}).  These are exactly the
relations one would expect for a quantum group matrix, despite the
fact that we have not seen any sign of an R-matrix yet.  To further
interpret what we have here, let us introduce the deformed complex
plane $\complex_q$ as the unital associative algebra generated by
$z,\bar{z}$ which satisfy $z\bar{z}=q^2 \bar{z}z$.  We define a left
coaction of $\A$ on $\complex_q$ as
\begin{eqnarray}
\AD (z):=e^{i\gamma}\otimes z+m\otimes 1_{\complex_q} ,&\AD (\bar{z})
:=e^{-i\gamma}\otimes\bar{z}+\bar{m}\otimes 1_{\complex_q} .
\end{eqnarray}
By introducing the column vectors $z^i :=\left( \begin{array}{c}
z\\1_{\complex_q}\end{array}\right)$ and $\bar{z}^i :=\left(
\begin{array}{c} \bar{z}\\1_{\complex_q}\end{array}\right)$, these can
be rewritten as
\begin{eqnarray}
\AD (z^i )=\matrix{E}{i}{j}\otimes z^j ,&\AD (\bar{z}^i )=
\matrix{\bar{E}}{i}{j}\otimes\bar{z}^j .
\end{eqnarray}
This suggests identifying $\A$ as a deformation of the 2-dimensional
Euclidean group, which we denote $E_q (2)$ (this is just a particular
example of an inhomogeneous quantum group \cite{SWW,Cast}).  We will
now make an explicit construction (following the methods in
\cite{Rosso}) of the dually paired *-Hopf algebra $\U$, which will be
identifiable with $U_q(e(2))$, the quantized universal enveloping
algebra of the 2-dimensional Euclidean algebra.  We choose span$\{
\gamma^a m^b\bar{m}^c |a,b,c=0,1,\ldots \}$ as a basis for $\A$, and
define $h$, $\mu$, and $\nu$ as the elements of $\U$ whose inner
products with these basis elements are
\begin{eqnarray}
\inprod{\mu}{\gamma^a m^b \bar{m}^c}:=\delta_{a,0}\delta_{b,1}
\delta_{c,0},&\inprod{\nu}{\gamma^a m^b \bar{m}^c}:=\delta_{a,0}
\delta_{b,0}\delta_{c,1},\nonumber\\
\inprod{h}{\gamma^a m^b \bar{m}^c}:=\delta_{a,1}\delta_{b,0}
\delta_{c,0}.
\end{eqnarray}
We require the two algebras to be dually paired; therefore, using the
coproduct on $\A$ to obtain the multiplication on $\U$ gives
\begin{equation}
\inprod{\nu^k \mu^{\ell} h^n}{\gamma^a m^b \bar{m}^c}=\quint{k}{q}!
\quint{\ell}{q^{-1}}!n!\delta_{n,a}\delta_{\ell ,b}\delta_{k,c}
\end{equation}
so $\{ \nu^k \mu^{\ell}h^n |k,\ell ,n=0,1,\ldots \}$ is a basis for
$\U$.  The rest of the *-Hopf algebra structure of $\U$ can be
similarly obtained:
\begin{eqnarray}
\comm{h}{\mu}=i\mu,&\comm{h}{\nu}=-i\nu,&\mu\nu=q^2 \nu\mu,
\nonumber\\
\Delta (\mu)=\mu\otimes q^{2ih}+\IU\otimes\mu,&\Delta (\nu)=\nu
\otimes q^{2ih}+\IU\otimes\nu,\nonumber\\
\Delta (h)=h\otimes\IU +\IU\otimes h,&\epsilon (\mu)=\epsilon
(\nu)=\epsilon (h)=0,\nonumber\\
S(\mu)= -\mu q^{-2ih},&S(\nu)=-\nu q^{-2ih},&S(h)=-h,
\nonumber\\
\theta(h)=-h,&\theta(\mu)=-q^2 \nu,&\theta(\nu)=-q^{-2}\mu.
\end{eqnarray}
Defining new operators $J$, $P_{\pm}$ as
\begin{eqnarray}
J:=ih,& P_+ :=q^{1-ih}\nu,&P_- := -\mu q^{-1-ih},
\end{eqnarray}
we find that they all have vanishing counit, $\theta(J)=J$, $\theta
(P_{\pm}) =P_{\mp}$, and
\begin{eqnarray}
\comm{J}{P_{\pm}}=\pm P_{\pm},&\comm{P_+}{P_-}=0,\nonumber\\
\Delta (P_{\pm})=P_{\pm}\otimes q^J +q^{-J}\otimes P_{\pm},&\Delta
(J)=J\otimes\IU +\IU\otimes J,\nonumber\\
S(J)=-J,&S(P_{\pm})=-q^{\pm 1} P_{\pm}.\label{gens-euclidean}
\end{eqnarray}
An interesting fact is that as a unital associative *-algebra, this is
the {\em undeformed} UEA of the classical algebra $e(2)$ \cite{CGST}.
However, it has a nontrivial *-Hopf algebra structure.

The inner products between these new generators of $\U$ and the basis
elements of $\A$ can be computed; they are
\begin{eqnarray}
\inprod{P_+^k P_-^{\ell} J^n}{\gamma^a m^b \bar{m}^c}&=&\frac{i^a
a!(-1)^{\ell +a-n}}{(a-n)!}q^{-\half (k-\ell )(k+\ell -1)+\ell
(k-1)}(\ln q^{k+\ell})^{a-n}\nonumber\\
&\times&\quint{k}{q}!\quint{\ell}{q^{-1}}!\Theta_{a,n}\delta_{\ell
,b}\delta_{k,c},
\end{eqnarray}
where $\Theta_{i,j}$ is 1 if $i\geq j$ and zero otherwise.  With these
inner products in hand, plus the coproducts given in
(\ref{gens-euclidean}), we may use (\ref{diffact}) to find the actions
of $J$ and $P_{\pm}$ on the basis elements of $\A$.  They are
\begin{eqnarray}
P_+ \trg \gamma^a m^b \bar{m}^c &=&q^{2b}\quint{c}{q}e^{-i\gamma}(\gamma
-i\ln q)^a m^b \bar{m}^{c-1},\nonumber\\
P_- \trg \gamma^a m^b \bar{m}^c &=&-\inv{q}\quint{b}{q^{-1}}e^{i\gamma}
(\gamma -i\ln q)^a m^{b-1}\bar{m}^c ,\nonumber\\
J\trg \gamma^a m^b \bar{m}^c &=&ia\gamma^{a-1}m^b \bar{m}^c .
\end{eqnarray}
Therefore, if $f(\gamma,m,\bar{m})$ is a function written in terms of
the basis elements of $\A$, we find
\begin{eqnarray}
P_+ \trg f(\gamma,m,\bar{m})&=&e^{-i\gamma}\frac{f(\gamma-i\ln q,q^2
m,q^2 \bar{m})-f(\gamma-i\ln q,q^2 m,\bar{m})}{q^2 -1}\inv{\bar{m}},
\nonumber\\
P_+ \trg f(\gamma,m,\bar{M})&=&-\inv{q}e^{i\gamma}\frac{f(\gamma-i\ln
q,q^{-2}m,q^{-2}\bar{m})-f(\gamma-i\ln q,m,q^{-2}\bar{m})}{q^{-2}-1}
\inv{m},\nonumber\\
J\trg f(\gamma,m,\bar{m})&=& i\partder{}{\gamma}f(\gamma,m,\bar{m}).
\end{eqnarray}
In the $q\rightarrow 1$ limit, we see that $(P_+\trg)\rightarrow
e^{-i\gamma}\partder{}{\bar{m}}$, $(P_-\trg)\rightarrow -e^{i\gamma}
\partder{}{m}$, and $(J\trg)\rightarrow i\partder{}{\gamma}$, which are
precisely what we'd expect.  However, for the $q\neq 1$ case, the
actions of $P_{\pm}$ give differences rather than derivatives.  It
might therefore be possible to use these expressions to consistently
regularize a 2-dimensional theory with Euclidean symmetry: the
expressions for $P_{\pm}\trg$ above involve differences between
neighboring points on a 3-dimensional lattice where the
$\gamma$-lattice spacing is $i\ln q$, and neighboring points in the
$m$- and $\bar{m}$-lattices differ by a factor of $q^{\pm 2}$ (the
action of $J$ can be treated classically).  In the $q\rightarrow 1$
limit, the lattice spacings will shrink to zero, but the finiteness of
the theory for $q\neq 1$ may allow us to control any divergences which
arise.

What are the commutation relations within the smash product for this
example?  Since we have the inner products and coproducts, we just
turn the crank and use (\ref{mult}) to find
\begin{eqnarray}
\comm{P_{\pm}}{\gamma}=-i\ln q\, P_{\pm},&&\comm{J}{\gamma}=i,\nonumber\\
\comm{P_+}{\bar{m}}=e^{-i\gamma}q^J ,&&\comm{P_-}{m}=-\inv{q}e^{i\gamma}
q^J ,
\end{eqnarray}
with all other commutators between $J,P_+ ,P_-$ and $\gamma,m,\bar{m}$
vanishing.  All commutation relations within $\smash$ may be obtained
from these.

\chapter{Quantum Lie Algebras}\label{chap-QLA}

\section{Basics of Quantum Lie Algebras}\label{chap-QLA-basics}

Let $\U$ be a Hopf algebra; we say that $\U$ is a {\em quantum Lie
algebra (QLA)} iff there exists a finite subspace $\g\subset\U$
(dim$\,\g=n$) which has the following properties:
\begin{enumerate}
\item As a vector space, $\U\equiv U_q (\g)$, \ie the universal
enveloping algebra (UEA) of $\g$ modulo commutation relations;
\item The adjoint action $\ad$ closes on $\g$, \ie $y\ad\chi\in\g$ for
all $y\in\U$ and $\chi\in\g$;
\item $\Delta(\chi )\in\U\otimes (\IU\oplus\g)$ for all $\chi\in\g$;
\item For all $\chi\in\g$, $\epsilon(\chi )=0$.
\end{enumerate}
(the $q$ subscript in (1) simply indicates that the commutation
relations may be deformed relative to the classical case.)  Let $\{
\gen{A}|A=1,2,\ldots,n \}$ be a basis for $\g$ \cite{Bernard}; (3)
therefore requires the coproduct to take the form
\begin{equation}
\Delta(\gen{A})=\gen{A}'\otimes\IU +\O{A}{B}
\otimes\gen{B},
\end{equation}
where $\gen{A}',\O{A}{B}\in\U$.  However, the fact that $\U$ is a Hopf
algebra requires
\begin{eqnarray}
\gen{A}&\equiv&(\id\otimes\epsilon)\Delta(\gen{A})\nonumber\\
&=&\gen{A}'\epsilon(\IU)+\O{A}{B}\epsilon(\gen{B}),
\end{eqnarray}
so (4) imposes the condition that
\begin{equation}
\Delta(\gen{A})=\gen{A}\otimes\IU+\O{A}{B}
\otimes\gen{B}.\label{cop-chi}
\end{equation}
Using this, and the other requirements for $\U$ to be a Hopf algebra,
we find the following:
\begin{eqnarray}
\Delta(\O{A}{B})=\O{A}{C}\otimes\O{C}{B},&
\epsilon(\O{A}{B})=\delta_{A}^{B},&\nonumber\\
S(\O{A}{B})=({\cal O}^{-1})_{A}{}^{B},&S(
\gen{A})=-S(\O{A}{B})\gen{B}.
\end{eqnarray}
Condition (2) allows us express the commutation relations between
elements of $\g$ (and therefore between all elements of $\U$) in a
more transparent form.  To see this, we define the $k$-numbers
$\bigR{AB}{CD}$ and $\struc{AB}{C}$
via
\begin{eqnarray}
\gen{A}\ad\gen{B}&:=&\struc{AB}{C}\gen{C},\nonumber\\
\O{C}{B}\ad\gen{D}&:=&\bigR{AB}{CD}\gen{A}.
\end{eqnarray}
$\bigR{}{}$ is referred to as the R-matrix of $\g$, and the $f$s are,
as in the classical case, just the structure constants of $\g$.
$\bigR{}{}$ is invertible, with $\bigR{-1}{}$ given by
\begin{equation}
S^{-1}(\O{D}{A})\ad\gen{C}=\matrix{(\bigR{-1}{})}{AB}{CD}\gen{B},
\end{equation}
and the matrix $\tilde{\real}$ (see Appendix \ref{chap-matrix}) is
given by
\begin{equation}
S(\O{C}{A} )\ad\gen{D}=\matrix{\tilde{\real}}{AB}{CD}\gen{B}.
\end{equation}

Now, note that for any Hopf algebra $\U$, we have the following
identity:
\begin{equation}
(x_{(1)}\ad y)x_{(2)}=x_{(1)}yS(x_{(2)})x_{(3)}=xy\label{adj-mult}
\end{equation}
for all $x,y\in\U$.  Therefore, using (\ref{cop-chi}),
\begin{equation}
\gen{A}\gen{B}=(\gen{A}\ad\gen{B})\IU+(\O{A}{D}\ad\gen{B})\gen{D}.
\end{equation}
When we use the explicit forms of the adjoint actions, we have
\begin{equation}
\gen{A}\gen{B}-\bigR{CD}{AB}\gen{C}\gen{D}=\struc{AB}{C}\gen{C},
\label{Lie-comm}
\end{equation}
which are the commutation relations between basis elements of $\g$.
Here we see explicitly the ``deformation'' of the algebra, via the
R-matrix.  In the classical case, $\bigR{AB}{CD} =\delta^A_D
\delta^B_C$, and the left-hand side of (\ref{Lie-comm}) reduces to the
commutator.  Thus, the commutation relations between the generators
are parametrized not only by the structure constants, but by the
R-matrix as well.  (However, this is not the full story; in the next
section, we will also see that there will have to be certain numerical
conditions between $\bigR{}{}$ and the $f$s to ensure consistency of
the algebra.)  The adjoint action is still given entirely by the
structure constants, though.

By continuing along the above lines, we find more commutation
relations, involving the $\cal O$s:
\begin{eqnarray}
\bigR{EF}{AB}\O{E}{C}\O{F}{D}&=&\O{A}{E}\O{B}{F}\bigR{CD}{EF},
\nonumber\\
\gen{A}\O{B}{C}-\bigR{DE}{AB}\O{D}{C}\gen{E}&=&\struc{AB}{D}
\O{D}{C}-\O{A}{D}\O{B}{E}\struc{DE}{C},\nonumber\\
\O{A}{B}\gen{C}&=&\bigR{DE}{AC}\gen{D}\O{E}{B}.\label{O-comm}
\end{eqnarray}
The last of these is a consistency condition the fact that
the elements of $\O{A}{B}$ are expressible in terms of the $\chi$s,
due to $\U=U_q(\g)$.  We also find that the self-consistency of these
relations requires
\begin{equation}
\bigR{}{12}\bigR{}{23}\bigR{}{12}=\bigR{}{23}\bigR{}{12}\bigR{}{23}.
\end{equation}
So even though we have not said anything at all about the
quasitriangularity of $\U$, we see that the R-matrix associated with a
QLA must satisfy a numerical Yang-Baxter equation.  However, as we
will see, this matrix is {\em not} the representation of the universal
R-matrix when $\U$ is in fact quasitriangular.

\section{The Adjoint Representation}

The closure of $\g$ under the adjoint action defines the adjoint
representation ad of $\U$ (with module $\g$) as
\begin{equation}
y\ad\gen{A}=\gen{B}\,\matrix{(\mbox{ad}(y))}{B}{A}.
\end{equation}
As is discussed in Chapter \ref{chap-Hopf}, this motivates the
introduction of elements $\bigA{A}{B}$ in the Hopf algebra
$\A$ dually paired with $\U$, given by
\begin{equation}
\matrix{\mbox{ad}(y)}{A}{B}=\inprod{y}{\bigA{A}{B}}.
\end{equation}
Therefore, we find
\begin{eqnarray}
\inprod{\gen{A}}{\bigA{C}{B}}=\struc{AB}{C},&
\inprod{\O{C}{B}}{\bigA{A}{D}}=\bigR{AB}{CD},&\nonumber\\
\inprod{S^{-1}(\O{D}{A})}{\bigA{B}{C}}=
\matrix{(\hat{\real}^{-1})}{AB}{CD},& \inprod{
S(\O{C}{A})}{\bigA{B}{D}}=\matrix{\tilde{\real}
}{AB}{CD}.&
\end{eqnarray}
By using the definition of the right coaction of $\A$ on $\U$ given in
Chapter \ref{chap-coactions}, we see that
\begin{equation}
\DA(\gen{A})=\gen{B}\otimes\bigA{B}{A}.\label{gens-coact}
\end{equation}
For consistency with the defining properties of the QLA, the adjoint
matrices $\bigA{}{}$ must satisfy the following:
\begin{eqnarray}
\Delta(\bigA{A}{B})=\bigA{A}{C}\otimes\bigA{C}{B},&
\epsilon(\bigA{A}{B})=\delta^A_B,\nonumber\\
S(\bigA{A}{B})=\matrix{(\bigA{-1}{})}{A}{B},&\bigR{}{}
\bigA{}{1}\bigA{}{2}=\bigA{}{1}\bigA{}{2}\bigR{}{},\nonumber\\
\struc{AB}{D}\bigA{C}{D}=\bigA{D}{A}\bigA{E}{B}\struc{DE}{C}.
\end{eqnarray}
Once again, even though we did not assume that $\U$ was quasitriangular,
the dual appearing here has a very quantum-grouplike structure to it.

We can use the above properties of $\bigA{}{}$ to find several
numerical relations among the R-matrix and structure constants; for
instance, if we take the inner product of $\bigA{M}{N}$ and
(\ref{Lie-comm}), we find the deformed version of the Jacobi identity:
\begin{equation}
\struc{AL}{M}\struc{BN}{L}-\bigR{CD}{AB}\struc{CL}{M}\struc{DN}{L}
=\struc{AB}{C}\struc{CN}{M}.
\end{equation}
Repeating this for the first of (\ref{O-comm}) just recovers the
numerical Yang-Baxter relation for $\bigR{}{}$; the other two give
\begin{eqnarray}
\bigR{DC}{BN}\struc{AD}{M}-\bigR{DE}{AB}\bigR{MC}{DF}\struc{EN}{F}
&=&\bigR{MC}{DN}\struc{AB}{D}-\bigR{DF}{BN}\bigR{ME}{AD}
\struc{EF}{C},\nonumber\\
\bigR{MB}{AD}\struc{CN}{D}&=&\bigR{DE}{AC}\bigR{FB}{EN}\struc{DF}{M}.
\end{eqnarray}
These are the numerical relations alluded to earlier which to specify
the QLA.

The commutation relations between the elements of $\U$ and the adjoint
matrices can be determined by using the inner products given above and
the smash product.  They take the form
\begin{eqnarray}
\gen{A}\bigA{B}{C}&=&\bigR{DE}{AC}\bigA{B}{D}\gen{E}+\struc{AC}{D}
\bigA{B}{D},\nonumber\\
\O{A}{B}\bigA{C}{D}&=&\bigR{EF}{AD}\bigA{C}{E}\O{F}{B}.
\end{eqnarray}

\section{Quasitriangular Quantum Lie Algebras}

Now, in the case where $\U$ is in fact quasitriangular, we can use the
contents of Chapter \ref{chap-Hopf-reps} to immediately obtain a QLA
\cite{Jurco}.  This is done as follows: let $\rho$ be a representation
of $\U$; we therefore have the matrices $L^{\pm}$ which satisfy
(\ref{LL1}) and (\ref{LL2}).  We define the matrix $Y$ by
\cite{RS,Leipzig}
\begin{equation}
Y:=L^+ S(L^- )\equiv (\rho\otimes\id)\R_{21}\R ;\label{Y-def}
\end{equation}
this matrix therefore satisfies
\begin{eqnarray}
L^+_1 Y_2 =R_{21}^{-1}Y_2 R_{21}L^+_1,&L^-_1 Y_2 =RY_2 R^{-1}L^-_1
,\nonumber\\
R_{21}Y_1 R Y_2 =Y_2 R_{21}Y_1 R.
\end{eqnarray}
$Y$ has coproduct, counit and antipode given by
\begin{eqnarray}
\Delta (\matrix{Y}{i}{j})=\matrix{(L^+ )}{i}{k}S(
\matrix{(L^- )}{\ell}{j})\otimes \matrix{Y}{k}{\ell},&
\epsilon(\matrix{Y}{i}{j})=\delta^i_j ,&\nonumber\\
S(\matrix{Y}{i}{j})=S^2 (\matrix{(L^-)}{k}{j})S(\matrix{(L^+)}{i}{k})
.&&
\end{eqnarray}
We would naturally like to know what the coactions of $\A$ are on $Y$;
since $L^{\pm}$ are left-invariant, so is $Y$.  The right coaction is
a bit more problematic, since we do not have explicit forms for the
right coactions of $L^{\pm}$.  However, we can get around this in the
following way:  for $a\in\A$, we define $\Upsilon_a \in\U$ as
\begin{equation}
\Upsilon_a :=\inprod{\R_{21}\R}{a\otimes\id}.
\end{equation}
Thus, by definition,
\begin{equation}
\matrix{Y}{i}{j}=\Upsilon_{\matrix{A}{i}{j}}.
\end{equation}
Now, we note that for $x\in\U$,
\begin{eqnarray}
x\ad\Upsilon_a &=& x\ad\inprod{\R_{21}\R}{a\otimes\id}\nonumber\\
&=&\inprod{(\IU\otimes x_{(1)})\R_{21}\R (\IU\otimes
S(x_{(2)}))}{a\otimes\id}\nonumber\\
&=&\inprod{(\IU\otimes x_{(1)})\R_{21}\R (S(x_{(3)})x_{(4)}\otimes
S(x_{(2)}))}{a\otimes\id}\nonumber\\
&=&\inprod{(\IU\otimes x_{(1)})\R_{21}\R\Delta (S(x_{(2)}))(x_{(3)}
\otimes\IU)}{a\otimes\id}\nonumber\\
&=&\inprod{(\IU\otimes x_{(1)})\Delta (S(x_{(2)}))\R_{21}\R
(x_{(3)}\otimes\IU)}{a\otimes\id}\nonumber\\
&=&\inprod{(S(x_{(3)})\otimes x_{(1)}S(x_{(2)}))\R_{21}\R
(x_{(4)}\otimes\IU)}{a\otimes\id}\nonumber\\
&=&\inprod{(S(x_{(1)})\otimes\IU)\R_{21}\R(x_{(2)}\otimes
\IU)}{a\otimes\id}\nonumber\\
&=&\inprod{S(x_{(1)})\otimes x_{(2)}}{a_{(1)}\otimes a_{(3)}}
\inprod{\R_{21}\R}{a_{(2)}\otimes\id}\nonumber\\
&=&\inprod{x}{S(a_{(1)})a_{(3)}}\Upsilon_{a_{(2)}},\label{closure}
\end{eqnarray}
where we have made ample use of the various properties of dually
paired Hopf algebras, and used the very important fact that
$\R_{21}\R$ commutes with all of $\Delta (\U )$.  Thus, from
(\ref{coact}),
\begin{eqnarray}
\DA (\Upsilon_a )&=&(e_i \ad\Upsilon_a )\otimes f^i \nonumber\\
&=&\inprod{e_i }{S(a_{(1)})a_{(3)}}\Upsilon_{a_{(2)}}\otimes f^i
\nonumber\\
&=&\Upsilon_{a_{(2)}}\otimes S(a_{(1)})a_{(3)}.
\end{eqnarray}
(Note the appearance of the adjoint coaction (\ref{adj-coact}) in this
equation.)  Therefore, we find that $\A$ right coacts on $Y$ as
\begin{equation}
\DA (\matrix{Y}{i}{j})=\matrix{Y}{k}{\ell}\otimes S(\matrix{A}{i}{k})
\matrix{A}{\ell}{j}.
\end{equation}
The above calculation has an added bonus: (\ref{closure}) tells us
that the adjoint action of $\U$ on any element in the subspace $\{
\Upsilon_a |a\in\A\}$ returns another element of the same subspace.
In particular,
\begin{equation}
x\ad\matrix{Y}{i}{j}=\inprod{x}{S(\matrix{A}{i}{k})
\matrix{A}{\ell}{j}}\matrix{Y}{k}{\ell},\label{closure-qla}
\end{equation}
which is simply a linear combination of the entries of $Y$.

Notice that in the classical limit, since $\R\rightarrow\IU\otimes
\IU$, $Y\rightarrow I\IU$; therefore, we can define the matrix $X$ by
\begin{equation}
X:=\frac{I\IU -Y}{\lambda},\label{X-def}
\end{equation}
where, as always, $\lambda=q-\inv{q}$.  Thus, in the $q\rightarrow 1$
limit, $X$ is well-defined.  However, the real reason for defining
this new matrix becomes apparent when we look at its properties which
follow from those of $Y$:  the relevant commutation relations are
\begin{eqnarray}
&L^+_1 X_2 =R_{21}^{-1}X_2 R_{21}L^+_1 ,\,\, L^-_1 X_2 =RX_2
R^{-1}L^-_1, &\nonumber\\
&R_{21}X_1 RX_2 -X_2 R_{21}X_1 R=\frac{1}{\lambda}(R_{21}RX_2 -X_2
R_{21}R),&\label{X-X}
\end{eqnarray}
and the Hopf algebra properties of $X$ are
\begin{eqnarray}
\Delta (\matrix{X}{i}{j})=\matrix{X}{i}{j}\otimes\IU +
\matrix{(L^+)}{i}{k}S(\matrix{(L^-)}{\ell}{j})\otimes
\matrix{X}{k}{\ell},&\epsilon (X)=0,& \nonumber\\
S(\matrix{X}{i}{j})=-S^2 (\matrix{(L^-)}{\ell}{j})S(
\matrix{(L^+)}{i}{k})\matrix{X}{k}{\ell}.
\end{eqnarray}
It follows immediately that $\A$ left and right coacts on $X$ exactly
as it does on $Y$, \ie
\begin{eqnarray}
\AD (\matrix{X}{i}{j})=\IA\otimes\matrix{X}{i}{j},&\DA
(\matrix{X}{i}{j})=\matrix{X}{k}{\ell}\otimes S(\matrix{A}{i}{k})
\matrix{A}{\ell}{j},\label{X-coacts}
\end{eqnarray}

The adjoint action of $x\in\U$ on an entry of $X$ is given by
(\ref{closure-qla}) with $Y$ replaced by $X$, and this returns an
element in $\g$, the subspace of $\U$ defined to be the span of the
entries of $X$ over $k$.  Furthermore, $\epsilon (\g )=0$; thus, the
UEA $U_q (\g )$ satisfies all criteria needed for a QLA.  We therefore
see that any quasitriangular Hopf algebra, together with a
representation, allows the construction of a QLA.  The connection to
the contents of Section \ref{chap-QLA-basics} is made by taking the
capital roman indices to correspond to pairs of small roman indices in
the present quasitriangular case.  To see how this is done, we compute
the adjoint action of an element of $X$ on another:
\begin{eqnarray}
\matrix{X}{i}{j}\ad\matrix{X}{k}{\ell}&=&\matrix{X}{i}{j}
\matrix{X}{k}{\ell}-\matrix{(R_{21}^{-1}X_2 R_{21}X_1 R)}{im}{n\ell}
\matrix{\tilde{R}}{nk}{jm}\nonumber\\
&=&\matrix{\tilde{R}}{sk}{jr}\matrix{R}{ri}{nm}\matrix{(R_{21}X_1 RX_2
-X_2 R_{21}X_1 R)}{mn}{s\ell}\nonumber\\
&=&\inv{\lambda}\left[ \delta^i_j \matrix{X}{k}{\ell}-\matrix{(
R_{21}^{-1}X_2 R_{21}R)}{im}{n\ell}\matrix{\tilde{R}}{nk}{jm}\right] .
\label{X-ad-X}
\end{eqnarray}
Comparison with (\ref{Lie-comm}) motivates the definitions of the
generators, $\cal O$s, and adjoint matrices as
\begin{eqnarray}
\gen{(ij)}:=\matrix{X}{i}{j},&\O{(ij)}{(k\ell )}:=\matrix{(L^+
)}{i}{k}S(\matrix{(L^- )}{\ell}{j}),&\bigA{(ij)}{(k\ell )}:=S(
\matrix{A}{k}{i})\matrix{A}{j}{\ell},\label{adjoint-rep}
\end{eqnarray}
and the R-matrix and structure constants as
\begin{eqnarray}
\bigR{(ab)(cd)}{(ij)(k\ell )}&:=&\matrix{\tilde{R}}{mk}{jn}
\matrix{\Rhat}{sd}{m\ell}\matrix{(\Rhat^{-1})}{ni}{ra}
\matrix{\Rhat}{rb}{sc},\nonumber\\
\struc{(ij)(k\ell )}{(rs)}&:=&\inv{\lambda}\left[ \delta^i_j
\delta^k_r \delta^s_{\ell} -\matrix{\tilde{R}}{mk}{jn}
\matrix{(\Rhat^{-1})}{ni}{tr}\matrix{(\Rhat^2 )}{ts}{m\ell}\right] .
\end{eqnarray}
However, notice that the universal R-matrix in this representation,
\ie $\matrix{\real}{AB}{CD}:=\inprod{\R}{\bigA{A}{C}\otimes
\bigA{B}{D}}$, is
\begin{eqnarray}
\matrix{\real}{(ab)(cd)}{(ij)(k\ell )}&=&\inprod{\R}{S(
\matrix{A}{i}{a})\matrix{A}{b}{j}\otimes S(\matrix{A}{k}{c})
\matrix{A}{d}{\ell}}\nonumber\\
&=&\matrix{\tilde{R}}{mk}{jn}\matrix{\Rhat}{sb}{m\ell}
\matrix{\Rhat}{ni}{rc}\matrix{(\Rhat^{-1})}{rd}{sa},
\end{eqnarray}
which is {\em not} equal to $\bigR{(cd)(ab)}{(ij)(k\ell )}$.

\subsection{Example:  $U_q(sl(2))$}

To provide a concrete example of the results of the previous
subsection, we consider the QLA $U_q(sl(2))$: define $\gen{1}$,
$\gen{+}$, $\gen{-}$ and $\gen{2}$ to be the entries of the $2\times
2$ matrix of generators $X$, \ie
\begin{equation}
X=\left( \begin{array}{cc}
\gen{1}&\gen{+}\\
\gen{-}&\gen{2}
\end{array} \right) .\label{chi-fund}
\end{equation}
Now, putting the R-matrix for $SL_q(2)$, \ie (\ref{R-sl2}), into the
expression for $\bigR{}{}$ from above, we find the $16\times 16$
matrix
\begin{equation}
\left( \begin{array}{rccccccccccccccl}
1&0&0&0&0&0&0&0&0&0&0&0&0&0&0&0\\
0&\frac{\lambda}{q}&0&0&1&0&0&0&0&0&0&0&0&0&0&0\\
0&0&-q\lambda&0&0&0&0&0&1&0&0&0&0&0&0&0\\
\frac{\lambda^2}{q^2}&0&0&-\lambda^2&0&0&-\frac{\lambda}{q}&0&0&
\frac{\lambda}{q}&0&0&1&0&0&0\\
0&\inv{q^2}&0&0&0&0&0&0&0&0&0&0&0&0&0&0\\
0&0&0&0&0&1&0&0&0&0&0&0&0&0&0&0\\
\frac{\lambda}{q}&0&0&-q\lambda&0&0&0&0&0&1&0&0&0&0&0&0\\
0&\frac{(q^4-1)\lambda}{q^3}&0&0&\frac{\lambda}{q}&0&0&-q\lambda
&0&0&0&0&0&q^2&0&0\\
0&0&q^2&0&0&0&0&0&0&0&0&0&0&0&0&0\\
-\frac{\lambda}{q}&0&0&q\lambda&0&0&1&0&0&0&0&0&0&0&0&0\\
0&0&0&0&0&0&0&0&0&0&1&0&0&0&0&0\\
0&0&0&0&0&0&0&0&-\frac{\lambda}{q^3}&0&0&\frac{\lambda}{q}&0&0&
\inv{q^2}&0\\
0&0&0&1&0&0&0&0&0&0&0&0&0&0&0&0\\
0&-\frac{\lambda}{q}&0&0&0&0&0&1&0&0&0&0&0&0&0&0\\
0&0&q\lambda&0&0&0&0&0&0&0&0&1&0&0&0&0\\
-\frac{\lambda^2}{q^2}&0&0&\lambda^2&0&0&\frac{\lambda}{q}&0&0&
-\frac{\lambda}{q}&0&0&0&0&0&1
\end{array}\right)
\end{equation}
where we have taken the ordered basis $\{\gen{1},\gen{+},\gen{-},
\gen{2} \}$.  The nonvanishing structure constants, also using the
expression given, are
\begin{equation}
\begin{array}{llll}
\struc{11}{1}=-\frac{\lambda}{q^2},&\struc{11}{2}=\frac{\lambda}{q^2},
&\struc{1+}{+}=\inv{q}+\inv{q^3}-q,&\struc{1-}{-}=-q,\\
\struc{12}{1}=\lambda ,&\struc{12}{2}=-\lambda,&\struc{2+}{+}=-q,&
\struc{2-}{-}=\inv{q},\\
\struc{+1}{+}=-\inv{q},&\struc{+-}{1}=\inv{q},&\struc{+-}{2}=
-\inv{q},&\struc{+2}{+}=q,\\
\struc{-1}{-}=\inv{q^3},&\struc{-+}{1}=-\inv{q},&\struc{-+}{2}
=\inv{q},&\struc{-2}{-}=-\inv{q}.
\end{array}
\end{equation}
Naturally, in the $q\rightarrow 1$ limit, we get $\bigR{AB}{CD}=
\delta^A_D \delta^B_C$ and the correct (antisymmetric) structure
constants for $sl(2)$.

So what commutation relations do these give?  They take the form
\begin{eqnarray}
\gen{1}\gen{+}=\gen{+}\gen{1}+\inv{q}\gen{+}-\frac{\lambda}{q}\gen{+}
\gen{2},&\gen{1}\gen{-}=\gen{-}\gen{1}-\inv{q}\gen{-}+
\frac{\lambda}{q}\gen{2}\gen{-},\nonumber\\
\gen{2}\gen{+}=q^2\gen{+}\gen{2}-q\gen{+},&\gen{2}\gen{-}=\inv{q^2}
\gen{-}\gen{2}+\inv{q}\gen{-},\nonumber\\
\gen{+}\gen{-}=\gen{-}\gen{+}+\inv{q}(\IU-\lambda\gen{2})(\gen{1}-
\gen{2}),&\gen{1}\gen{2}=\gen{2}\gen{1}.
\label{chi-comm}
\end{eqnarray}
However, for the purposes of the next subsection, it becomes
convenient to change bases by defining the generators $\gen{0}:=
\gen{1}-\gen{2}$ and $\Chi:=\gen{1}+\inv{q^2}\gen{2}$.  The usefulness
of these is apparent when we consider the adjoint actions in this new
basis: we find
\begin{eqnarray}
\Chi\ad\Chi=0,&\gen{i}\ad\Chi=0,&\Chi\ad\gen{i}=-\inv{q}(q^2-
\inv{q^2})\gen{i}
\end{eqnarray}
(where $i=0,+,-$), as well as
\begin{eqnarray}
\gen{0}\ad\gen{0}=-\inv{q}(q^2-\inv{q^2})\gen{0},&&\gen{0}\ad\gen{\pm}
=\pm q^{\mp 1}(1+\inv{q^2})\gen{\pm}.
\end{eqnarray}
It is also interesting to note that by using (\ref{chi-comm}), we
discover that $\Chi$ is central in the algebra; we will have more to
say on this a bit later.

\section{The Killing Metric}\label{chap-QLA-Killing}

\subsection{The Killing Form for a Quasitriangular Hopf Algebra}

Let $\U$ be a quasitriangular Hopf algebra with R-matrix $\R$, and
$\rho :\U\rightarrow M_N (k)$ be an $N\times N$ matrix representation
of $\U$ with entries in $k$.  We may define the bilinear map $\eta^{(
\rho)}:\U\otimes\U\rightarrow k$, the {\em Killing form} associated
with the representation $\rho$, as
\begin{equation}
\killing{\rho}{x}{y}:=\trace{\rho}{xy}
\end{equation}
where $x,y\in\U$, $\tr_{\rho}$ is the trace over the given
representation, and $u$ is the generator of the square of the antipode
(see Appendix \ref{chap-matrix}).  $\eta^{(\rho)}$ has the following
properties:
\begin{eqnarray}
\killing{\rho}{y}{x}&=&\killing{\rho}{x}{S^2 (y)},\nonumber\\
\killing{\rho}{(z_{(1)}\ad x)}{(z_{(2)}\ad y)}&=&
\killing{\rho}{x}{y}\epsilon(z),\label{inv-inf}
\end{eqnarray}
for all $x,y,z\in\U$.  The first of these identities expresses the
``symmetry'' of $\eta^{(\rho)}$, and immediately follows from the
properties of $u$; the second is a statement of the invariance of the
Killing form under the adjoint action of $\U$ on itself, and comes
from the fact that
\begin{eqnarray}
\trace{\rho}{(x\ad y)}&=&\trace{\rho}{x_{(1)}yS(x_{(2)})}\nonumber\\
&=&\killing{\rho}{x_{(1)}}{yS(x_{(2)})}\nonumber\\
&=&\killing{\rho}{yS(x_{(2)})}{S^2 (x_{(1)})}\nonumber\\
&=&\trace{\rho}{yS(S(x_{(1)})x_{(2)})}\nonumber\\
&=&\trace{\rho}{y}\epsilon(x).
\end{eqnarray}

The invariance under the adjoint action may be thought of as how the
Killing form behaves under an ``infinitesimal'' transformation; as
remarked in Chapter \ref{chap-coactions}, the ``finite''
transformation is given by the right coaction (\ref{coact}) of the
dually paired Hopf algebra $\A$ on $\U$, and the Killing form has the
property
\begin{eqnarray}
\killing{\rho}{x^{(1)}}{y^{(1)}}x^{(2)'}y^{(2)'}&=&\killing{\rho}{(e_i
\ad x)}{(e_j \ad y)}f^i f^j \nonumber\\
&=&\killing{\rho}{((e_i)_{(1)}\ad x)}{((e_i)_{(2)}\ad y)}f^i \nonumber\\
&=&\killing{\rho}{x}{y}\epsilon (e_i )f^i \nonumber\\
&=&\killing{\rho}{x}{y}\IA.\label{fin-inv}
\end{eqnarray}
This is therefore the ``finite'' version of the invariance of
$\metric{\rho}{}$.

\subsection{The Killing Metric for a Quantum Lie Algebra}

In the case when $\U$ is not only quasitriangular, but also a QLA with
generators $\{ \gen{A} \}$, we can define the {\em Killing metric}
associated with the representation $\rho$ as
\begin{equation}
\metric{\rho}{AB}:=\killing{\rho}{\gen{A}}{\gen{B}}.
\end{equation}
It is now convenient to define the quantity $\Irep{\rho}{A}:=
-\trace{\rho}{\gen{A}}$ (the sign is merely a convention); from the
results of the previous subsection,
\begin{eqnarray}
\trace{\rho}{(\gen{A}\ad\gen{B})}&=&-\struc{AB}{C}\Irep{\rho}{C}
=0,\nonumber\\
\trace{\rho}{(\O{A}{B}\ad\gen{C})}&=&-\bigR{DB}{AC}\Irep{\rho}{D}=
-\delta^A_B \Irep{\rho}{C},\nonumber\\
\trace{\rho}{\gen{B}}\bigA{B}{A}&=&-\Irep{\rho}{B}\bigA{B}{A}=
-\Irep{\rho}{A}\IA .\label{I-relations}
\end{eqnarray}
The first of (\ref{I-relations}) implies that if we multiply
(\ref{Lie-comm}) by $u$ and trace over a representation $\rho$, we
find that the Killing metric satisfies
\begin{equation}
\metric{\rho}{AB}=\bigR{CD}{AB}\metric{\rho}{CD},
\end{equation}
which gives the ``symmetry'' of the Killing metric\footnote{This
equation also explicitly shows the existence of an eigenvector of
$\hat{\real}$ with eigenvalue 1, so the frequently occuring
combination $\hat{\real}-{{\Bbb I}}$ is noninvertible.}.  We can also
obtain the ``total antisymmetry'' of the structure constants in a
similar way; since the counit of all the generators vanish,
(\ref{inv-inf}) requires that
\begin{eqnarray}
0&=&\killing{\rho}{(\gen{C(1)}\ad\gen{A})}{(\gen{C(2)}\ad\gen{B})}
\nonumber\\ &=&\killing{\rho}{(\gen{C}\ad\gen{A})}{\gen{B}}+
\killing{\rho}{(\O{C}{D}\ad\gen{A})}{(\gen{D}\ad\gen{B})}\nonumber\\
&=&\struc{CA}{D}\killing{\rho}{\gen{D}}{\gen{B}}+\bigR{ED}{CA}
\struc{DB}{F}\killing{\rho}{\gen{E}}{\gen{F}},
\end{eqnarray}
so we find that
\begin{equation}
\struc{CA}{D}\metric{\rho}{DB}+\bigR{ED}{CA}\struc{DB}{F}
\metric{\rho}{EF}=0.
\end{equation}

If we use (\ref{gens-coact}), together with (\ref{fin-inv}), we see
that the invariance of the Killing metric under finite rotations takes
the form
\begin{equation}
\metric{\rho}{CD}\bigA{C}{A}\bigA{D}{B}=\metric{\rho}{AB}\IA .
\label{inv-finite}
\end{equation}

\subsubsection{Quadratic Casimirs}

Now, suppose that $\metric{\rho}{AB}$ is invertible, \ie there exists
a numerical matrix $\invmet{\rho}{AB}$ such that
\begin{equation}
\metric{\rho}{AC}\invmet{\rho}{CB}=\invmet{\rho}{BC}\metric{\rho}{CA}
=\delta^B_A .
\end{equation}
Then (\ref{inv-finite}) implies that $\bigA{A}{C}\bigA{B}{D}
\invmet{\rho}{CD}=\invmet{\rho}{AB}\IA$; this in turn indicates that
the {\em quantum quadratic Casimir} defined by
\begin{equation}
\cas{\rho}:=\invmet{\rho}{AB}\gen{A}\gen{B}
\end{equation}
is central.  Why?  Firstly, note that $\cas{\rho}$ is right-invariant:
\begin{eqnarray}
\DA (\cas{\rho})&=&\invmet{\rho}{AB}\DA (\gen{A})\DA (\gen{B})
\nonumber\\
&=&\gen{C}\gen{D}\otimes\invmet{\rho}{AB}\bigA{C}{A}\bigA{D}{B} \nonumber\\
&=&\gen{C}\gen{D}\invmet{\rho}{CD}\otimes\IA\nonumber\\
&=&\cas{\rho}\otimes\IA .
\end{eqnarray}
Now, recall (\ref{adjoint-coact}) and (\ref{adj-mult}); the first of
these states that if $x$ is right-invariant, $y\ad x=\epsilon (y)x$
for all $y\in\U$.  The second gives $yx=(y_{(1)}\ad x)y_{(2)}$, so the
two together imply that $xy=yx$, namely, any right-invariant element
of $\U$ is central.  Since we have just shown right-invariance of
$\cas{\rho}$, it follows that the quantum quadratic Casimir commutes
with everything in the algebra, just as in the classical case.

\subsection{Examples}

We now present some explicit examples of some of the results in the
previous subsections.  These will hopefully illustrate many of the
concepts we have just encountered.

\subsubsection{Fundamental Representations of $GL_q(2)$ and $SL_q(2)$}

As our first example, we compute what the Killing metrics for the
fundamental representations of $GL_q (2)$ and $SL_q(N)$.  These may be
considered together due to the fact that their R-matrices differ only
by a factor of $q^{-\half}$.  In the basis $(\gen{1},\gen{+},\gen{-},
\gen{2})$) given by (\ref{chi-fund}), we find
\begin{equation}
\matrix{(\gen{ij})}{k}{\ell}=\left(\frac{I-\Rhat^2}{\lambda}\right)^{ik
}{}_{j\ell},
\end{equation}
where $\Rhat$ is given through (\ref{R-sl2}) for $SL_q(2)$, and
$r^{-1}=q^{\half}$ times this for $GL_q(2)$.  However, it is somewhat
more useful to use $\Chi$ and $\gen{0}$ rather than $\gen{1}$ and
$\gen{2}$; when we do this, we find that
\begin{eqnarray}
&&\Chi =\inv{\lambda}\left( 1+\inv{q^2}-r^2(q^2+\inv{q^2})\right)
\left( \begin{array}{cc}
1&0\\0&1\end{array}\right) ,\,
\gen{+}=-r^2 \left( \begin{array}{cc}
0&0\\1&0\end{array}\right) ,\nonumber\\
&&\gen{-}=-r^2 \left( \begin{array}{cc}
0&1\\0&0\end{array}\right) ,\,\gen{0}=r^2\left( \begin{array}{cc}
-q&0\\0&\inv{q}\end{array}\right) ,
\end{eqnarray}
and also
\begin{equation}
u=\inv{rq^3}\left( \begin{array}{cc}
1&0\\0&q^2\end{array}\right) .
\end{equation}
Thus, when we compute $\eta^{(\rho)}$ in the basis $(\Chi,\gen{+},
\gen{-},\gen{0})$ with the appropriate value for $r$ stuck in, we find
\begin{eqnarray}
\eta^{(\mbox{\scriptsize fund $GL$}_q (2))}&=&\inv{q^2}\left(
\begin{array}{cccc}
q^2(q+\inv{q})&0&0&0\\0&0&q&0\\0&\inv{q}&0&0\\0&0&0&q+\inv{q}
\end{array}\right) ,\nonumber\\
\eta^{(\mbox{\scriptsize fund $SL$}_q(2))}&=&q^{-\frac{7}{2}}\left(
\begin{array}{cccc}
\inv{q^2}(\frac{q^3-1}{q+1})^2(q+\inv{q})&0&0&0\\
0&0&q&0\\0&\inv{q}&0&0\\0&0&0&q+\inv{q}\end{array}\right) .
\label{killing-fund}
\end{eqnarray}
Except for an overall factor, we see that the lower right-hand
$3\times 3$ matrices are the same, whereas the upper left-hand entry
vanishes in the classical limit for $SL_q(2)$.  This is not
surprising, since it corresponds to the fact that classical $SL(2)$
has only three generators, not four.

For $q\neq 1$, both of these Killing metrics are invertible, and thus
the quadratic Casimirs can be found.  When we do the calculations for
these representations, we find
\begin{equation}
Q^{(\mbox{\scriptsize fund $G/SL$}_q(2))}_2=rq\quint{2}{q}\left(
\begin{array}{cc}
1&0\\0&1\end{array}\right) ,
\end{equation}
so, as we'd expect, it is proportional to the $2\times 2$ identity
matrix.

\subsubsection{Adjoint Representation of $SL_q(2)$}

Using the structure constants for $U_q(sl(2))$ from before, we find
the generators in the adjoint representation:
\begin{eqnarray}
\Chi=-\frac{\lambda}{q}\left( \begin{array}{cccc}
\inv{q}&-q&0&0\\
-\inv{q}&q&0&0\\0&0&q+\inv{q}&0\\0&0&0&q+\inv{q}\end{array}\right) ,\,
\gen{+}=\left( \begin{array}{cccc}
0&0&0&\inv{q}\\0&0&0&-\inv{q}\\-\inv{q}&q&0&0\\0&0&0&0
\end{array}\right) ,&&\nonumber\\
\gen{-}=\inv{q^2}\left( \begin{array}{cccc}
0&0&-q&0\\0&0&q&0\\0&0&0&0\\
\inv{q}&-q&0&0
\end{array}\right) ,\,
\gen{0}=\inv{q}\left( \begin{array}{cccc}
-\frac{\lambda}{q}&q\lambda &0&0\\
\frac{\lambda}{q}&-q\lambda&0&0\\0&0&1+\inv{q^2}&0\\0&0&0&-(1+q^2)
\end{array}\right) ,&&
\end{eqnarray}
and $u$ is
\begin{equation}
u=\left( \begin{array}{cccc}
1-\inv{q^2}+\inv{q^4}&\frac{\lambda}{q}&0&0\\
\frac{\lambda}{q^3}&\inv{q^2}&0&0\\0&0&\inv{q^2}&0\\0&0&0&\inv{q^6}
\end{array}\right) .
\end{equation}
The Killing metric, which is just $\struc{AC}{D}\struc{BD}{C}$ in this
representation, is
therefore
\begin{eqnarray}
\lefteqn{\eta^{(\mbox{\scriptsize adj $SL_q(2)$})}_{AB}=}\nonumber\\
&&\frac{q+\inv{q}}{q^6}\left( \begin{array}{cccc}
\frac{\lambda^2}{q^2}(q+\inv{q})\quint{3}{q}&0&0&0\\
0&0&q(q^2+\inv{q^2})&0\\0&\inv{q}(q^2+\inv{q^2})&0&0
\\0&0&0&(q^2+\inv{q^2})(q+\inv{q})
\end{array}\right) .\label{killing-adj}
\end{eqnarray}
Once again, for $q\neq 1$, this is invertible, and the quadratic
Casimir comes out to be
\begin{equation}
Q^{(\mbox{\scriptsize adj $SL_q(2)$})}_2 =\frac{q^4\quint{2}{q}
}{\quint{3}{q}}\left( \begin{array}{cccc}
1&-q^2&0&0\\-1&q^2&0&0\\0&0&\quint{2}{q}&0\\0&0&0&\quint{2}{q}
\end{array}\right) .
\end{equation}
This matrix has a zero eigenvalue and three degenerate eigenvalues of
$q^4\quint{2}{q}^2 / \quint{3}{q}$, so it can be block-diagonalized
into a $(1\times 1)\oplus (3\times 3)$ matrix.  (This is the first
indication that the adjoint representation for $SL_q(2)$ is reducible,
and we will come back to this point shortly.)

Up to multiplicative factors, the lower right-hand $3\times 3$
submatrices of (\ref{killing-fund}) and (\ref{killing-adj}) are the
same.  However, recall that there is a general theorem for compact Lie
algebras in the classical case: for a given basis of generators, all
Killing metrics computed from irreducible representations are
proportional.  The appearance of the same matrix in the three cases
considered above is an indication that perhaps there is an analagous
theorem for the deformed case as well.  In fact, consider the
classical case of $SU(N)$; up to an overall normalization, the
quadratic Casimir in the fundamental representation is proportional to
$N^2 -1$, and for the adjoint representation, it is the same constant
times $N$, so the ratio between the former and the latter is
$\frac{N^2 -1}{N}$.  For the $SL_q(2)$ cases we have just studied,
this ratio is
\begin{equation}
Q_2^{(\mbox{\scriptsize fund $SL_q(2)$})}/Q_2^{(\mbox{\scriptsize adj
$SL_q(2)$})}=q^{-\frac{7}{2}}\frac{\quint{3}{q}}{\quint{2}{q}},
\end{equation}
which agrees exactly with the classical case in the $q\rightarrow 1$
limit for $N=2$.

We also note that if we instead choose the basis $(\Chi ,\gen{-},(q+
\inv{q})^{-\half}\gen{0}, \gen{-})$, this $3\times 3$ matrix would be
proportional to
\begin{equation}
\left( \begin{array}{ccc}
0&0&\inv{q}\\0&1&0\\q&0&0\end{array}\right) ,\nonumber
\end{equation}
which is the metric for $SO_{q^2}(3)$.  There is prior evidence for
the equivalence of this quantum group with $SL_q(2)$ (just as in the
classical case) \cite{SO-SL}, and our result here supports this.

\section{Some Comments on the Adjoint Representation}

To conclude this chapter, we examine some of the implications of
(\ref{I-relations}).  Notice that unless $\Irep{\rho}{A}$ vanishes
identically for {\em all} representations, we are able to deduce the
existence of another numerical object $\D{A}$ which satisfies
\begin{equation}
\struc{AB}{C}\D{B}=0.
\end{equation}
Why should this quantity exist?  From the last of (\ref{I-relations}),
$\Irep{\rho}{A}\IA$ is an algebra-valued eigenvector of $\bigA{t}{}$
with eigenvalue unity.  The transpose of any matrix has the same
eigenvalues as the original, so this implies the existence of a
numerical quantity $\D{A}$ such that $\D{A}\IA$ is the algebra-valued
eigenvector of $\bigA{}{}$ with unit eigenvalue, \ie
\begin{equation}
\bigA{A}{B}\D{B}=\D{A}\IA .\label{D-defn}
\end{equation}
This in turn implies that
\begin{eqnarray}
\struc{AB}{C}\D{B}=0,&&\bigR{CA}{BD}\D{D}=\delta^A_B \D{C}.
\end{eqnarray}
The first of these equations implies that $\D{A}$ is a nonzero null
eigenvector for {\em all} the generator matrices in the adjoint
representation, so if there does indeed exist a representation for
which $\trace{\rho}{\gen{A}}$ does not vanish, the adjoint
representation is reducible.  (In fact, when we computed the quadratic
Casimir for $SL_q(2)$ in this representation, there were already hints
of this result.)  Since we know that the adjoint is irreducible for
the classical compact Lie algebras, this indicates that as $q
\rightarrow 1$, $\trace{\rho}{\gen{A}}\rightarrow 0$ for all
representations, so tracelessness of the generators is recovered.

Another consequence is that the quantity $\D{A}\gen{A}$ is central,
for precisely the same reason that $\cas{\rho}$ is, namely, it is
right-invariant.  This follows immediately from the definition of
$\D{A}$ given above in (\ref{D-defn}).

The normalizations of $\Irep{\rho}{A}$ and $\D{A}$ are not fixed by
their definitions; they are both arbitrary up to multiplicative
factors.  However, if we wanted to, we could eliminate one of these
factors in terms of the other by fixing the product $\Irep{\rho}{A}
\D{A}$ to be some convenient number.

For an explicit example, we look at the case in which the QLA in
question is one constructed from a quasitriangular Hopf algebra.
Consider the third of equations (\ref{I-relations}); the explicit form
of the adjoint matrices $\bigA{}{}$ in (\ref{adjoint-rep}) implies
that if a nonvanishing $\Irep{\rho}{A}$ exists, the matrix
$\matrix{{\cal I}}{i}{j}:={\cal I}^{\rho}_{(ij)}$ must satisfy $A{\cal
I}={\cal I}A$.  The only matrices which satisfy this relation are
multiples of the identity; it is easily shown that such matrices also
satisfy the first two of (\ref{I-relations}) as well.  Therefore, for
such QLAs, we choose the canonical form $\Irep{\rho}{(ij)}:=\kappa
\delta^i_j$, and compute $\kappa$ accordingly.  We also find that
${\cal D}_{(ij)}$ must be proportional to $\matrix{(D^{-1}) }{j}{i}$,
so that $\Irep{\rho}{A}\D{A}\propto\tr (D^{-1})$.  This expression for
$\D{A}$ also indicates that $\D{A}\gen{A}=\tr (D^{-1}X)$, which we
know from Appendix \ref{chap-matrix-u} is right-invariant, and
therefore commutes with every element of the QLA.  In fact, for the
$SL_q(2)$ case, this is just proportional to $\Chi$ from the previous
section, which we saw from the explicit commutation relations was
indeed central.

The fundamental representations of the quantum Lie groups in Appendix
\ref{chap-matrix} satisfy the above criteria, \ie the quantities ${\cal
I}^{(\mbox{\scriptsize fund})}_{(ij)}$ are all nonzero, provided
$q\neq 1$.  The values of $\kappa$ therefore can be computed, and are:
\begin{eqnarray}
\kappa (GL_q(N))&=&1,\nonumber\\
\kappa (SL_q(N))&=&q^{-\inv{N}}\left( 1-\quint{\inv{N}}{q}
\quint{N}{\inv{q}}\right) ,\nonumber\\
\kappa (SO_q(N)/SP_q(\half N))&=&q^{N-\epsilon}-q^{\epsilon -N}
\end{eqnarray}
(where we have combined the orthogonal and symplectic groups by using
the quantity $\epsilon=\pm 1$).  Looking at the values of $\kappa$
given above, we see that they vanish in the classical limit for
$SL_q(N)$, $SO_q(N)$, and $SP_q(\half N)$.  This must happen, since we
know that in the classical case, the adjoint representation is
irreducible.  $\kappa (GL_q(N))$ is nonzero for {\em all} values of
$q$, but this is not surprising, since $GL(N)$ is not compact, and its
adjoint representation is indeed reducible.

\chapter{Cartan Calculus on Hopf Algebras and Quantum Lie Algebras}
\label{chap-Cartan}

The purpose of this chapter is to generalize the classical case, and
it builds upon the structure of the universal differential calculus
associated with a Hopf algebra.  (For readers unfamiliar with the
classical Cartan calculus, Appendix \ref{chap-classical} contains the
basic background material and references.)  The basics of this
approach are discussed in Appendix \ref{chap-UDC}, and the reader
unfamiliar with the subject should look therein before proceeding, if
only to familiarize him- or herself with the notation we use here.
Our method of attack will be to start with the UDC $(\UDE,\dg)$ of a
Hopf algebra $\A$, and introduce Lie derivatives and inner derivations
which act on $\UDE$.  Our ``deformed'' version presented here will
allow for possible noncommutativity of the elements of $\UDE$, unlike
the classical case.  However, as in the latter, we need specify only
how the derivations act on and commute with 0- and 1-forms; the
extension to arbitrary $p$-forms in $\UDE$ follows immediately.

\section{Universal Cartan Calculus}

We begin with two dually paired Hopf algebras $\A$ and $\U$, and the
UDC associated with $\A$.  As always, $\U$ is to be thought of as an
algebra of left-invariant generalized derivations which act on
elements of $\A$ in the manner described in Section
\ref{chap-coactions-actions}.  We now associate with each $x\in\U$ a
new object, the {\em Lie derivative} $\lie_x$; it is a linear function
of $x$, has the same transformation properties as $x$ under
$\A$-coactions (\ie $\lie_x\mapsto\IA\otimes\lie_x$ and $\lie_x\mapsto
\lie_{x^{(1)}}\otimes x^{(2)'}$), and is a linear map taking $\UDE$
into itself such that $p$-forms map to $p$-forms.  Furthermore, we
require that
\begin{equation}
\lie_x\dg =\dg\lie_x.\label{Lie-der}
\end{equation}
This relation allows us to uniquely recover the action of $\lie_x$ on
all of $\UDE$ from its action on $\A$, \ie 0-forms.  Just as in the
classical case, the action of the Lie derivative on $a\in\A$ is
defined to be the same as that of the corresponding differential
operator, \ie
\begin{equation}
\lie_x (a)=x\trg a=a_{(1)}\inprod{x}{a_{(2)}},
\end{equation}
and its commutation relations with 0-forms is the same as that given
in $\smash$:
\begin{equation}
\lie_x a =a_{(1)}\inprod{x_{(1)}}{a_{(2)}}\lie_{x_{(2)}}
=\lie_{x_{(1)}}(a)\lie_{x_{(2)}}.\label{XA}
\end{equation}
{}From (\ref{Lie-der}) and (\ref{XA}) we can find the action on and
commutation relation with any 1-form $a\dg(b)$:
\begin{eqnarray}
\lie_x(a\dg(b))&=&a_{(1)}\dg(b_{(1)})\inprod{x}{a_{(2)}b_{(2)}},
\nonumber\\
\lie_x a\dg(b)&=&a_{(1)}\dg(b_{(1)})\inprod{x_{(1)}}{a_{(2)}b_{(2)}}
\lie_{x_{(2)}}=\lie_{x_{(1)}}(a\dg(b))\lie_{x_{(2)}}.\label{XDA}
\end{eqnarray}

At this point we introduce for each $x\in\U$ the corresponding {\em
inner derivation} $\I_x$.  The guideline for this generalization of
the classical case will be the classical Cartan identity
\begin{equation}
\lie_x =\I_x \dg +\dg\I_x \label{Cartan}
\end{equation}
(so $\I_x$ is linear in $x$).  To find the action of $\I_x$ on $\UDE$
we can now attempt to use (\ref{Cartan}) in the identity $\lie_x(a)=
\I_x(\dg(a))+\dg(\I_x(a))$.  We take as an assumption that the action
of $\I_x$ on 0-forms like $a$ vanishes; therefore, we obtain
\begin{equation}
\I_x(\dg(a))= a_{(1)}\inprod{x}{a_{(2)} }.
\label{incomplete}
\end{equation}
However, this cannot be true for any $x\in\U$ because $\dg (1)=0$.
{}From (\ref{incomplete}), $\I_x(\dg(1))=1\epsilon (x)$, which does not
necessarily vanish identically (as we require).  We see that the
trouble arises when dealing with those $x\in\U$ with nonzero counit.
This apparent inconsistency can be dealt with by noting that fof any
$x$, the counit of $x-\IU\epsilon(x)$ does vanish identically; thus,
we modify equation (\ref{incomplete}) to read
\begin{equation}
\I_x(\dg(a))=a_{(1)}\inprod{x-\IU \epsilon(x)}{a_{(2)}},
\label{ida}
\end{equation}
so that $\I_x(\dg(1))$ does indeed vanish for all $x$.  Also note that
this requires the consistency condition
\begin{equation}
\I_{\IU}\equiv 0.
\end{equation}
To define $\I_x$ for all $x\in\U$, therefore, we also need to modify
equation (\ref{Cartan}) to
\begin{equation}
\lie_{x-\IU\epsilon(x)}=\I_x \dg +\dg\I_x,
\end{equation}
or, in view of (\ref{XA}), defining $\lie_{\IU}:=\bid$, and using the
linearity of the Lie derivative,
\begin{equation}
\lie_x = \I_x \dg +\dg\I_x +\epsilon(x)\bid\label{uCi}
\end{equation}
(here $\bid$ is the identity map on $\UDE$, and therefore the unit in
the algebra of generalized derivations, defined to be invariant under
left- and right-coactions).  We call this the {\em universal Cartan
identity}.  From this, it is apparent that $\A$ must coact on $\I_x$
in the same way as on $\lie_x$.

To find the complete commutation relations of $\I_x$ with elements of
$\UDE$ rather than just its action on them, we need only determine how
$\I_x$ moves through 0- and 1-forms.  Both of these can be found by
commuting $\lie_x$ through a function $a\in\A$, using (\ref{XA}) and
(\ref{uCi}).  The left-hand side of the former gives (using the
Leibniz rule)
\begin{equation}
\lie_x a=\I_x \dg (a)+\I_x a\dg +\epsilon (x)a+\dg\I_x a
\end{equation}
and the right-hand side gives
\begin{eqnarray}
\lefteqn{a_{(1)}\inprod{x_{(1)}}{a_{(2)}}\lie_{x_{(2)}} =} \nonumber \\
& &a_{(1)}\inprod{x_{(1)}}{a_{(2)}}\dg \I_{x_{(2)}} \\
& &+a_{(1)}\inprod{x}{a_{(2)}}+a_{(1)}\inprod{x_{(1)}}{a_{(2)}}\I_{x_{(2)}}
\dg.\nonumber
\end{eqnarray}
Equating the two and using (\ref{Leibniz}), (\ref{Lie-der}),
(\ref{ida}), and $\I_x (a)=0$, we obtain
\begin{equation}
\I_x\dg(a)-\I_x(\dg(a))+\lie_{x_{(1)}}(\dg(a))\I_{x_{(2)}}=
\acomm{-\I_xa+\I_x(a)+\lie_{x_{(1)}}(a)\I_{x_{(2)}}}{\dg}.\label{anti}
\end{equation}
Therefore, we propose the commutation relation
\begin{equation}
\I_x\phi=\I_x(\phi)+(-1)^p\lie_{x_{(1)}}(\phi)\I_{x_{(2)}}\label{IXA}
\end{equation}
for any $p$-form $\phi$, so that both sides of (\ref{anti}) vanish.

Missing in our list are commutation relations of Lie derivatives with
themselves and inner derivations.  To find the $\lie$-$\lie$
relations, we the identity (\ref{adj-mult}), and, as before, we extend
the properties of the elements of $\U$ to those of the corresponding
Lie derivatives.  Therefore,
\begin{equation}
\lie_x\lie_y=\lie_{(x_{(1)}\tad y)}\lie_{x_{(2)}},
\end{equation}
and therefore, using (\ref{uCi}),
\begin{equation}
\lie_x\I_y=\I_{(x_{(1)}\tad y)}\lie_{x_{(2)}}.\label{Lie-in}
\end{equation}
(It would seem that (\ref{uCi}) could also give the relation
\begin{equation}
\I_x\lie_y=\lie_{(x_{(1)}\tad y)}\I_{x_{(2)}}+\I_{(x-\IU\epsilon(x))
\tad y},
\end{equation}
but this is inconsistent with the commutation relation (\ref{IXA}).)

After all these derivations (pun intended), it is probably convenient
to pause for a while and recap our results from this section.  Here is
a summary of the actions of the Lie derivatives and inner derivations
with 0- and 1-forms:
\begin{eqnarray}
\lie_x(a)&=&a_{(1)}\inprod{x}{a_{(2)}},\nonumber\\
\lie_x(\dg(a))&=&\dg(a_{(1)})\inprod{x}{a_{(2)}},\nonumber\\
\I_x(a)&=&0,\nonumber\\
\I_x(\dg(a))&=&a_{(1)}\inprod{x}{a_{(2)}}-\epsilon(x)a,
\end{eqnarray}
where, as usual, $x\in\U$, $a\in\A$.  These allow the actions of
$\lie$ and $\I$ on an arbitrary $p$-form $\phi\in\UDE$ to be found
iteratively.  Once this has been done, the commutation relations
between the derivations and elements of $\UDE$ are therefore
\begin{eqnarray}
\lie_x\phi &=&\lie_{x_{(1)}}(\phi)\lie_{x_{(2)}},\nonumber\\
\I_x\phi&=&\I_x(\phi)+(-1)^p\lie_{x_{(1)}}(\phi)\I_{x_{(2)}},
\end{eqnarray}
(The actions and commutation relations for $\dg$ were already given
when the UDC was introduced.)  Finally, here are the relations between
the derivations themselves:
\begin{eqnarray}
\acomm{\dg}{\dg}&=&0,\nonumber\\
\comm{\dg}{\lie_x}&=&0,\nonumber\\
\acomm{\dg}{\I_x}&=&\lie_x -\epsilon(x)\bid,\nonumber\\
\lie_x\lie_y&=&\lie_{(x_{(1)}\tad y)}\lie_{x_{(2)}}\nonumber\\
\lie_x\I_y&=&\I_{(x_{(1)}\tad y)}\lie_{x_{(2)}}\label{der-comm}
\end{eqnarray}

Note that at this point we do not have $\I$--$\I$ commutation
relations, which may at first seem a bit worrisome.  However, this is
not unexpected; $\I_x\I_y$ and $\I_y\I_x$ are simply elements of the
calculus whose action on and commutation relations with $p$-forms are
perfectly well-defined, in precisely the same way that $\dg(a)\dg(b)$
and $\dg(b)\dg(a)$ are simply elements of $\UDE$.  We have not assumed
relations such as $\dg(a)\dg(b)+\dg(b)\dg(a)\equiv 0$ (unlike the
``classical'' case), so it is not surprising that we do not have any
similar relations between the $\I$s.  However, later in this chapter
we will see that such restrictions between elements of $\UDE$ may be
imposed in some cases, and we will comment on the possibility of
$\I$--$\I$ commutation relations.

\subsection{Cartan-Maurer Forms}

The most general left-invariant 1-form can be written
\cite{Woronowicz1}
\begin{equation}
\omega_a :=S(a_{(1)})\dg(a_{(2)})=-\dg(S(a_{(1)}))a_{(2)};
\end{equation}
we will refer to such an element of $\UDE$ as the {\em Cartan-Maurer
form} corresponding to the function $a\in\A$.  This once again follows
the familiar terminology: if $\A$ is an $m\times m$ matrix
representation of some Lie group with $\Delta (g)=g\dot{\otimes}g$,
$S(g)=g^{-1}$ and $\epsilon (g)=I$ for $g\in\A$, then $\omega_g
=g^{-1}\dg(g)$, \ie $\omega_g$ is the well-known left-invariant
classical Cartan-Maurer form.  The exterior derivative of $\omega_a$
has a particularly nice form, given by
\begin{eqnarray}
\dg(\omega_a )&=&\dg(S(a_{(1)}))\dg(a_{(2)})\nonumber\\
&=&\dg(S(a_{(1)}))a_{(2)}S(a_{(3)})\dg(a_{(4)})\nonumber \\
&=&-\omega_{a_{(1)}}\omega_{a_{(2)}}.
\end{eqnarray}
The Lie derivative of $\omega_a$ is
\begin{eqnarray}
\lie_x(\omega_a)&=&\lie_{x_{(1)}}(S(a_{(1)}))\lie_{x_{(2)}}(\dg(
a_{(2)}))\nonumber\\
&=&\inprod{x_{(1)}}{S(a_{(1)})} S(a_{(2)})\dg(a_{(3)})
\inprod{x_{(2)}}{a_{(4)}}\nonumber\\
&=&\omega_{a_{(2)}} \inprod{x}{S(a_{(1)})a_{(3)}}.\label{XOM}
\end{eqnarray}
The contraction of left-invariant forms with $\I_x$ gives a number in
the field $k$, rather than a function in $\A$ (as was the case for
$\dg(a)$):
\begin{eqnarray}
\I_x(\omega_a)&=&-\I_x(\dg(S(a_{(1)}))a_{(2)})\nonumber\\
&=&-\I_x(\dg(S(a_{(1)})))a_{(2)}\nonumber\\
&=&-\inprod{x-\IU\epsilon(x)}{S(a_{(1)})} S(a_{(2)})a_{(3)}
\nonumber\\
&=&(-\inprod{x}{S(a)}+\epsilon(x)\epsilon(a))1.\label{IOM}
\end{eqnarray}
As an exercise, as well as a demonstration of the consistency of our
results, we will compute the same expression in a different way:
\begin{eqnarray}
\I_x(\omega_a)&=&\I_x(S(a_{(1)})\dg(a_{(2)}))\nonumber\\
&=&\inprod{x_{(1)}}{S(a_{(1)})}S(a_{(2)})\I_{x_{(2)}}(\dg(a_{(2)}))
\nonumber\\
&=&\inprod{x_{(1)}}{S(a_{(1)})}S(a_{(2)})a_{(3)}
\inprod{x_{(2)}-\IU\epsilon(x_{(2)})}{a_{(4)}}\nonumber\\
&=&\inprod{x_{(1)}}{S(a_{(1)})}\inprod{x_{(2)}-\IU
\epsilon(x_{(2)})}{a_{(2)}}1\nonumber\\
&=&(-\inprod{x}{S(a)}+\epsilon(x)\epsilon(a))1.
\end{eqnarray}
This result is a consequence of the fact that $\U$ was interpreted as
an algebra of left-invariant differential operators, so $\I_x
(\omega_a)$ must be a left-invariant 0-form, \ie proportional to $1$.

As a final observation, if $\{e_i\}$ and $\{f^i\}$ are, respectively,
(countable) bases of $\U$ and $\A$ with $\inprod{e_i}{f^j}=
\delta_i^j$, the action of $\dg$ on functions $a\in\A$ may be
expressed as
\begin{equation}
\dg (a) =\lie_{e_i}(a)\omega_{f^i}=-\omega_{S^{-1}(f^i)}\lie_{e_i}(a);
\end{equation}
so that the Cartan-Maurer forms form a left-invariant basis for $\G$.

\subsection{General Cartan Calculus}

So far, we have only considered the case of the universal differential
calculus of a Hopf algebra $\A$, as described in Appendix
\ref{chap-UDC-univ}, in which there is no \`a priori assumption of any
commutation relations between 1-forms.  However, in most cases which
will appear in a physics context, we will want to consider situations
in which there are such relations, \ie the {\em general} differential
calculus described in Appendix \ref{chap-UDC-gen}.  So the question
is, how do we incorporate our Cartan calculus into this scheme?  We
start by assuming that we already have a general differential calculus
on a Hopf algebra $\A$, and we define a subspace $\T\subset\U$, given
by
\begin{equation}
\T:=\{ x\in \U |\epsilon(x)=0;\, \inprod{x}{S(m)}=0,\, m\in\M\}.
\end{equation}
It is easily seen that the defining properties for $\M$ imply,
respectively\footnote{The converse is also true, \ie we could start by
defining $\T$ as having the above properties, and taking $\M$ to be
that subalgebra of $\A$ whose inner product with $S(\T )$ vanishes.},
\begin{enumerate}
\item $\IU\not\in\T$,
\item $\Delta (\T)\subseteq\U\otimes (\T\oplus\IU)$,
\item $\U\ad\T\subseteq\T$.
\end{enumerate}
(These properties of $\T$ should remind the reader of the definition
of a QLA in Section \ref{chap-QLA-basics}.  As we shall shortly see,
this is not a coincidence.)  Note that for $x\in\T$ and $a\in\A$,
\begin{equation}
\I_x (\omega_a )=-\inprod{x}{S(a)}.\label{dual}
\end{equation}
Suppose this vanishes; then either $x=0$, $a=\IA$, or $a\in\M$.
Therefore, if we restrict $a$ to be in $\KM$, then the vanishing of
(\ref{dual}) implies that $x=0$ or $a=0$, \ie the inner product
$\dinprod{\,}{\,}:\T\otimes\KM\rightarrow k$ defined by
\begin{equation}
\dinprod{x}{a}:=-\inprod{x}{S(a)}\label{dinprod}
\end{equation}
is nondegenerate.  Hence, $\T$ and $\KM$ are dual to one another.  The
nondegeneracy of (\ref{dinprod}) guarantees that the map from
$\KM\rightarrow\T^*$ given by $a\mapsto\omega_a$ is bijective,
insuring that $\GN$ is the space of all 1-forms over $\A$.  Therefore,
to consistently define our Cartan calculus on all of $\GDE$, we must
restrict the arguments of the Lie derivative and inner derivation from
$\U$ to $\T$, and the argument of $\omega$ from $\A$ to $\KM$.  As an
example of how this works, note that for $x\in\T$ and $a
\omega_m \in\N$,
\begin{eqnarray}
\lie_x a\omega_m &=& a_{(1)}\omega_{m_{(2)}}\inprod{x_{(1)}}{a_{(2)}
S(m_{(1)})m_{(3)}}\lie_{x_{(2)}},\nonumber\\
\I_x a\omega_m &=&-a_{(1)}\omega_{m_{(2)}}\inprod{x_{(1)}}{a_{(2)}
S(m_{(1)})m_{(3)}}\I_{x_{(2)}}.
\end{eqnarray}
Property (3) of $\M$ guarantees that $a_{(1)}\omega_{m_{(2)}}
\inprod{x}{a_{(2)}S(m_{(1)})m_{(3)}}\in\N$ for all $x\in\U$, so both
sides of the two preceding equations are $\simeq 0$ in $\GN$.

Note that we have not yet found a method for expressing any $\I$-$\I$
relations in a form depending manifestly on $\M$, \ie in the manner of
$\omega_{m_{(1)}}\omega_{m_{(2)}}\simeq 0$.  However, in specific
cases we can find such relations; this will be shown explicitly in the
next chapter.

\section{Cartan Calculus for Quantum Lie Algebras}
\label{chap-cartan-QLA}

If our Hopf algebra $\U$ is a QLA, then the subspace $\g$ satisfies
precisely the same relations that $\T$ does; this is of course the
motivation for the definition of a QLA.  Since the existence of the
subspace $\T$ implies the existence of the subalgebra $\M$ (and vice
versa), we are dealing implicitly with the general, rather than the
universal, case.

The first three of (\ref{der-comm}) look the same, and the
second-to-last is simply (\ref{Lie-comm}) with the generators replaced
by their corresponding Lie derivatives.  The remaining commutation
relation may be expressed using the explicit forms for the adjoint
actions given in Section \ref{chap-QLA-basics}:
\begin{eqnarray}
\lie_A\I_B&=&\I_{\gen{A(1)}\tad\gen{B}}\lie_{\gen{A(2)}}\nonumber\\
&=&\I_{\gen{A}\tad\gen{B}}\lie_{\IU}+\I_{\O{A}{D}\tad\gen{B}}\lie_D
\nonumber\\
&=&\struc{AB}{C}\I_C+\bigR{CD}{AB}\I_C\lie_D .
\end{eqnarray}
Once again, we see that in the $q\rightarrow 1$ limit, this becomes
the familiar relation $\comm{\lie_A}{\I_B}=\struc{AB}{C}\I_C$.

\subsection{The Quasitriangular Case}\label{chap-cartan-quasi}

We now apply the results of the previous subsection to the case where
our QLA is one derived from a quasitriangular Hopf algebra $\U$, with
$\A$ being the dually paired Hopf algebra defined by a representation
$\rho$ of $\U$ in the manner which the reader is certainly accustomed
to by now.

We introduce the Lie derivative matrix $\lie_X$ and inner derivation
matrix $\I_X$ as follows: $X$ is the matrix of elements of $\U$
defined by (\ref{X-def}), and
\begin{eqnarray}
\matrix{(\lie_X)}{i}{j}:=\lie_{\matrix{X}{i}{j}},&&\matrix{(\I_X
)}{i}{j}:=\I_{\matrix{X}{i}{j}}.
\end{eqnarray}
These are of course related by the universal Cartan identity
(\ref{uCi}), \ie
\begin{equation}
\lie_X=\I_X\dg+\dg\I_X,\label{uCi-X}
\end{equation}
where the term involving $\bid$ does not appear because $\epsilon
(X)=0$.  The induced coactions of $\A$ on both these matrices are
taken to be the same as those of $X$ itself.

The 0-forms in $\UDE$ are taken to be the elements of $\A$, as usual,
and the basis for the 1-forms are the elements of the matrix $\dg (A)$
(with coefficients in $\A$).  However, as discussed previously, we
will instead use the entries of the Cartan-Maurer matrix $\Omega$,
given by
\begin{equation}
\matrix{\Omega}{i}{j}:=\matrix{(S(A)\dg (A))}{i}{j}.
\end{equation}
This matrix will figure prominently in the next chapter.

What do (\ref{der-comm}) look like in this formulation?  The first
four are just $\dg^2=0$, (\ref{uCi-X}), $\dg\lie_X =\lie_X\dg$, and
(\ref{X-X}) with $X\rightarrow\lie_X$, but by using the explicit forms
of $\hat{\real}$ and the structure constants, the last can be written
using the numerical R-matrix:
\begin{equation}
R_{21}\lie_{X_1}R\I_{X_2}-\I_{X_2}R_{21}\lie_{X_1}R=\inv{\lambda}
(R_{21}R\I_{X_2}-\I_{X_2}R_{21}R).
\end{equation}
We also have the commutation relations with the 0- and 1-forms:
\begin{eqnarray}
\lie_{X_1}A_2  &=&A_2 R_{21}\lie_{X_1}R+A_2\left( \frac{I-R_{21}R}{
\lambda}\right) ,\nonumber\\
R_{21}\lie_{X_1}R\Omega_2 -\Omega_2 R_{21}\lie_{X_1}R&=&\inv{\lambda}
(R_{21}R\Omega_2 -\Omega_2 R_{21}R),\nonumber\\
\I_{X_1}A_2 &=&A_2 R_{21}\I_{X_1}R,\nonumber \\
R_{21}\I_{X_1}R\Omega_2 +\Omega_2 R_{21}\I_{X_1}R&=&\frac{I-
R_{21}R}{\lambda}.\label{inner}
\end{eqnarray}

We can introduce another matrix of 1-forms, $\Omega'$, which is
defined in terms of the exterior derivative on 0-forms:
\begin{equation}
\dg (a)\equiv\tr (D^{-1}\Omega' \lie_X (a)),\label{def-prime}
\end{equation}
where $D$ is the numerical matrix defined in Appendix
\ref{chap-matrix-u}\footnote{We include the $D^{-1}$ in the trace so
that $\Omega'$ is left-invariant and right-covariant under the usual
coactions.}.  If we take $a$ as an entry of $A$, and require that the
Leibniz rule holds, \ie
\begin{equation}
\tr(D^{-1}\Omega'\lie_X)\,A=\dg(A)+A\,\tr(D^{-1}\Omega'\lie_X),
\end{equation}
then by using the first of (\ref{inner}), we find that
\begin{eqnarray}
\Omega'_1A_2&=&A_2R^{-1}\Omega'_1R_{21}^{-1},\nonumber\\
\Omega&=&\inv{\lambda}\left[ \tr_1 (D_1^{-1}R^{-1}\Omega'_1
R_{21}^{-1})-\tr(D^{-1}\Omega')I\right] ,\label{prime}
\end{eqnarray}
Since the defining properties of a QLA are equivalent to starting with
a general differential calculus, it is no surprise that we obtain
these commutation relations between 0- and 1-forms.  Computing the
entries of $\Omega$ in terms of those of $\Omega'$ is of course
immediate if we have the R-matrix; however, it may not always be
possible to do the reverse, namely, to express these commutation
relations in terms of the Cartan-Maurer forms rather than $\Omega'$,
since the second of the above equations may not be invertible.
Whether or not this can be done will depend on the characteristic
equation of the numerical R-matrix.

\chapter{The Linear Quantum Groups $GL_q(N)$ and
$SL_q(N)$}\label{chap-linear}

\section{The Quantum Plane and the Quantum Determinant}

In Appendix \ref{chap-matrix-proj}, we defined the projectors
associated with a given numerical $N^2 \times N^2$-dimensional
R-matrix.  Now, we define the {\em N-dimensional quantum hyperplane}
\cite{Manin1,WZ,Z1} as follows: let $\{x^i |i=1,\ldots,N\}$ be
coordinates and $\{dx^i \}$ be the associated differentials of a
vector space on which the quantum group $\A$ associated with $\Rhat$
coacts as
\begin{eqnarray}
x^i \mapsto\matrix{A}{i}{j}\otimes x^j,& dx^i \mapsto\matrix{A}{i}{j}
\otimes dx^j.&\label{plane-transf}
\end{eqnarray}
Furthermore, the quantum hyperplane is given a unital algebra
structure by specifying commutation relations:
\begin{eqnarray}
\matrix{(P_a)}{ij}{k\ell}x^k x^{\ell}&=0, &a\in\J ,\nonumber\\
\matrix{(P_a)}{ij}{k\ell}dx^k dx^{\ell}&=0, &a\in\J' ,
\end{eqnarray}
where $\{ P_a |a=1,\ldots,m\}$ are the projectors, and $\J$ and $\J'$
are certain subsets of $\{ 1,\ldots,m\}$.  Since $A$ satisfies
(\ref{RAA}), these commutation relations are consistent with the
transformations (\ref{plane-transf}).  There must of course be further
commutation relations between $x^i$ and $dx^i$, consistent not only
with the above but also with the interpretation of $dx^i$ as the
exterior derivative of $x^i$, but the form of these will depend on the
characteristic equation of $\Rhat$.

Once the commutation relations between the differentials are
specified, we can define $\epsilon_q$, the deformed version of the
Levi-Civita tensor.  This is done in the same way as in the undeformed
case, \ie
\begin{equation}
dx^{i_1}dx^{i_2} \ldots dx^{i_N}=\epsilon_q^{i_1 i_2 \ldots i_N}
dx^1 dx^2 \ldots dx^N .\label{L-C}
\end{equation}
Once we have this, we can define $\det{A}$, the quantum determinant of
the matrix $A$, again in analogy with the classical case:
\begin{equation}
\matrix{A}{i_1}{j_1}\ldots\matrix{A}{i_N}{j_N}\epsilon_q^{j_1 \ldots
j_N}= \epsilon_q^{i_1 \ldots i_N} \det{A}.\label{det}
\end{equation}

The cases we are most interested in in this chapter are $GL_q(N)$ and
$SL_q(N)$; for these cases, there are the two projectors $P_{\pm}$,
and we take the coordinates to ``commute'' and their differentials to
``anticommute'', \ie
\begin{equation}
\matrix{(P_-)}{ij}{k\ell}x^k x^{\ell}=\matrix{(P_+)}{ij}{k\ell}dx^k
dx^{\ell}=0.
\end{equation}
In R-matrix notation, these commutation relations take the form
\begin{eqnarray}
x^j x^i &=&(rq)^{-1}\matrix{R}{ij}{k\ell}x^k x^{\ell},\nonumber\\
dx^j dx^i &=&-\frac{q}{r}\matrix{R}{ij}{k\ell}dx^k dx^{\ell}.
\label{plane-comm}
\end{eqnarray}
The mixed commutation relation is then
\begin{equation}
x^j dx^i =\frac{q}{r}\matrix{R}{ij}{k\ell}dx^k x^{\ell}.
\end{equation}
This is obviously covariant under the coaction of $\A$, but it also
respects the action of the exterior derivative; applying $d$ to this
equation (with the Leibniz rule and $d^2=0$) just gives the second of
(\ref{plane-comm}).

Now, we go ahead and use the $dx$--$dx$ commutation relation above to
find $\epsilon_q$ from (\ref{L-C}); note that since $r^{-1}\Rhat$ is
the same numerical matrix for both $GL_q(N)$ and $SL_q(N)$,
$\epsilon_q$ is independent of $r$, so it is the same for both
$GL_q(N)$ and $SL_q(N)$.  Furthermore, it satisfies the relations
\begin{eqnarray}
&\matrix{(R_{0N}\ldots R_{02}R_{01})}{i_0 i_1 i_2 \ldots i_N}{j_0
j_1 j_2 \ldots j_N}\epsilon_q ^{j_1 j_2 \ldots j_N}&=\nonumber\\
&\matrix{(R_{10}R_{20} \ldots R_{N0})}{i_0 i_1 i_2 \ldots i_N}{j_0
j_1 j_2 \ldots j_N}\epsilon_q^{j_1 j_2 \ldots j_N}&=\nonumber\\
& qr^N \delta^{i_0}_{j_0} \epsilon_q^{i_1 i_2 \ldots i_N}.
&\label{R-L-C}
\end{eqnarray}
This relation implies an extremely important result which follows from
(\ref{RAA}), namely, $\det{A}$ commutes with all elements of $A$, and
thus the entire Hopf algebra $\A$.

\noindent {\bf Note:} This centrality of $\det{A}$ is by no means a
result unique to $GL_q(N)$ and $SL_q(N)$; it turns out to be true for
many other cases of interest as well.  For example, for $SO_q (N)$,
the commutation relations between the differentials are defined not
only to be ``antisymmetric'', but also so that their contraction with
the metric vanishes.  In other words,
\begin{equation}
\matrix{(P_1)}{ij}{k\ell}dx^k dx^{\ell}=\matrix{(P_0)}{ij}{k\ell}dx^k
dx^{\ell}=0.
\end{equation}
When we put this into R-matrix notation, we find that the commutation
relations for the differentials take the form of the second of
(\ref{plane-comm}) with $r^{-1}R$ replaced by the $SO_q (N)$ R-matrix,
and therefore (\ref{R-L-C}) holds as well (with $qr^N$ replaced by 1,
that is).  Thus, since (\ref{RAA}) still holds, the centrality of the
quantum determinant follows.  It is this fact that allows us to give
meaning to the ``$S$'' in $SL_q(N)$, $SO_q(N)$, and $SP_q(\half N)$,
since we could not interpret the matrices in these quantum groups as
having unit determinant if it were not central.  However, there are
cases where $\det{A}$ is {\em not} central, most notably in
multiparametric deformations such as $GL_{pq}(2)$ \cite{Schirr2}.

\subsection{The Cartan Calculus for $GL_q(N)$}

In order to apply the results of the previous section to the quantum
group $GL_q(N)$, we first note that there does indeed exist a
subalgebra $\M$ which satisfies the three criteria given in Appendix
\ref{chap-UDC-gen}, namely the one generated by the $N^4$ elements
\begin{equation}
\matrix{m}{ij}{k\ell}:=\matrix{(A_1 A_2 -A_2 -R^{-1}A_1 R_{21}^{-1}
+R^{-1}R_{21}^{-1})}{ij}{k\ell}.\label{GL-ideal}
\end{equation}
In checking that these elements generate $\M$, we must explicitly use
the fact that the R-matrix for $GL_q(N)$ satisfies the quadratic
characteristic equation
\begin{equation}
\Rhat^2 -\lambda\Rhat -I=0.\label{char-GL}
\end{equation}
When we impose $\omega_m \simeq 0$ on these elements, we obtain the
$A$--$\dg (A)$ commutation relations \cite{Manin2,Mal,Schirr,Sud}
\begin{equation}
\dg (A)_1 A_2 =R^{-1}A_2 \dg (A)_1 R_{21}^{-1}.
\end{equation}
Upon differentiation, the $\dg (A)$--$\dg (A)$ relations follow:
\begin{equation}
\dg (A)_1 \dg (A)_2 +R^{-1}\dg (A)_2 \dg (A)_1 R_{21}^{-1}=
0.\label{diff}
\end{equation}
(Alternatively, we could have taken $\M$ as generated by
\begin{equation}
\matrix{m}{ij}{k\ell}=\matrix{(A_1A_2-A_2-R_{21}A_1R+R_{21}
R)}{ij}{k\ell}.
\end{equation}
The resulting $A$--$\dg (A)$ commutation relations are
\begin{equation}
A_1 \dg (A)_2 =R^{-1}\dg (A)_2 A_1 R_{21}^{-1},
\end{equation}
but the $\dg (A)$--$\dg (A)$ relations do not change.)  The
Cartan-Maurer matrix $\Omega=S(A)\dg (A)$ therefore satisfies the
following relations:
\begin{eqnarray}
\Omega_1 A_2-A_2 R^{-1}\Omega_1 R_{21}^{-1}&=&0,\nonumber \\
\Omega_1 \dg(A)_2 +\dg (A)_2 R^{-1} \Omega_1 R&=&0, \nonumber \\
\Omega_1 R_{21}^{-1} \Omega_2 R_{21}+R_{21}^{-1} \Omega_2 R^{-1}
\Omega_1 &=&0.\label{omega}
\end{eqnarray}

To relate $\Omega$ to $\Omega'$ from the previous chapter, we merely
use (\ref{char-GL}) and the R-matrix trace relations from Appendix
\ref{chap-matrix-u}.  We find the simple relation $\Omega'=-\alpha
\Omega$, which of course is consistent with the commutation relations
immediately above\footnote{Note that in some treatments of this
subject, such as \cite{Leipzig}, the 1-form matrix used is actually
$\Omega'$ and not $\Omega$.}.

We know from Appendix \ref{chap-UDC-hopf} that the Cartan-Maurer forms
are left-invariant and right-covariant, so the coactions of $GL_q(N)$
on $\Omega$ are the same as (\ref{X-coacts}) with $X$ replaced by
$\Omega$.  Therefore, the 1-form $\xi$ defined by taking the invariant
trace of $\Omega$, namely
\begin{equation}
\xi\equiv -\alpha\tr(D^{-1}\Omega),
\end{equation}
is left- and right-invariant.  There is more than this to $\xi$,
however; as a consequence of (\ref{omega}), (\ref{char-GL}), and the
various trace properties of $D$, we find
\begin{eqnarray}
\dg (A) = \lambda^{-1}\comm{\xi}{A}, & \dg (\Omega)=-\Omega^{2}=
\lambda^{-1}\acomm{\xi}{\Omega},
\end{eqnarray}
so $\xi$ is in fact the generator of the exterior derivative.  These
imply that the exterior derivative of any $p$-form $\phi$ is given by
\begin{equation}
\dg (\phi )=\lambda^{-1}\gcomm{\xi}{\phi}=\lambda^{-1}(\xi\phi
-(-1)^p \phi\xi ).
\end{equation}
This may seem a bit weird, since in the classical case there is no
such 1-form which generates the exterior derivative, but notice that
the $\lambda$ in the above equation goes to zero in the classical
limit.  Since $\dg (\phi )$ still exists in this limit, this just
implies that $\xi$ (anti)commutes with everything.

Now, we consider $\det{A}$; it is a 0-form, and the above equations
imply that
\begin{eqnarray}
\Omega\,\det{A}=q^{-2}\det{A}\,\Omega, \nonumber \\
\dg (\det{A})=-q^{-1}\det{A}\,\xi=-q\xi\,\det{A}.
\end{eqnarray}
(A consequence of these equations is that both $\dg (\xi)$ and $\xi^2$
vanish.)

Now, we bring in the Lie derivatives and inner derivations, which we
have already shown must satisfy (\ref{inner}) for {\em any} matrices
$A$ and $\Omega$, including those from $GL_q(N)$.  However, since we
are now dealing with a specific Hopf algebra, with an R-matrix,
commutation relations, and the works, it's no surprise that we have
some more identities.  For instance, by using (\ref{char-GL}), the
combination $\frac{I-R_{21}R}{\lambda}$ could be replaced by $-\Rhat$
if we wanted.  In fact, this allows us to use the last of
(\ref{inner}) to obtain the action of $\I_X$ on the Cartan-Maurer
matrix, which turns out to be
\begin{equation}
\I_{X_1}(\Omega_2 )=-\alpha D_2 P.
\end{equation}
In addition, it can be shown that $\xi$ and $\det{A}$ satisfy the
following:
\begin{eqnarray}
\lie_X \xi=\xi \lie_X, &&\I_{X} \xi +\xi \I_{X}=I,\nonumber\\
\lie_X \,\det{A}=q^2\det{A}\,\lie_X -q\det{A},&&\I_X \,\det{A}=
q^2 \det{A}\,\I_{X}.
\end{eqnarray}
But perhaps the most meaningful results we obtain by considering a
specific case are the commutation relations between the inner
derivation matrices, as promised in the last chapter.  They are
reminiscent of the ones for $\Omega$, not surprisingly, and take the
form
\begin{equation}
R^{-1}\I_{X_1}R\I_{X_2}+\I_{X_2}R_{21}\I_{X_1}R=0.
\label{RiRi}
\end{equation}

Many of these relations take a much simpler form if we introduce a new
matrix $\Y$, which corresponds to the matrix $Y$ from (\ref{Y-def}) in
the same way $\lie_X$ corresponds to $X$, \ie
\begin{equation}
\Y=I\bid -\lambda\lie_X .
\end{equation}
$\Y$ now is an operator within our differential calculus, and we
obtain
\begin{eqnarray}
\Y\dg &=&\dg\Y,\nonumber\\
R_{21}\Y_1 R\I_{X_2}&=&\I_{X_2}R_{21}\Y_1 R,\nonumber\\
R_{21}\Y_1 R\Y_2 &=&\Y_2 R_{21}\Y_1 R,\nonumber\\
\Y_1 A_2 &=&A_2 R_{21}\Y_1 R,\nonumber \\
R_{21}\Y_1 R\Omega_2 &=&\Omega_2 R_{21}\Y_1 R,\label{Y}
\end{eqnarray}
as well as
\begin{eqnarray}
\Y\xi=\xi\Y,&&\Y\,\det{A}=q^2 \det{A}\,\Y.
\end{eqnarray}
However, $\Y$ is useful for more than making our equations prettier.
Since its leading term is unity, it is invertible.  More importantly,
we can define a quantity $\Det{\Y}$, which we identify as the
determinant of $\Y$, satisfying
\begin{equation}
\Y\,\Det{\Y}=\Det{\Y}\,\Y.
\end{equation}
This quantity is defined through
\begin{equation}
(\Y_{1 \ldots N}^{(1)} \ldots \Y_{1 \ldots N}^{(N)})^{i_1 \ldots
i_N}{}_{j_{1} \ldots j_{N}} \epsilon_{q} ^{j_{1} \ldots j_{N}}=
\epsilon_{q}^{i_1 \ldots i_N}\Det{\Y} , \label{Dety}
\end{equation}
where
\begin{equation}
\Y^{(k)}_{1 \ldots N}=\left \{
	\begin{array}{ll}
	(R_{kN} \ldots R_{k(k+1)})^{-1}\Y_{k}(R_{kN} \ldots R_{k(k+1)}) &
	\mbox{for $k=1,\ldots ,N-1$,} \\
	\Y_{N} & \mbox{for $k=N$.}
	\end{array}
	\right.
\end{equation}
This determinant is invariant under transformations of $\Y$, and
satisfies the following as a consequence of the above:
\begin{eqnarray}
\dg\,\Det{\Y}&=&\Det{\Y}\,\dg ,\nonumber\\
\Det{\Y}\,\I_X &=&\I_X\,\Det{\Y},\nonumber\\
\Det{\Y}\,A&=&q^2 A\,\Det{\Y},\nonumber\\
\Det{\Y}\,\Omega &=&\Omega\,\Det{\Y},\nonumber\\
\Det{\Y}\,\xi &=&\xi\,\Det{\Y},\nonumber\\
\Det{\Y}\,\det{A}&=&q^{2N}\det{A}\,\Det{\Y}.
\end{eqnarray}
The above equations for $\Det{\Y}$ suggest the definition of an
operator $\H$ as
\begin{equation}
\Det{\Y}\equiv q^{2\H}=I\bid +q\lambda\quint{\H}{q}.
\end{equation}
$\H$ defined in this way commutes with $\Y$, $\dg$, $\I_X$, $\Omega$,
and $\xi$, and satisfies
\begin{eqnarray}
\comm{\H}{A}=A, & \comm{\H}{\det{A}}=N(\det{A}).\label{Hcomm}
\end{eqnarray}
This operator will be important in the next section.

\section{$SL_q(N)$}
\subsection{The Quantum Group $SL_q(N)$}

There seems to be an obvious way to specify the quantum group $SL_q
(N)$: take the matrix $A$ and set its determinant to unity.  This
seems reasonable; $\det{A}$ is central, so within the context of the
Hopf algebra, this would appear to be the right thing to do.
Unfortunately, this doesn't work.  True, $\det{A}$ commutes with the
entries of $A$, but it does {\em not} commute with such quantities as
$\Omega$ and $\Y$.  Therefore, to restrict $GL_q(N)$ to $SL_q(N)$,
instead of {\em imposing} $\det{A}=\IA$ we {\em define} matrices $T$
as
\begin{equation}
T=(\det{A})^{-\inv{N}}A.\label{defT}
\end{equation}
With $\det{T}$ defined as in (\ref{det}), the centrality of $\det{A}$
automatically gives $T$ determinant unity.  Furthermore, we also find
that $\Delta (T)=T\dot{\otimes}T$, $\epsilon(T)=I$, and $S(T)=T^{-1}$.
Therefore, this matrix $T$ is what we identify as an element of the
defining representation of $SL_q(N)$, since it also satisfies
(\ref{RAA}) with $A$ replaced by $T$.  However, as we will see in the
next section, it becomes convenient to introduce the matrix
\begin{equation}
\slR =q^{-\inv{N}}R,
\end{equation}
which we identify as the R-matrix for $SL_q(N)$.  Thus, we shall
write (\ref{RAA}) as
\begin{equation}
\slR T_1 T_2 =T_2 T_1 \slR .
\end{equation}

\subsection{The Calculus for $SL_q(N)$}

The exterior derivative on $SL_q(N)$ can be taken to be the same as
that introduced on $GL_q(N)$; this is because $T$ is a function of
elements of $A$, so its differentials are still given by
\begin{equation}
\dg (T)=\lambda^{-1}\comm{\xi}{T}.
\end{equation}
Note that this implies that the Cartan-Maurer form $\tilde{\Omega}$
for $SL_q(N)$ is related to that of $GL_q(N)$ by
\begin{equation}
\tilde{\Omega}:=S(T)\dg (T)=q^{\frac{2}{N}}\Omega +q
\quint{\inv{N}}{q}\xi.\label{SLform}
\end{equation}
In the classical limit $q\rightarrow 1$, $\tilde{\Omega}$ is
traceless, giving the appropriate reduction from $N^2$ to $N^2 -1$
independent entries in $\tilde{\Omega}$, which of course agrees with
the number of 1-forms of the classical group $SL(N)$.  However, for
$q\neq 1$, we have no such reduction, and we do indeed have $N^2$
linearly independent 1-forms for $SL_q(N)$.

We have thus found a way to set the determinant of our $SL_q(N)$
matrices to unity; for the calculus of the group, we must do something
similar, namely impose a constraint so that the number of independent
differential operators is reduced from $N^2$ to $N^2 -1$.  In a way,
we have already done this, because (\ref{Hcomm}) and (\ref{defT})
together imply $\comm{\H}{T}=0$, so that $\H$ commutes with everything
of interest in $SL_q(N)$, \ie matrices, forms, exterior derivative,
{\it etc.} Thus, within the context of $SL_q(N)$, $\H$ may be
consistently set to zero, reducing the number of generators from $N^2$
to $N^2 -1$, as desired.  Explicitly, this restriction is accomplished
by defining a new Lie derivative valued operator $\Z$ by\footnote{When
restricted to acting on 0-forms, this operator is identical to the
operator $Y$ in \cite{Leipzig}.}
\begin{equation}
\Z:=q^{-\frac{2\H}{N}}\Y .
\end{equation}
Note that the determinant of $\Z$, computed using (\ref{Dety}), is
unity (or, to be more precise, $\bid$, which is the unit of the
differential calculus).  This is equivalent to the introduction of a
set of $N^2$ ``vector fields'' $\matrix{V}{i}{j}$ through
$\Z=I\bid-\lambda
\lie_V$, so that
\begin{equation}
\lie_V =\lie_X +q^{-1}\quint{\frac{\H}{N}}{q^{-1}}I-q^{-1}\lambda
\lie_X \quint{\frac{\H}{N}}{q^{-1}}.
\end{equation}
The fact that $\Det{\Z}=\bid$ implies that only $N^2 -1$ of the
elements of $\lie_V$ are actually independent, which is precisely what
we require for $SL_q(N)$.  In the classical limit, $\H =-\tr(\lie_X
)$, so $\lie_V$ becomes traceless; thus, $V$ contains only $N^2 -1$
linearly independent vector fields, as we'd expect.

Now that we have obtained all these quantities, we want to find the
various relations they satisfy.  As a starting point, note that the
commutation relations between $\Omega$ and $T$ are given by
\begin{equation}
\Omega_1 T_2 =q^{\frac{2}{N}}T_2 R^{-1}\Omega_1 R_{21}^{-1}
=T_2 \slR^{-1}\Omega_1 \slR_{21}^{-1}.
\end{equation}
Here we see the appearance of the $SL_q(N)$ R-matrix $\slR$, as
promised.  In fact, there is a general pattern: by using the
substitutions $A\rightarrow T$, $R\rightarrow\slR$, and $\lie_X
\rightarrow\lie_V$, we obtain most of the corresponding relations for
$SL_q(N)$.  This only goes so far, though; we do {\em not} generally
make the substitution $\Omega\rightarrow\tilde{\Omega}$ or $\I_X
\rightarrow\I_V$.  This would seem to be a result of the fact that
many of the relations satisfied by these quantities arise as a result
of going from the universal differential calculus to the general.  In
any case, $\lie_V$ satisfies
\begin{eqnarray}
\slR_{21}\lie_{V_1}\slR\lie_{V_2}-\lie_{V_2}\slR_{21}\lie_{V_1}
\slR &=&\lambda^{-1}(\slR_{21}\slR\lie_{V_2}-\lie_{V_2}\slR_{21}
\slR),\nonumber\\
\slR_{21}\lie_{V_1}\slR\I_{X_2}-\I_{X_2}\slR_{21}\lie_{V_1}\slR
&=&\lambda^{-1}(\slR_{21}\slR\I_{X_2}-\I_{X_2}\slR_{21}\slR),
\end{eqnarray}
and we also find the relations between our new differential operators
and the 0- and 1-forms of $SL_q(N)$:
\begin{eqnarray}
\lie_{V_1}T_2 &=&T_2 \slR_{21}\lie_{V_1}\slR+T_2 \left(
\frac{I-\slR_{21}\slR}{\lambda}\right) ,\nonumber\\
\slR_{21}\lie_{V_1}\slR\Omega_2 -\Omega_2 \slR_{21}\lie_{V_1}\slR
&=&\lambda^{-1}(\slR_{21}\slR\Omega_2 -\Omega_2 \slR_{21}\slR),
\nonumber\\
\I_{X_1}T_2 &=&T_2 \slR_{21}\I_{X_1}\slR,\nonumber\\
\slR_{21}\I_{X_1}\slR\tilde{\Omega}_2 +\tilde{\Omega}_2
\slR_{21}\I_{X_1}\slR &=&\frac{I-\slR_{21}\slR}{\lambda}.
\end{eqnarray}
Since $\dg$ still commutes with $\lie_V$ and is still generated by
$\xi$, we obviously also have
\begin{equation}
\lie_V \xi =\xi\lie_V .
\end{equation}
The relations for $\Z$ corresponding to (\ref{Y}) are easily obtained
by using $\lie_V=\frac{I\bid -\Z}{\lambda}$ in all of the above
equations.

\chapter{Conclusions}

In this work, we have presented a general approach to the analysis of
the differential geometry of Hopf algebras and quantum groups,
primarily through the introduction of the Lie algebra of derivations
and the resulting Cartan calculus.  The concentration has been on
those cases which often arise in physical problems, namely, the
deformed versions of the groups $GL(N)$ and $SL(N)$, although we have
tried to develop methods and approaches which will be useful for other
cases as well.  In this we have been largely successful, but there are
still several avenues which have not been dealt with entirely, and in
this final chapter, we will address some of them.

\section{The Killing Metric}

In Chapter \ref{chap-QLA-Killing}, the Killing form for an arbirtary
quasitriangular Hopf algebra was introduced.  When the Hopf algebra
was also a QLA, we could define the deformed version of the Killing
metric, which we found had many of the same properties and uses as the
classical one.  However, there were also hints that it had more in
common with the classical Killing metric than was first thought.  The
appearance of the same $3\times 3$ submatrix for the fundamental and
adjoint representations of $SL_q(2)$ (up to a factor) was immediately
reminiscent of the classical case, where the Killing metric of an
irreducible representation of a compact Lie algebra is proportional to
some canonical form.  This similarity was even more suggested by the
fact that the quadratic Casimirs for these two representations had
eigenvalues whose ratio had the correct classical limit.

All this evidence seems to point to the possibility that most, or
perhaps even all, of the properties of the classical Killing metric
have analogues in the deformed case.  There may indeed exist some
canonical form of the Killing metric; if so, this would allow the
definition of the level of a representation as the proportionality
constant between its Killing metric and the canonical one.  The
classification of irreducible representations of QLAs by the values of
their deformed quadratic Casimirs also seems to be a distinct
possibility.  Many of the objects in classical physics depend on such
representation-dependent quantities (\eg the QCD $\beta$-function's
dependence on the $SU(N_f)$ Casimir), so studying the properties of
the Killing metric may be extremely fruitful in the context of a
deformed field theory.

\section{Inner Derivations for a General Cartan Calculus}

When we developed the Cartan calculus in Chapter \ref{chap-Cartan},
the initial lack of any $\I$--$\I$ commutation relations was not a
surprise; after all, we were dealing purely with the universal
differential calculus, which was itself missing any commutation
relations between 1-forms.  Since the inner derivations and
Cartan-Maurer forms are in a sense dual to each other (\ie
$\I_x(\omega_a)$ is proportional to the unit), this was expected.

Unfortunately, the same cannot be said for the general differential
calculus.  There we do in fact have commutation relations between 0-
and 1-forms, given by the vanishing of the Cartan-Maurer form on the
subalgebra $\M$.  We even have the dual version of this condition, the
existence of the subspace $\T$.  It was shown that the previously
existing Cartan calculus could easily accomodate the general
differential calculus, by restricting the arguments of the Lie
derivative and the inner derivation to $\T$, but it should also follow
that there are some sort of $\I$--$\I$ relations dual to the
$\omega$--$\omega$ relations in $\GDE$.  Such (anti)commutation
relations have yet to be found, and this remains one of the glaring
faults in the treatment of the Cartan calculus in this work.

\section{$SO_q(N)$ and $SP_q(\half N)$}

In Chapter \ref{chap-linear}, we considered the application of the
results of Chapter \ref{chap-Cartan} to the specific cases of
$GL_q(N)$ and $SL_q(N)$, by using the fact that we knew the numerical
R-matrix explicitly, and could find the subalgebra $\M$ needed to
restrict from the universal differential calculus to the general, from
(\ref{GL-ideal}).  We also found that there were consistent sets of
equations, still given in terms of $R$, which gave the deformed
anticommutation relations between the inner derivations.  The
differential geometry of $SL_q(N)$ was then shown to be obtainable
from that of $GL_q(N)$ via the restriction (\ref{defT}).

Our success in doing so begs the question: can we do the same, or
something similar, with the other quantum Lie algebras $SO_q(N)$ and
$SP_q(\half N)$?  There are certainly still some relations which will
hold for the differential calculus of {\em all} quasitriangular Lie
algebras, namely those discussed in Chapter\ref{chap-cartan-quasi}.
For instance, the construction of the smash product for the
quasitriangular case was completely general, so one may introduce the
differential operators $X$ and therefore the 1-forms $\Omega'$ via the
exterior derivative (\ref{def-prime}); this was done explicitly
(albeit in somewhat different notation) in \cite{CW}.  In fact, since
the characteristic equations for $SO_q(N)$ and $SP_q(\half N)$ are
explicitly known, we can even find the relation between $\Omega$ and
$\Omega'$ given in (\ref{prime}):
\begin{equation}
\matrix{\Omega}{i}{j}=\matrix{(\Rhat^{-1}-q^{\epsilon-N}I)}{\ell
n}{kj}g^{ik}g_{\ell m}\Omega'{}^m{}_n.
\end{equation}

However, this only goes so far; recall that, in the fundamental
representation, the quantum matrices $A$ for these quantum groups are
not only restricted by a determinant condition, but also the metric
condition
\begin{equation}
g^{k\ell}\matrix{A}{i}{k}\matrix{A}{j}{\ell}=g^{ij},\label{met-cond}
\end{equation}
where $g^{ij}$ is the appropriate numerical matrix from Appendix
\ref{chap-matrix-Lie}.  Consistency between this and (\ref{RAA})
implies the numerical relations
\begin{equation}
\matrix{R}{ij}{k\ell}=g^{im}\matrix{\tilde{R}}{nj}{m\ell}g_{nk}
=g^{jm}\matrix{(R^{-1})}{in}{km}g_{n\ell},
\end{equation}
and these, together with (\ref{LA}), imply similar conditions on
$L^{\pm}$:
\begin{equation}
g^{k\ell}\matrix{(L^{\pm})}{j}{\ell}\matrix{(L^{\pm})}{i}{k}=g^{ij}.
\end{equation}
In terms of the matrix $Y:=L^+S(L^-)$, this translates into the
condition
\begin{equation}
\matrix{(\Rhat^{-1}Y_2\Rhat Y_2)}{ij}{k\ell}g^{k\ell}=g^{ij}.
\label{Y-met}
\end{equation}
Is it possible to start from the $GL_q(N)$ case, which we know well,
and impose (\ref{met-cond}) and (\ref{Y-met}) somehow?  The answer
would seem to be no, since there is no obvious way to construct an $A$
satisfying (\ref{met-cond}) from a general $GL_q(N)$ matrix.  Finding
the differential geometry on these groups may therefore only be
tractable by starting from scratch, by determining what subalgebra
$\M$ not only satisfies the three criteria in Appendix
\ref{chap-UDC-gen}, but also respects the metric condition.  At this
point, such a subalgebra has not been found (at least not to our
knowledge), so this remains an open problem.

\section{Fiber Bundles and Deformed Gauge Theories}

Finally, we reconsider the main motivation behind this work, namely,
the formulation of a deformed gauge field theory.  As alluded to in
the Introduction, the main reason for including Appendix
\ref{chap-classical} in this work was not only to remind the reader of
what the classical Cartan calculus is, but also to present the general
method by which topological properties (such as continuity of a
function) are related to algebraic concepts (\eg the unital
associative algebra $\Fun (M)$).  At least in the case where the
manifold in question is a topological group, and therefore the
corresponding function algebra a quantum group, we have largely
succeeded, with the linear cases $G/SL_q(N)$ being the most fully
realized.  If we also limit ourselves to deformations of {\em
principal} bundles, we are even closer to our goal, because then the
entire bundle (as well as the structure group) is described by a Hopf
algebra.  Others have already had much success in this case
\cite{Brz,Pflaum}, and this is generally the one of interest for most
physical systems.

What remains is to consider the cases where the bundle is not
principle.  The structure group continues to be treated as a Hopf
algebra, so the results herein still apply, but the base manifold and
fiber are treated as unital associative algebras (or more precisely,
collections of unital associative algebras, together with transition
functions relating each algebra, {\it \`a la} sheaf theory).  How our
conclusions can be applied to these cases is as yet unknown, but there
seems to be no reason to assume that our techniques would be utterly
useless, so we are still (perhaps na\"ively) optimistic that deformed
gauge theories are within reach.

\appendix

\chapter{Numerical R-Matrix Relations}\label{chap-matrix}

\section{The Element $u$}\label{chap-matrix-u}

Suppose $\U$ is a quasitriangular Hopf algebra with universal R-matrix
$\R$; then there exists an invertible element $u$ defined by
\cite{Drinfeld}
\begin{equation}
u:=m((S\otimes\id)(\R_{21}))=S(r^{\alpha})r_{\alpha}.
\end{equation}
$u$ has counit $\Ik$, and its inverse and coproduct are
\begin{eqnarray}
u^{-1}&=&r^{\alpha}S^2 (r_{\alpha}),\nonumber\\
\Delta (u)&=&(\R_{21}\R)^{-1}(u\otimes u)=(u\otimes u)(\R_{21}\R)^{-1}.
\end{eqnarray}
This element generates the square of the antipode via an inner
automorphism:
\begin{equation}
S^2 (x)=uxu^{-1}\label{uxu}
\end{equation}
for all $x\in\U$.  A consequence of this is that the element
$c:=uS(u)$ is central in $\U$.

Suppose we have a faithful $N\times N$ matrix representation $\rho$ on
$\U$, and $A$ is the associated matrix of dual elements in $\A$ (see
Chapter \ref{chap-Hopf-reps}).  We define the numerical matrix $D$ to
be equal to $u$ in this representation, up to an overall
multiplicative constant $\alpha$:
\begin{equation}
\matrix{D}{i}{j}:=\alpha\inprod{u}{\matrix{A}{i}{j}}.
\end{equation}
Several results follow immediately:  first of all, an explicit
computation using the definition of $u$ leads to the result
\begin{equation}
I=\alpha\tr_1 (D_1 ^{-1}\Rhat^{-1})=\alpha^{-1} \tr_2 (D_2
\Rhat),
\end{equation}
where $\tr_J$ is shorthand for the contraction over the $J$th pair of
indices, \eg the $\matrix{}{i}{j}$th element of the rightmost
expression in the above equation is $\alpha^{-1}\matrix{D}{m}{n}
\matrix{\Rhat}{in}{jm}$.  These relations can be ``inverted'' in the
sense of solving them for $D$ and $D^{-1}$; to do this, we introduce
for any $N^{2}\times N^{2}$ matrix $K$ a matrix $\tilde{K}=[(
K^{t_{1}})^{-1}] ^{t_{1}}$ ($t_J$ denotes transposing with respect to
the $J$th pair of indices).  When this matrix exists, it satisfies
\begin{equation}
\matrix{K}{im}{n\ell}\matrix{\tilde{K}}{nk}{jm}=
\matrix{K}{mi}{\ell n}\matrix{\tilde{K}}{kn}{mj}=\delta^i_j \delta^k
_{\ell}.
\end{equation}
With this in hand, we find
\begin{eqnarray}
D=\alpha\tr_2 (P\tilde{R}),&D^{-1}=\alpha^{-1}\tr_2 (P
\tilde{(R^{-1})}).&
\end{eqnarray}
Since the representation is faithful (by assumption), $c$ must be
proportional to the unit matrix in the representation.  We therefore
define the constant $\beta$ by means of the identity
\begin{equation}
\inprod{c}{\matrix{A}{i}{j}}=(\alpha\beta)^{-1}\delta^i _j .
\end{equation}
Using the explicit forms of $c$ and $u$ gives
\begin{equation}
I=\beta^{-1}\tr_1 (D_1^{-1} \Rhat)=\beta\tr_2 (D_2 \Rhat^{-1}),
\end{equation}
or, if we ``invert'',
\begin{eqnarray}
D=\beta^{-1}\tr_1 (P\tilde{(R^{-1})}),&D^{-1}=\beta\tr_1 (P
\tilde{R}).&
\end{eqnarray}

{}From the fact that $(S^2 \otimes S^2)(\R)=\R$, we find the numerical
relation
\begin{equation}
D_1 D_2 R=RD_1 D_2 .
\end{equation}
The dual version in $\A$ of (\ref{uxu}) is
\begin{equation}
S^2 (A)=DAD^{-1}.
\end{equation}
The definition of the $D$-matrix, together with (\ref{RAA}), gives
\begin{equation}
(D^{-1})^t A^t D^t S(A)^t =S(A)^t (D^{-1})^t A^t D^t =\IA,
\label{DADA}
\end{equation}
(\ref{RAA}) and (\ref{DADA}) together then imply the identities
\begin{equation}
\tilde{R}=D_1 ^{-1}R^{-1}D_1 =D_2 R^{-1}D^{-1}_2 .
\end{equation}
All of the above give the following important results: if $M$ is an
$N\times N$ matrix, then
\begin{eqnarray}
\matrix{\tr_1 (D_1 ^{-1}R^{-1}M_1 R)}{i}{j}&=&\matrix{\tr_1 (D_1
^{-1}R_{21}M_1 R_{21}^{-1})}{i}{j}\nonumber \\
&=&\tr(D^{-1}M) \delta^{i}_{j}.
\end{eqnarray}
Also, if the elements of $M$ commute with the elements of $A$,
\begin{equation}
\tr(D^{-1}S(A)MA)=\tr(D^{-1}M).
\end{equation}
In particular, if $M$ is a matrix on which $\A$ right coacts via $\DA
(\matrix{M}{i}{j})=\matrix{M}{k}{\ell}\otimes S(\matrix{A}{i}{k})
\matrix{A}{\ell}{j}$, then (\ref{DADA}) implies
\begin{equation}
\DA(\tr(D^{-1}M))=\tr(D^{-1}M)\otimes\IA.
\end{equation}
For this reason, $\tr(D^{-1}M)$ is called the {\em invariant trace of}
$M$.

\section{R-Matrices for the Simple Lie
Algebras}\label{chap-matrix-Lie}

In their seminal work \cite{RTF}, Reshetikhin, Takhtadzhyan and
Faddeev give the numerical R-matrices for the quantum versions of the
fundamental representations of the Lie algebras $A_n$, $B_n$, $C_n$
and $D_n$; here, we review these forms, and include other results for
the particular cases.

\subsection{R- and D-Matrices}

We take $E_{IJ}$ to be the $N\times N$ numerical matrix whose only
nonzero entry is a 1 at $(I,J)$.  Furthermore, the tensor product
which appears is that between numerical spaces; specifically, the
$N^2\times N^2$-dimensional matrix $E_{IJ}\otimes E_{KL}$ has entries
\begin{equation}
\matrix{(E_{IJ}\otimes E_{KL})}{ij}{k\ell}=\delta^i_I \delta^j_J
\delta^K_k \delta^L_{\ell}.
\end{equation}
The $D$ matrices take the standard diagonal form
\begin{equation}
D=\sum_I q^{-2\rho_I }E_{II},
\end{equation}
where the $N$ values of $\rho_I$ will be given in each case.  Where
primed indices appear, they are defined to be $I'=N+1-I$.

\noindent 1. $A_n =SL_q (n+1=N)$:
\begin{eqnarray}
R&=&q^{-\frac{1}{N}}(q\sum_I E_{II}\otimes E_{II}+ \sum_{I\neq
J}E_{II}\otimes E_{JJ}+\lambda \sum_{I>J}E_{IJ}\otimes E_{JI}),
\nonumber\\
(\rho_1,\ldots,\rho_N )&=&(0,-1,\ldots,-n).
\end{eqnarray}

\noindent 2. $B_n =SP_q (n=\half N)$:
\begin{eqnarray}
R&=&\sum_I (qE_{II}\otimes E_{II}+q^{-1}E_{I'I'}\otimes E_{II})+
\sum_{I\neq J,J'}E_{II}\otimes E_{JJ}\nonumber\\
&&+\lambda\sum_{I>J}(E_{IJ}\otimes E_{JI}-q^{\rho_I -\rho_J}\epsilon_I
\epsilon_J E_{IJ}\otimes E_{I'J'}),\nonumber\\
(\rho_1,\ldots,\rho_N )&=&(n,n-1,\ldots,1,-1,\ldots,-(n-1),-n),
\nonumber\\
\epsilon_I &=&\left\{ \begin{array}{cl}
+1&I=1,\ldots,n,\\-1&I=(n+1),\ldots,N.
\end{array}\right.
\end{eqnarray}

\noindent 3. $C_n =SO_q (2n+1=N)$:
\begin{eqnarray}
R&=&\sum_{I\neq n+1}(qE_{II}\otimes E_{II}+q^{-1}E_{I'I'}\otimes
E_{II})+E_{n+1,n+1}\otimes E_{n+1,n+1}\nonumber\\
&&+\sum_{I\neq J,J'}E_{II}\otimes E_{JJ}+\lambda\sum_{I>J}(E_{IJ}
\otimes E_{JI}-q^{\rho_I -\rho_J}E_{IJ}\otimes E_{I'J'}),\nonumber\\
(\rho_1,\ldots,\rho_N)&=&(n-\half,n-\frac{3}{2},\ldots,\half,0,
-\half,\ldots,-(n-\frac{3}{2}),-(n-\half)).
\end{eqnarray}

\noindent 4. $D_n =SO_q (2n=N)$:
\begin{eqnarray}
R&=&\sum_I (qE_{II}\otimes E_{II}+q^{-1}E_{I'I'}\otimes E_{II})+
\sum_{I\neq J,J'}E_{II}\otimes E_{JJ}\nonumber\\
&&+\lambda\sum_{I>J}(E_{IJ}\otimes E_{JI}-q^{\rho_I -\rho_J}E_{IJ}
\otimes E_{I'J'}),\nonumber\\
(\rho_1,\ldots,\rho_N )&=&(n-1,n-2\ldots,1,0,0,1,\ldots,-(n-2),
-(n-1)).
\end{eqnarray}

\subsection{Characteristic Equations and Trace Relations}\label{knot}

The matrices $\Rhat$ satisfy certain characteristic equations; in the
context of knot theory, these are viewed as skein relations, relating
particular sequences of strand crossings.  For the deformed Lie
algebras we consider here, these characteristic equations are of two
types: the R-matrices of $GL_q (N)$ and $SL_q (N)$ satisfy the
quadratic equation
\begin{equation}
(\Rhat-rqI)(\Rhat+rq^{-1}I)=0,
\end{equation}
where $r=1$ for $GL_q (N)$ and $r=q^{-\frac{1}{N}}$ for $SL_q (N)$.
The R-matrices for $SO_q (N)$ and $SP_q (\half N)$, on the other hand,
satisfy the cubic equation
\begin{equation}
(\Rhat-qI)(\Rhat+q^{-1}I)(\Rhat-\epsilon q^{\epsilon -N}I)=0,
\end{equation}
where $\epsilon=+1$ for $SO_q (N)$ and $\epsilon=-1$ for $SP_q (\half
N)$.

Now, we have all the trace relations discussed in the first section of
this appendix; since the R- and D-matrices have been given, we are now
in a position to compute the constants $\alpha$ and $\beta$, as well
at the trace of $D$ itself.  Just like the characteristic equations,
these computations split up into two types:  for $GL_q (N)$ and $SL_q
(N)$, we find
\begin{eqnarray}
\alpha=rq^{2N-1},&\beta=rq,&\tr D=\frac{q^{2N}-1}{q^2 -1},
\end{eqnarray}
and for $SO_q (N)$ and $SP_q (\half N)$,
\begin{eqnarray}
\alpha=\beta=q^{N-\epsilon},&&\tr D=\frac{q^{N-\epsilon}-
q^{-(N-\epsilon)}}{q-q^{-1}}+\epsilon.
\end{eqnarray}
($\tr D^{-1}$ can be obtained simply by replacing $q$ with $q^{-1}$ in
the expression for $\tr D$.)

\subsection{Projectors}\label{chap-matrix-proj}

Suppose we have a $N^2 \times N^2$-dimensional numerical R-matrix
which satisfies a characteristic equation of the form
\begin{equation}
\prod_{a=1}^m (\Rhat -\mu_a I)=0,
\end{equation}
where $\{ \mu_a |a=1,\ldots,m \}$ are the $m$ distinct eigenvalues of
$\Rhat$.  We may therefore define $m$ projectors $P_i$ as
\begin{equation}
P_a :=\prod_{b\neq a}\left( \frac{\mu_b I-\Rhat}{\mu_b -\mu_a}\right)
{}.
\end{equation}
These projectors satisfy
\begin{eqnarray}
P_a P_b =\delta_{ab} P_a,&&\Rhat^M =\sum_a \mu_a^M P_a .
\end{eqnarray}

Now we turn our attention to the specific case where the R-matrix is
that of one of the simple quantum Lie groups discussed above; for
$GL_q (N)$ and $SL_q (N)$, there are two projectors $P_+$ and $P_-$,
given by
\begin{eqnarray}
P_+ =\frac{q^{-1}I+r^{-1}\Rhat}{q+q^{-1}},&&P_- =\frac{qI-r^{-1}\Rhat
}{q+q^{-1}},
\end{eqnarray}
corresponding to the symmetrizer and antisymmetrizer respectively.

For $SO_q (N)$ and $SP_q (\half N)$, there are three projectors:
\begin{eqnarray}
P_1 =\frac{(q^{-1}I+\Rhat)(\epsilon q^{\epsilon -N}I-\Rhat)}{(q+
q^{-1})(\epsilon q^{\epsilon -N}-q)},& &P_2 =\frac{(qI-\Rhat)(\epsilon
q^{\epsilon -N}I-\Rhat)}{(q+q^{-1})(\epsilon q^{\epsilon -N}+q^{-1})}
,\nonumber\\
P_0 =\frac{(qI-\Rhat)(q^{-1}I+\Rhat)}{(q-\epsilon q^{\epsilon
-N})(\epsilon q^{\epsilon -N}+q^{-1})}.
\end{eqnarray}
Remember that there exists a metric for both these cases; it is
defined as that matrix $g^{ij}$ which satisfies
\begin{eqnarray}
\matrix{(P_1)}{ij}{k\ell}g^{k\ell}&=&\matrix{(P_2)}{ij}{k\ell}
g^{k\ell}=0,\nonumber\\
\matrix{(P_0)}{ij}{k\ell}g^{k\ell}&=&g^{ij}.
\end{eqnarray}
For $SO_q (N)$, $P_1$ projects out the symmetric traceless subspace,
$P_2$ the antisymmetric subspace, and $P_0$ the trace; for $SP_q
(\half N)$, the metric is now antisymmetric, and $P_1$ and $P_2$
switch roles.  $g^{ij}$ is invertible (with inverse $g_{ij}$), and in
terms of the unit matrices in the fundamental representation takes the
form
\begin{equation}
g=\sum_I q^{-\rho_I}E_{II'},
\end{equation}
so that $\matrix{D}{i}{j}=g^{ik}g_{jk}$.

\chapter{Classical Differential Geometry}\label{chap-classical}

We present here a quick review of the ``Cartan calculus'' on a
classical differentiable manifold; for a more in-depth treatment of
the subject, there are several texts which the reader may refer
him/herself to \cite{Schutz,Warn,Glock}.

\section{The Tangent Space}

Let $M$ be a $C^{\infty}$ N-dimensional real differentiable manifold;
the unital associative algebra $\Fun (M)$ is the space of all
$C^{\infty}$ functions which map $M$ to $\real$, with addition,
multiplication, and unit given by
\begin{eqnarray}
(f+g)(m)=f(m)+g(m),&(fg)(m)=f(m)g(m),&1(m)=1.
\end{eqnarray}
for $m\in M$, $f,g\in\Fun (M)$.

For each subset $S\subseteq M$ we can define an equivalence relation
$\germ$ on $\Fun (M)$: two functions $f,g\in\Fun (M)$ satisfy $f\germ
g$ if there exists an open set $U\subseteq M$ containing $S$ such that
$f|_U =g|_U$.  $f_S$, the {\em germ of} $f$ {\em on} $S$, is defined
as the equivalence class of $f$ under $\germ$, and $\F_S$ is the space
of all such classes (also a unital associative algebra).  We may then
introduce $T_S(M)$, the {\em tangent space of} $M$ {\em on} $S$, as
the vector space over $\real$ consisting of derivations on $\F_S$,
\ie linear maps from $\F_S$ into itself such that for $X_S \in T_S
(M)$ and $f_S , g_S \in\F_S$,
\begin{eqnarray}
X_S (f_S g_S)=X_S (f_S )g_S +f_S X_S (g_S ),&&X_S (1_S )\equiv 0.
\end{eqnarray}
This space is a Lie algebra, with the commutator being defined through
\begin{equation}
\comm{X_S}{Y_S}(f_S ):=X_S\left( Y_S (f_S )\right) -Y_S \left( X_S
(f_S )\right) .
\end{equation}

As an example, take $S$ to be the single point $m$, and let $\{
x^{\mu}|\mu =1,\ldots ,N\}$ be a local coordinate system at $m$.
$T_m(M)$ is a vector space over $\F_m$ with basis $\{ (\partial_{\mu}
)_m \}$, where
\begin{equation}
(\partial_{\mu})_m(f_m):=\left( \partder{f}{x^{\mu}}\right)_m.
\end{equation}
Thus, $T_m(M)$ consists of differential operators on functions at $m$.

\section{The Exterior Derivative}

We now assume the existence of a linear map $\epsilon_S :\F_S
\rightarrow\real$ (for instance, $\epsilon_S(f_S):=\int_S f$ is such a
map).  The existence of such a map allows the definition of a linear
map $\dg_0:\F_S\rightarrow T^*_S(M)$ as
\begin{equation}
\dual{\dg_0(f_S )}{X_S}:=\epsilon_S(X_S (f_S )),\label{class-inprod}
\end{equation}
where $T^*_S(M)$, the {\em cotangent space on} $S$, is the dual of
$T_S(M)$, and $\dual{\,}{\,}$ is the inner product pairing the two
spaces.  Notice that this inner product is degenerate; if either $f_S
\propto 1_S$ or $\epsilon_S\left(X_S(f_S)\right)$ vanishes, then the
inner product above will be zero as well.  However, if we define
$\F^1_S:=\ker\epsilon_S$ and $\F^2_S:=\{ f_S\in\F^1_S|X_S(f_S)\in
\F^1_S\,\,\forall X_S\in T_S(M)\}$, then (\ref{class-inprod}) will
vanish for $f_S\in\F^1_S/\F^2_S$ iff either $f_S\equiv 0$ or
$X_S\equiv 0$.  Thus, when $\dual{\,}{\,}$ is restricted to $\dg_0(
\F^1_S/\F^2_S)\otimes T_S(M)\rightarrow\real$, the inner product is
nondegenerate, and $\dg_0:\F^1_S/\F^2_S\rightarrow T^*_S(M)$ is a
bijective map.

We define the $\F_S$-bimodule $\Gamma_S$ to be the space spanned by
elements of the form $f_S\dg_0(g_S)$ and $\dg_0(f_S)g_S$ with
$f_S,g_S\in\F_S$, where
\begin{eqnarray}
\dual{f_S\dg_0(g_S)}{X_S}:=\epsilon_S(f_SX_S(g_S)),&&\dual{\dg_0
(f_S)g_S}{X_S}:=\epsilon_S(X_S(f_S)g_S).
\end{eqnarray}
These lead to the following important result:
\begin{eqnarray}
\dual{\dg_0(f_Sg_S)}{X_S}&=&\epsilon_S(X_S(f_Sg_S ))\nonumber\\
&=&\epsilon_S(X_S(f_S)g_S+f_SX_S(g_S))\nonumber\\
&=&\dual{\dg_0(f_S)g_S+f_S\dg_0(g_S)}{X_S},
\end{eqnarray}
so $\dg_0$ satisfies the familiar Leibniz rule on functions.  Note
that this implies that $\dg_0(f_S)g_S\equiv\dg_0(f_Sg_S)-f_S\dg_0
(g_S)$, so even though $\Gamma_S$ was defined as a bimodule, it may be
thought of as $\F_S\dg_0(\F_S^1/\F_S^2)$.

We can define the space of $p$-forms $\bigwedge_S^p$ to be the
span over $\real$ of elements of the form $f_S^{(0)}\dg_0(f_S^{(1)})
\wedge\ldots\wedge\dg_0(f_S^{(p)})$, where $f_S^{(0)}\in\F_S$,
$f_S^{(k)}\in\F^1_S/\F^2_S,k=1,\ldots,p$ (so $\bigwedge_S^0\equiv
\F_S$ and $\bigwedge_S^1\equiv\Gamma_S$) and the {\em wedge product}
$\wedge$ is defined such that $\dg_0(f_S)\wedge\dg_0(g_S)\in(T_S^*(M)
)^{\otimes 2}$ satisfies
\begin{equation}
\dual{\dg_0(f_S)\wedge\dg_0(g_S)}{X_S\otimes Y_S}=\dual{\dg_0(f_S)
}{X_S}\dual{\dg_0(g_S )}{Y_S}-(X_S\leftrightarrow Y_S),
\end{equation}
and analogously for forms of higher degree (\ie total
antisymmetrization with respect to the vector fields).  $\dg_p:
\bigwedge_S^p \rightarrow\bigwedge_S^{p+1}$ is then defined as the
linear map
\begin{equation}
\dg_p (f_S^{(0)}\dg_0(f_S^{(1)})\wedge\ldots\wedge\dg_0(f_S^{(p)})):=
\dg_0(f_S^{(0)})\wedge\dg_0(f_S^{(1)})\wedge\ldots\wedge\dg_0
(f_S^{(p)}),
\end{equation}
so $\dg_p$ satisfies $\dg_{p+1}\dg_p\equiv 0$.  (From this point
onward, we will suppress the wedge product, with its presence being
assumed whenever we multiply any $p$-forms together.)

We can now define the {\em exterior algebra on} $S$ to be the direct
sum of each of the spaces of $p$-forms, \ie $\bigwedge_S(M):=
\bigoplus_p\bigwedge_S^p$.  The multiplication between elements of
this space is just the wedge product, so the product of a $p$-form
$\phi_S$ and a $q$-form $\psi_S$ is a $(p+q)$-form.  We may extend
$\dg_0$ to a linear map $\dg$ on all of $\bigwedge_S(M)$ by defining
\begin{eqnarray}
\dg (1_S)&=&0,\nonumber\\
\dg^2 (\phi_S)&=&0,\nonumber\\
\dg (\phi_S\psi_S)&=&\dg (\phi_S)\psi_S+(-1)^p\phi_S\dg (\psi_S),
\end{eqnarray}
where $\phi_S$ is a $p$-form.  This map is called the {\em exterior
derivative on} $\bigwedge_S(M)$, and is a derivation of degree +1 (a
map of degree $d$ maps $\bigwedge_S^p$ into $\bigwedge_S^{p+d}$).

\section{The Inner Derivation}

We now introduce another linear map from $\bigwedge_S(M)$ into itself,
the {\em inner derivation} $\I$, which in effect ``undoes'' the
exterior derivative, in the sense that it maps $p$-forms to
$(p-1)$-forms.  It takes as an argument a vector field $X_S\in
T_S(M)$, and is first defined on 0- and 1-forms:
\begin{eqnarray}
\I_{X_S}(f_S):=0&&\I_{X_S}(f_S\dg_0(g_S)):=f_SX_S(g_S).
\end{eqnarray}
Its action on a $p$-form $\phi_S$ for $p>1$ is given via
\begin{eqnarray}
\lefteqn{\dual{\I_{X_S}(\phi_S)}{Y^{(1)}_S\otimes\ldots\otimes
Y^{(p-1)}_S}=}\nonumber\\
&\dual{\phi_S}{X_S\otimes Y^{(1)}_S\otimes\ldots\otimes Y^{(p-1)}_S}
-\dual{\phi_S}{Y^{(1)}_S\otimes X_S\otimes\ldots\otimes Y^{(p-1)}_S}+
\ldots\nonumber\\
&+(-1)^{p-1}\dual{\phi_S}{Y^{(1)}_S\otimes\ldots\otimes Y^{(p-1)}_S
\otimes X_S},
\end{eqnarray}
so, for example,
\begin{equation}
\I_{X_S}(f_S^{(0)}\dg_0(f^{(1)}_S)\dg_0(f^{(2)}_S))=f_S^{(0)}X_S(
f^{(1)}_S)\dg_0(f^{(2)}_S)-f^{(0)}_S\dg_0(f^{(1)}_S)X_S(f^{(2)}_S).
\end{equation}
{}From the above, it is easily seen that the inner derivation is a
derivation of degree -1.

\section{The Lie Derivative and the Graded Derivation Algebra}

It is now straightforward to define the {\em Lie Derivative} $\lie$ as
a linear map from the exterior algebra into itself which takes
$p$-forms to $p$-forms.  It is defined on a $p$-form $\phi_S$ by
\begin{equation}
\lie_{X_S}(\phi_S):=\dg(\I_{X_S}(\phi_S))+\I_{X_S}(\dg(\phi_S)),
\label{lie-classical}
\end{equation}
and is therefore a derivation of degree 0.  From this definition, it
immediately follows that $\lie_{X_S}$ acts as $X_S$ on 0-forms, and on
1-forms as
\begin{equation}
\lie_{X_S}(f_S\dg_0(g_S))=X_S(f_S)\dg_0(g_S)+f_S\dg_0(X_S(g_S)).
\end{equation}

The utility of introducing the Lie derivative via the definition
(\ref{lie-classical}) lies in the fact that with its inclusion, the
three derivations generate a graded algebra, the {\em Cartan
calculus}, whose (anti)commutation relations are
\begin{eqnarray}
\comm{\dg}{\lie_{X_S}}=0,&\acomm{\dg}{\dg}=0,&\acomm{\dg}{\I_{X_S}}=
\lie_{X_S},\nonumber\\
\comm{\lie_{X_S}}{\I_{Y_S}}=\I_{\comm{X_S}{Y_S}},&
\acomm{\I_{X_S}}{\I_{Y_S}}=0,&\comm{\lie_{X_S}}{\lie_{Y_S}}=
\lie_{\comm{X_S}{Y_S}}.
\end{eqnarray}
These relations, plus the actions of each of the derivations on 0- and
1-forms, completely specify the differential geometry of the manifold
$M$.

\chapter{Differential Calculus on Hopf Algebras}\label{chap-UDC}

We present here a review of the standard way to introduce a
differential calculus on an arbitrary Hopf algebra.  The concepts of
noncommutative geometry underlying this treatment were first examined
by Connes \cite{Connes}; a good treatment of these ideas for
physicists may be found in \cite{Coquereaux}.  The extension of this
general structure to the case of a bicovariant Hopf algebra was dealt
with in great detail by Woronowicz \cite{Woronowicz1}.  (Much of the
material herein will of course be very reminiscent of the classical
treatment given in Appendix \ref{chap-classical}.)

\section{The Universal Differential Calculus}\label{chap-UDC-univ}

Let $\A$ be a unital associative algebra over a field $k$, and $\G$ an
$\A$-bimodule such that there exists a linear map $\dg :\A\rightarrow
\G$ which satisfies the following:
\begin{eqnarray}
\dg (\IA)&=&0,\nonumber \\
\dg (ab)&=&\dg (a) b + a \dg (b),\label{gamma}
\end{eqnarray}
where $\IA$ is the unit in $\A$, and $a,b\in \A$.  Note that the
latter of these conditions implies that $\G$ is the span of elements
of the form $a\dg (b)$.

As an example of this, take $\G\subset\A\otimes\A$ as the kernel of
the multiplication on $\A$, \ie the span of elements of the form
$\sum_i a_i\otimes b_i$ where $\sum_i a_i b_i =0$.  $\G$ is made into
an $\A$-bimodule by defining left and right multiplication by $\A$ to
be $c (\sum_i a_i \otimes b_i)=\sum_i (ca_i )\otimes b_i$ and $(\sum_i
a_i \otimes b_i)c=\sum_i a_i \otimes (b_i c)$, $c \in \A$.  The map
$\dg$ which satisfies all the needed conditions is given by $\dg (a)
:=\IA\otimes a-a \otimes\IA$.

We now introduce the {\em differential envelope associated with} $\A$,
denoted by $\UDE$; it is the algebra which is spanned by elements of
$\A$, together with formal products of elements of $\G$ modulo the
relations (\ref{gamma}), namely, elements of the form $a_0 \dg
(a_1)\dg (a_2)\ldots \dg (a_p)$.  Such elements are called $p$-forms
(\eg 0-forms are elements of $\A$, 1-forms elements of $\G$, {\it
etc.} $\UDE$ is easily seen to be associative and unital (with unit
$1=\IA$); furthermore, $\dg$ can be extended to a linear map $\dg
:\UDE\rightarrow\UDE$ by requiring
\begin{eqnarray}
\dg (1)&=&0,\nonumber \\
\dg^2 (\phi)&=&0,\nonumber \\
\dg (\phi\psi)&=&\dg (\phi) \psi +(-1)^p \phi\dg
(\psi),\label{Leibniz}
\end{eqnarray}
where $\phi,\psi\in\UDE$, $\phi$ a $p$-form.  Thus, $\dg$ maps
$p$-forms to $(p+1)$-forms.  $\dg$ is the {\em exterior derivative} on
$\UDE$, and we call $(\UDE,\dg)$ the {\em universal differential
calculus (UDC) associated with} $\A$.

If $\C$ is a Hopf algebra which coacts on $\A$ as explained in Chapter
\ref{chap-coactions-coacts}, we may extend the right coaction to
$\UDE$ by requiring that $\Delta_{\C}:\UDE\rightarrow\UDE\otimes\C$
satisfies
\begin{eqnarray}
\Delta_{\C}(\dg(a))&=&(\dg\otimes\id)\Delta_{\C}(a),\nonumber\\
\Delta_{\C}(\phi\psi)&=&\Delta_{\C}(\phi)\Delta_{\C}(\psi),
\end{eqnarray}
for $a\in\A$, $\phi ,\psi\in\UDE$.  The left coaction may be
extended analogously.

\subsection{The Universal Differential Calculus of a Hopf
Algebra}\label{chap-UDC-hopf}

Up to this point, we have said nothing about $\A$ being anything more
than a unital associative algebra.  However, if we now allow $\A$ to
be a Hopf algebra, the UDC acquires more structure \cite{Woronowicz1}.
For instance, we may extend the natural coactions $\DA=\AD=\Delta$ of
$\A$ on itself as described above.  Thus, for each $a\in\A$, the {\em
Cartan-Maurer form} $\omega_a:= S(a_{(1)})\dg(a_{(2)})$ is both
left-invariant and right-covariant under these coactions, \ie
\begin{eqnarray}
\AD(\omega_a )=\IA\otimes\omega_a ,&&\DA(\omega_a )=\omega_{a_{(2)}}
\otimes S(a_{(1)})a_{(3)}.
\end{eqnarray}
(The latter of these shows explicitly the appearance of the adjoint
coaction (\ref{adj-coact}).)  Furthermore, its differential has the
particularly nice form
\begin{equation}
\dg (\omega_a )=-\omega_{a_{(1)}}\omega_{a_{(2)}}.\label{om-om}
\end{equation}

A further consequence of $\A$ being a Hopf algebra is that $\A\equiv
k\IA\oplus\K$, where $k\IA$ is shorthand for the subspace of all
elements proportional to the unit, and $\K :=\ker\epsilon$.  $\dg$ and
$\omega$ thus vanish on $k\IA$, so both may be restricted to acting on
$\K$ only.  Since $a\dg(b)=a b_{(1)}\omega_{b_{(2)}}$, any element of
$\G$ may be written in terms of Cartan-Maurer forms with coefficients
in $\A$.  Therefore, we take $\UDE$ to be the span of $p$-forms $a_0
\omega_{a_1 }\ldots\omega_{a_p }$, with $a_0\in\A$ and $a_i \in\K$,
$i=1,\ldots ,p$.

It is possible to impose a *-Hopf algebra structure on the UDC when
$\A$ itself is a *-Hopf algebra \cite{Schlieker}.  This is
accomplished by giving the following compatibility conditions: if
$\phi\in\UDE$ is a $p$-form,then
\begin{eqnarray}
\Delta (\dg\phi)&=&(\dg\otimes\id+(-1)^p \id\otimes\dg )\Delta
(\phi),\nonumber\\
\epsilon (\dg\phi)&=&0\nonumber\\
S(\dg\phi)&=&\dg(S(\phi)),\nonumber\\
\theta(\dg\phi)&=&\dg(\theta(\phi)),
\end{eqnarray}
provided that the multiplication on the tensor product
$\UDE\otimes\UDE$ is $\integer_2$-graded:
\begin{equation}
(\phi_1 \otimes\phi_2 )(\psi_1 \otimes\psi_2 )=(-1)^{pq}
(\phi_1 \psi_1 \otimes\phi_2 \psi_2 ),
\end{equation}
where $\phi_2$ and $\psi_1$ are $p$- and $q$-forms, respectively.  It
is easily shown that these conditions are consistent with both the
defining relations of a *-Hopf algebra and those of the UDC.  An
interesting result of these relations is that the coproduct of a
Cartan-Maurer form is related to the right and left coactions by
\begin{equation}
\Delta(\omega_a )=(\AD +\DA) (\omega_a)
\end{equation}
(where the elements of $\A$ appearing in the right-hand side of this
relation are to be taken as 0-forms in $\UDE$, of course).

\section{General Differential Calculus}\label{chap-UDC-gen}

So far, the only commutation relations we have in $\UDE$ are those
which follow from (\ref{Leibniz}); we assume nothing else.  Here we
review the standard method of introducing nontrivial commutation
relations into the differential envelope which maintains the
covariance properties we have chosen (\eg left-invariance of
$\omega_a$).

Suppose that $\A$ is a Hopf algebra such that there exists a
subalgebra $\M\subset\A$ satisfying
\begin{enumerate}
\item $\M\subseteq\K$,
\item $\M\A\subseteq\M$,
\item $\Ad (\M)\subseteq\M\otimes\A$.
\end{enumerate}
We define the submodule $\N\subseteq\G$ as the space spanned by
1-forms of the form $a\omega_m$, where $a\in\A$ and $m\in\M$.  The
above properties of $\M$ imply properties of $\N$: (1) and (2) give
$\N\A\subseteq\N$, and (3) gives $\DA(\N)\subseteq\N\otimes\A$.  Such
an $\M$ always exists; $\{ 0\}$ and $\K$ both satisfy all three
conditions.

With $\M$ as above, we can construct the $\A$-module $\GN := \G / \N$.
When $\M=\{ 0\}$, and therefore $\N=\{ 0\}$, the only commutation
relations between elements of $\A$ and $\GN$ are those allowed by the
Leibniz rule, and we recover the UDC; when $\M=\K$, $\N=\G$, so
$\GN=\{ 0\}$, and we end up with a trivial differential calculus.
However, if there exists an $\M$ in between these two extreme cases,
then there exist additional commutation relations between elements of
$\GN$, namely those given by $\omega_m\simeq 0$ for $m\in\M$ ($\simeq$
being the equivalence relation in $\GN$).  Furthermore, we find
explicit commutation relations between elements of $\GN$ by using
(\ref{om-om}) and the properties of $\M$, \ie $\omega_{m_{(1)}}
\omega_{m_{(2)}}\simeq 0$.  Therefore, we no longer have a UDC, but
rather a differential envelope with nontrivial commutation relations
which is constructed using $\A$ and $\GN$; we refer to this envelope
as $\GDE$, and the pair $(\GDE,\dg)$ is referred to as the {\em
general differential calculus (GDC) associated with} $\A$ {\em and}
$\M$.

\end{document}